\documentclass[prx,reprint,superscriptaddress]{revtex4-1}

\usepackage[english]{babel}
\usepackage{amsmath,amssymb,bbm,mathrsfs,mathtools,multirow,diagbox,xcolor,%
  graphicx,tabularx,comment,textcomp,amsfonts,dsfont,lettrine,units,comment}
\usepackage[colorlinks,citecolor=blue,urlcolor=blue]{hyperref}

\usepackage[normalem]{ulem}

\AtBeginDocument{%
  \newwrite\bibnotes
  \def\bibnotesext{Notes.bib}
  \immediate\openout\bibnotes=\jobname\bibnotesext
  \immediate\write\bibnotes{@CONTROL{REVTEX41Control}}
  \immediate\write\bibnotes{@CONTROL{%
      apsrev41Control,author="08",editor="1",pages="1",title="0",year="1",eprint=""}}
  \if@filesw
  \immediate\write\@auxout{\string\citation{apsrev41Control}}%
  \fi
}%

\newcommand{\hc}{\mathrm{H.c.}}
\newcommand{\id}{\mathbbm{1}}

\newcommand{\N}{\mathbb{N}}
\newcommand{\Z}{\mathbb{Z}}
\newcommand{\C}{\mathbb{C}}
\newcommand{\R}{\mathbb{R}}

\newcommand{\e}{\mathrm{e}}
\newcommand{\imag}{\mathrm{i}}
\newcommand{\ket}[1]{\lvert #1 \rangle}
\newcommand{\bra}[1]{\langle #1 \rvert}

\newcommand{\Rtop}{\mathcal{R}_{\mathrm{top}}}
\newcommand{\Rtr}{\mathcal{R}_{\mathrm{tr}}}
\newcommand{\Gedge}{\mathcal{G}_{\mathrm{edge}}}
\newcommand{\Gtr}{\mathcal{G}_{\mathrm{tr}}}
\newcommand{\Iedge}{\mathcal{I}_{\mathrm{edge}}}
\newcommand{\Itr}{\mathcal{I}_{\mathrm{tr}}}

\makeatletter
\newcommand*{\transpose}{%
  {\mathpalette\@transpose{}}%
}
\newcommand*{\@transpose}[2]{%
  \raisebox{\depth}{$\m@th#1\intercal$}%
}
\makeatother

\newcommand{\abs}[1]{\left\lvert #1 \right\rvert}

\newcommand{\tr}{\mathop{\mathrm{tr}}}
\newcommand{\pf}{\mathop{\mathrm{pf}}}
\newcommand{\sgn}{\mathop{\mathrm{sgn}}}
\newcommand{\atanh}{\mathop{\mathrm{atanh}}}
\renewcommand{\Re}{\mathop{\mathrm{Re}}}
\renewcommand{\Im}{\mathop{\mathrm{Im}}}

\makeatletter
\DeclareFontFamily{OMX}{MnSymbolE}{}
\DeclareSymbolFont{MnLargeSymbols}{OMX}{MnSymbolE}{m}{n}
\SetSymbolFont{MnLargeSymbols}{bold}{OMX}{MnSymbolE}{b}{n}
\DeclareFontShape{OMX}{MnSymbolE}{m}{n}{
    <-6>  MnSymbolE5
   <6-7>  MnSymbolE6
   <7-8>  MnSymbolE7
   <8-9>  MnSymbolE8
   <9-10> MnSymbolE9
  <10-12> MnSymbolE10
  <12->   MnSymbolE12
}{}
\DeclareFontShape{OMX}{MnSymbolE}{b}{n}{
    <-6>  MnSymbolE-Bold5
   <6-7>  MnSymbolE-Bold6
   <7-8>  MnSymbolE-Bold7
   <8-9>  MnSymbolE-Bold8
   <9-10> MnSymbolE-Bold9
  <10-12> MnSymbolE-Bold10
  <12->   MnSymbolE-Bold12
}{}

\let\llangle\@undefined
\let\rrangle\@undefined
\DeclareMathDelimiter{\llangle}{\mathopen}%
                     {MnLargeSymbols}{'164}{MnLargeSymbols}{'164}
\DeclareMathDelimiter{\rrangle}{\mathclose}%
                     {MnLargeSymbols}{'171}{MnLargeSymbols}{'171}
\makeatother

\newcommand{\kket}[1]{\lvert #1 \rrangle}
\newcommand{\bbra}[1]{\llangle #1 \rvert}
\newcommand{\bbraket}[1]{\llangle #1 \rrangle}

\newcommand{\done}[1]{}

\begin{document}

\title{Entanglement Spectrum Crossings Reveal non-Hermitian Dynamical Topology}

\author{Sharareh Sayyad}

\email{sharareh.sayyad@neel.cnrs.fr}

\affiliation{University Grenoble Alpes, CNRS, Grenoble INP, Institut N\'{e}el,
  38000 Grenoble, France}
 
\author{Jinlong Yu}

\affiliation{Center for Quantum Physics, University of Innsbruck, 6020
  Innsbruck, Austria}

\affiliation{Institute for Quantum Optics and Quantum Information of the
  Austrian Academy of Sciences, 6020 Innsbruck, Austria}
 
\author{Adolfo G. Grushin}

\affiliation{University Grenoble Alpes, CNRS, Grenoble INP, Institut N\'{e}el,
  38000 Grenoble, France}

\author{Lukas M. Sieberer }

\email{lukas.sieberer@uibk.ac.at}

\affiliation{Center for Quantum Physics, University of Innsbruck, 6020
  Innsbruck, Austria}

\affiliation{Institute for Quantum Optics and Quantum
  Information of the Austrian Academy of Sciences, 6020 Innsbruck, Austria}

\affiliation{Institute for Theoretical Physics, University of Innsbruck, 6020
  Innsbruck, Austria}

\date{\today}

\begin{abstract}
  The development of non-Hermitian topological band theory has led to
  observations of novel topological phenomena in effectively classical, driven
  and dissipative systems. However, for open quantum many-body systems, the
  absence of a ground state presents a challenge to define robust signatures of
  non-Hermitian topology. We show that such a signature is provided by crossings
  in the time evolution of the entanglement spectrum. These crossings occur in
  quenches from the trivial to the topological phase of a driven-dissipative Kitaev
  chain that is described by a Markovian quantum master equation in Lindblad
  form. At the topological transition, which can be crossed either by changing
  parameters of the Hamiltonian of the system or by increasing the strength of
  dissipation, the time scale at which the first entanglement spectrum crossing
  occurs diverges with a dynamical critical exponent of $\epsilon = 1/2$. We
  corroborate these numerical findings with an exact analytical solution of the
  quench dynamics for a spectrally flat postquench Liouvillian. This exact
  solution suggests an interpretation of the topological quench dynamics as a
  fermion parity pump. Our work thus reveals signatures of non-Hermitian
  topology which are unique to quantum many-body systems and cannot be emulated
  in classical simulators of non-Hermitian wave physics.
\end{abstract}

\maketitle

\begin{figure}
  \centering
  \includegraphics[width=\linewidth]{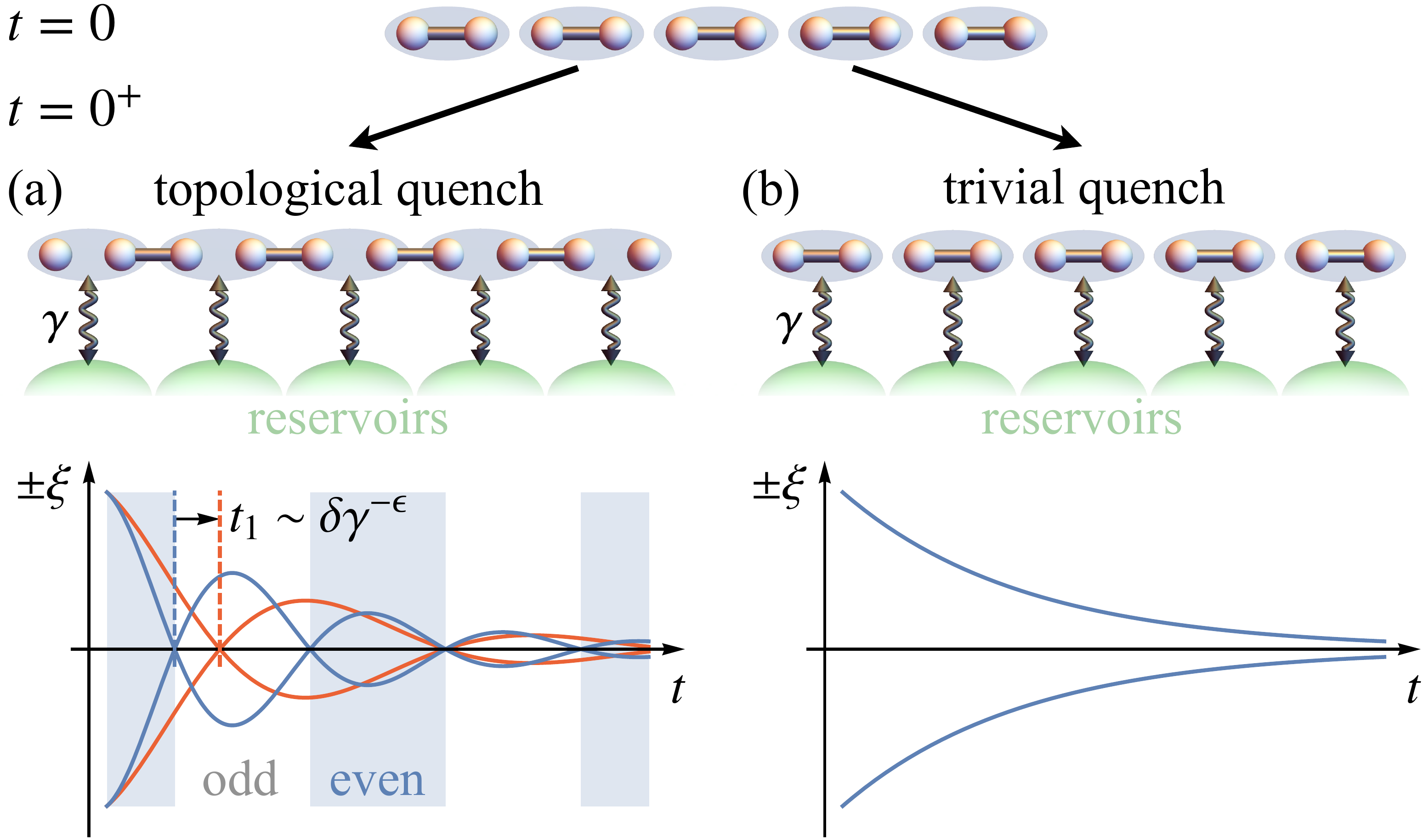}
  \caption{Quench dynamics of a driven-dissipative Kitaev chain. At $t = 0$, the
    system is prepared in the topologically trivial ground state of the Kitaev
    chain, where Majorana fermions (spheres) pair locally (gray bonds) at each
    site of the chain (blue ellipses). Then, at $t = 0^+$, the system is driven
    out of its initial state by an abrupt change of the chemical
    potential. Simultaneously, the chain is connected to Markovian reservoirs,
    which induce dissipation at a rate $\gamma$. (a) For quenches to the
    topological phase of the postquench Liouvillian, the entanglement spectrum
    exhibits oscillatory decay with repeated zero crossings (only a single pair
    of entanglement eigenvalues $\pm \xi$ is shown as blue lines in the
    figure). At each zero crossing, the fermion parity of the entanglement
    ground state switches between even and odd. Red lines show the evolution
    of $\pm \xi$ for a quench closer to the topological phase boundary at
    $\gamma_{\mathrm{c}}$. For
    $\delta \gamma = \abs{\gamma - \gamma_{\mathrm{c}}} \to 0$, the time $t_1$
    at which the first crossing occurs diverges
    as $t_1 \sim \delta \gamma^{-\epsilon}$ with a universal critical exponent
    $\epsilon = 1/2$. (b) For quenches to a trivial phase, the entanglement
    eigenvalues $\pm \xi$ decay without crossing zero.}
  \label{fig:schematic}
\end{figure}

\section{Introduction}
\label{sec:introduction}

Non-Hermitian topological band theory~\cite{Esaki2011, 
  Malzard2015, Leykam2017, 
  Shen2018, Gong2018, Yao2018a,
  Kunst2018, Kawabata2019, Zhou2019,
  Liu2019,
  Yokomizo2019, Borgnia2020, Ashida2020}
allows to derive a topological classification of the complex spectra of
quadratic Liouvillians~\cite{Lieu2019, Dangel2018, Lieu2019a, Song2019,
  Minganti2019, Liu2020, Okuma2020a}, which describe the dynamics of
noninteracting, driven and open quantum many-body systems. For complex spectra,
the very notions of a ground state, as well as of filled and empty bands, are
ill-defined. In the topological band theory for isolated and noninteracting
fermionic systems, however, these concepts are central to predict robust
topological features, such as the quantized response of a system in its ground
state~\cite{Hasan2010,Qi2011,Chui2016}. Moreover, the existence of a ground
state allows to generalize topological phases to interacting many-body
systems~\cite{Chui2016}. These observations raise the question that motivates
our work: Which robust signatures of non-Hermitian topology survive in driven
and open quantum many-body systems? Here we propose that the time evolution of
the entanglement spectrum can provide such a signature.

The entanglement spectrum is a powerful tool to detect the presence of
topological order in the pure ground state $\ket{\psi_0}$ of the Hamiltonian $H_0$ of an isolated
Hermitian system~\cite{Li2008, Pollmann2010, Fidkowski2010, Thomale2010,
  Cirac2011, Chandran2011, Qi2012, Schuch2013, Regnault2017}. For a subsystem
$A$ and its complement $A^{\mathrm{c}}$, the many-body entanglement spectrum is
defined as the spectrum of the reduced density matrix
$\rho_A = {\tr}_{A^{\mathrm{c}}}(\rho_0)$, where
$\rho_0 = \ket{\psi_0} \bra{\psi_0}$ is the density matrix that is associated
with the pure state $\ket{\psi_0}$. The bulk-edge correspondence of the
entanglement spectrum~\cite{Chandran2011, Qi2012} establishes a one-to-one
correspondence between the low-energy edge states of the physical Hamiltonian at an open boundary of
subsystem $A$, and the entanglement eigenstates of $\rho_A$ that have the highest
entanglement eigenvalues, i.e., that are most entangled with the rest of the
system. This correspondence holds both for noninteracting and interacting
systems, and has even been applied to systems which are driven out of
equilibrium: After a sudden quench from $H_0$ to $H$, the time-evolved state $\ket{\psi(t)} = \e^{-\imag H t} \ket{\psi_0}$ is the ground state of the time-dependent parent Hamiltonian $H_{\mathrm{p}}(t) = \e^{-\imag H t} H_0 \e^{\imag H t}$, and the entanglement spectrum of $\ket{\psi(t)}$ reflects the topology of $H_{\mathrm{p}}(t)$~\cite{Gong2017a, Chang2018, Lu2019, Pastori2020}.

Despite the strikingly universal applicability of the entanglement spectrum as a
diagnostic for topological order, it is unclear whether the non-Hermitian
topology of the complex spectrum of a Liouvillian $\hat{\mathcal{L}}$ leaves signatures in the entanglement spectrum of an associated quantum state. For example, the time-evolved density matrix
$\rho(t) = \e^{\hat{\mathcal{L}} t} \rho_0$ is, in general, a mixed
state. Therefore, it cannot be interpreted as the ground state of any putative
non-Hermitian parent Hamiltonian $H(t)$. Intriguingly, establishing the
connection between the entanglement spectrum of $\rho(t)$ and the topology of
$\hat{\mathcal{L}}$ could open up the possibility to characterize the
non-Hermitian topology of both noninteracting and interacting open quantum
many-body systems.

In this paper, we show that non-Hermitian dynamical topology of driven and open
quantum many-body systems is revealed by crossings in the time evolution of the
entanglement spectrum after a quench. We consider a Kitaev chain, which is
subject to Markovian gain and dissipation~\cite{Klett2017, Menke2017, Li2018, Kawabata2018a, VanCaspel2019} as illustrated in
Fig.~\ref{fig:schematic}. The topology of the corresponding Liouvillian is
characterized by a non-Hermitian winding number $W$~\cite{Esaki2011, Kawabata2019}. A
quench from a trivial to a topological phase leads to stable crossings in the
entanglement spectrum of the time-evolved density matrix $\rho(t)$ as shown in Fig.~\ref{fig:schematic}(a). Instead, as
illustrated in Fig.~\ref{fig:schematic}(b), the dynamics after a quench to a
trivial phase does not feature such crossings. We show that topological
entanglement spectrum crossings can be traced back to the reversal of the
fermion parity in individual entanglement eigenstates. Thus, our work
establishes that entanglement spectrum crossings and the concomitant pumping of
fermion parity are robust signatures of non-Hermitian dynamical topology of the
Liouvillian.

Surprisingly, we obtain these results for systems which heat up to a featureless
infinite-temperature state at late times. By contrast, much previous work
focused on inducing nontrivial topological features in the steady state
$\rho_{\mathrm{ss}} = \lim_{t \to \infty} \rho(t)$, which can be achieved by
properly designing purely dissipative dynamics~\cite{Diehl2011, Bardyn2013,
  Goldstein2018, Shavit2020, Tonielli2019}. Adopting this point of view, one is
led to the conclusion that the systems we study are trivial. Instead, the
perspective taken in this work, which is inspired by non-Hermitian band
theory~\cite{Esaki2011,
  Malzard2015, Leykam2017,
  Shen2018, Gong2018, Yao2018a,
  Kunst2018, Kawabata2019, Zhou2019,
  Liu2019,
  Yokomizo2019, Borgnia2020, Ashida2020}, focuses on
the topological properties of the Liouvillian. As explained above, these properties
are revealed in the dynamics, i.e., in the approach to the steady state. The
time evolution of the entanglement spectrum reveals a sharp dynamical
topological phase transition with associated dynamical criticality, reflected in
a divergence of the time scale on which entanglement spectrum crossings occur as
illustrated in Fig.~\ref{fig:schematic}(a). These features highlight the stark
contrast between the two complementary approaches to the topology of driven and
open quantum many-body systems: the one based on
$\rho_{\mathrm{ss}}$~\cite{Diehl2011, Bardyn2013, Goldstein2018, Shavit2020,
  Tonielli2019}, and the one that we propose, which is based on the entanglement
spectrum.

We finally emphasize that the characterization of topology in terms of
entanglement spectrum crossings is unique to quantum many-body systems and
cannot be emulated by photonic~\cite{Zeuner2015, Poli2015, Weimann2017,
  Xiao2017, Zhao2018, Zhou2018, Parto2018, Xiao2020, Pickup2020, Wang2020,
  Fedorova2020} or mechanical~\cite{Zhu2018} simulators of non-Hermitian wave
physics. Moreover, by using the entanglement spectrum as a tool to diagnose
non-Hermitian topology, our work paves the way to characterize non-Hermitian
dynamical topology also of interacting open quantum many-body systems.

This paper is organized as follows: We present our main results in
Sec.~\ref{sec:summary-main-results}. The model and formalism on which our study
is based are introduced in Sec.~\ref{sec:models}, and we summarize the
non-Hermitian topological band theory of the driven-dissipative Kitaev chain in
Sec.~\ref{sec:non-hermitian-band-theory}. The relationship between
entanglement spectrum crossings and non-Hermitian topology of the Liouvillian
depends on whether the jump operators, which describe the coupling of the system
to Markovian reservoirs, are Hermitian or
non-Hermitian. Section~\ref{sec:sys-hermi-jump-operators} is devoted to systems
with Hermitian jump operators. In particular, we discuss the time evolution of
entanglement spectra, the connection between entanglement spectrum crossings and
parity pumping, and dynamical criticality. We consider systems non-Hermitian
jump operators in Sec.~\ref{sec:sys-non-hermitian-jump-ops}. Our conclusions are
presented in Sec.~\ref{sec:outlook}, together with an outlook on future
perspectives. Technical details are deferred to Appendices~\ref{sec:quasiloc-jump-operator}--\ref{sec:ferm-parity-entanglement-gs}.

\section{Summary of main results }
\label{sec:summary-main-results}

\begin{figure}
  \centering
  \includegraphics[width=\linewidth]{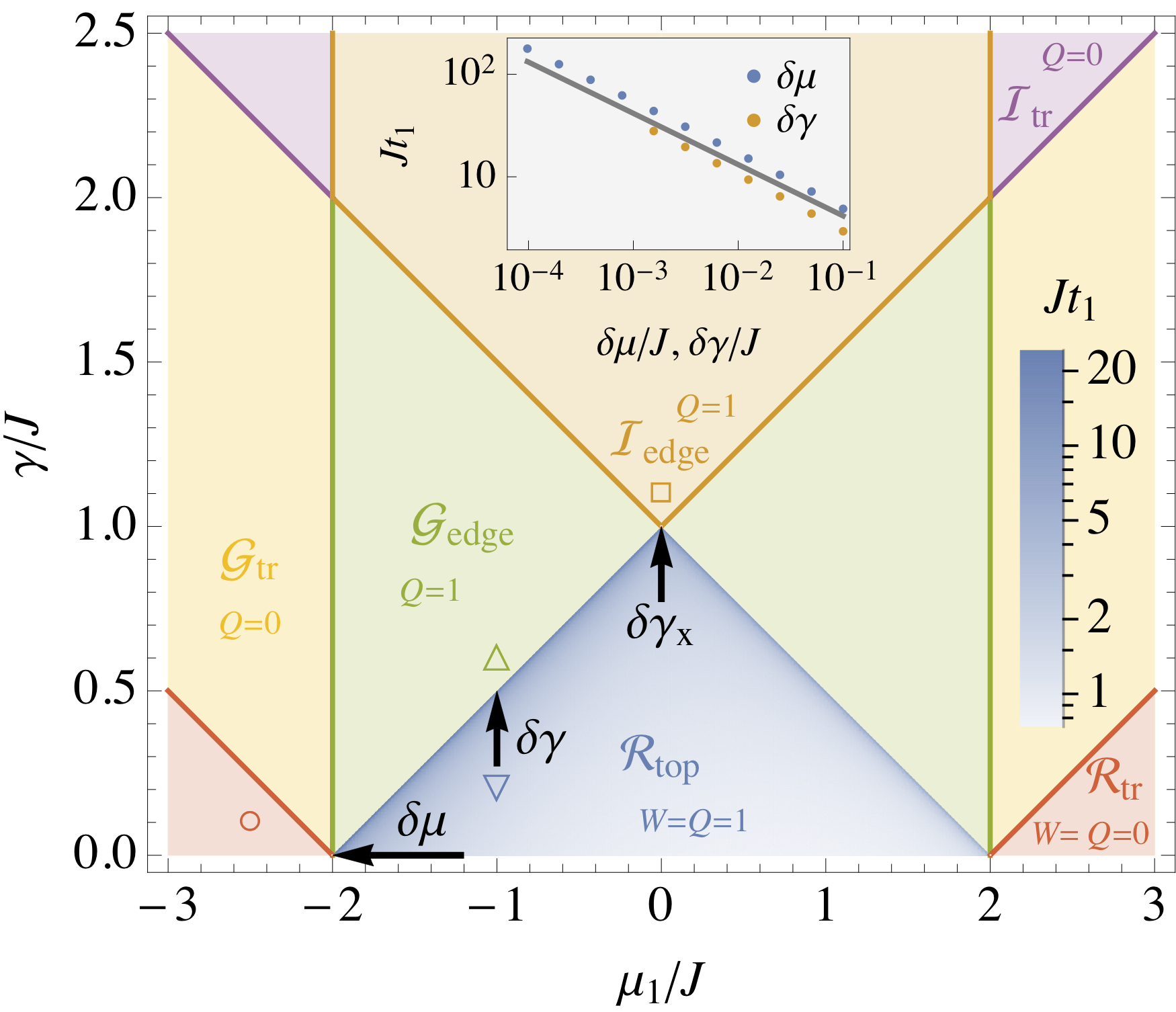}
  \caption{Dynamical topological phase diagram and dynamical criticality of
    entanglement spectrum crossings. The non-Hermitian winding number $W$ (for
    phases with real line gaps) and the global Berry phase $Q$ distinguish six
    dynamical phases: the real-line gapped phases with nontrivial ($\Rtop$, with
    $W=Q=1$) and trivial ($\Rtr$, with $W=Q=0$) topology, the imaginary-line
    gapped phases with ($\Iedge$, with $Q=1$) and without ($\Itr$, with $Q = 0$)
    edge states, and the gapless phases with ($\Gedge$, with $Q=1$) and without
    ($\Gtr$, with $Q=0$) edge states. Entanglement spectrum crossings occur for
    quenches from the trivial phase of the isolated Kitaev chain to $\Rtop$. The
    logarithmic color scale in $\Rtop$ encodes the time $t_1$ of the first
    entanglement spectrum zero crossing for a quench with $\mu_0 = -3 J$ and in
    a system of size $N = 100$. Inset: $t_1$ diverges at the boundary of $\Rtop$
    with a critical exponent of $\epsilon = 1/2$. The data shown corresponds to
    approaching the phase boundary along the black arrows labeled by
    $\delta \mu$ and $\delta \gamma$ in the main panel:
    $\mu_1 = - 2 J + \delta \mu$, $\gamma = 0$ for the blue dots, and
    $\mu_1 = - J$, $\gamma = 0.5 J - \delta \gamma$ for the orange dots. In the
    former case, longer evolution times can be reached due to the absence of
    decay. The system size is $N = 1000$. For comparison, the gray solid line
    shows a square-root singularity.}
  \label{fig:phase-diagram-LDM}
\end{figure}

We consider the following model and quench protocol: At time $t = 0$, a Kitaev
chain is prepared in its ground state $\ket{\psi_0}$ with a chemical potential
$\mu_0$, which is then changed abruptly to $\mu_1$ at $t = 0^+$.
Simultaneously, the system is coupled with strength $\gamma$ to local Markovian
reservoirs. The resultant driven-dissipative dynamics is described by a quantum
master equation for the density matrix $\rho$ of the system,
$\mathrm{d} \rho/\mathrm{d} t = \hat{\mathcal{L}} \rho$. Here, the non-Hermitian
Liouville superoperator $\hat{\mathcal{L}}$ comprises both coherent dynamics
induced by the Hamiltonian $H$ of the system, and the coupling to reservoirs
through quantum jump operators. We study the time evolution of the entanglement
spectrum, i.e., the spectrum of the reduced density matrix
$\rho_A(t) = {\tr}_{A^{\mathrm{c}}}(\rho(t))$, where
$\rho(t) = \e^{\hat{\mathcal{L}} t} \rho_0$ is the state of the system at time
$t$, and $\rho_0 = \ket{\psi_0} \bra{\psi_0}$ is the initial state.

We first discuss systems with Hermitian jump operators, for which the steady
state is the trivial fully mixed infinite-temperature state
$\rho_{\mathrm{ss}} = \rho_{\infty} = \id/D$, where $D$ is the Hilbert-space
dimension. For such systems, our main results are as follows:

\textit{Entanglement spectrum crossings reveal non-Hermitian dynamical
  topology.}  We find that entanglement spectrum crossings occur exclusively for
quenches from the ground state of the Kitaev chain in the topologically trivial
phase to the nontrivial phase of the postquench Liouvillian $\hat{\mathcal{L}}$,
which is designated as $\Rtop$ in the dynamical topological phase diagram in
Fig.~\ref{fig:phase-diagram-LDM}. In $\Rtop$, the two bands of complex
eigenvalues of the Liouvillian are separated by a real line gap, and each band
can be characterized by a non-Hermitian winding number that takes the value
$W = 1$~\cite{Esaki2011, Kawabata2019}. Entanglement spectrum crossings do not
occur for quenches to $\Rtr$ with $W = 0$, the gapless phases $\Gedge$ and
$\Gtr$, and the phases $\Iedge$ and $\Itr$ with an imaginary line
gap. Therefore, the presence of entanglement spectrum crossings is directly
connected to the nontrivial non-Hermitian topology of the Kitaev chain, and can
be used to map out the sharp phase boundary of the topological phase $\Rtop$
both at zero and nonzero strengths of dissipation.

A second topological invariant, the global Berry phase $Q$ that pertains to the
full Liouvillian and not to individual bands~\cite{Liang2013, Lieu2018}, can be
defined even in the gapless phases $\Gedge$ and $\Gtr$. The global Berry phase
takes the value $Q = 1$ and predicts the existence of edge modes of the
Liouvillian in the phases $\Gedge$, $\Rtop$, and $\Iedge$, while $Q = 0$ in the
phases $\Gtr$, $\Rtr$, and $\Itr$ without edge modes. Thus, the existence of
edge modes can be understood as a global property of the
Liouvillian~\cite{Lieu2018}, while, as we show in our work, the presence of
entanglement spectrum crossings requires nontrivial non-Hermitian topology of
individual bands.

\textit{Dynamical criticality of the entanglement spectrum at the topological
  phase boundary is governed by a universal critical exponent $\epsilon = 1/2$.}
At the phase boundary of $\Rtop$, the dynamics of the entanglement spectrum
exhibits critical behavior. This is shown in Fig.~\ref{fig:phase-diagram-LDM},
where the color scale within $\Rtop$ encodes the time $t_1$ of the first
entanglement spectrum crossing, which diverges at the phase boundary as
$t_1\sim\delta\gamma^{-\epsilon}$ and $t_1\sim\delta\mu^{-\epsilon}$. Through an
exact analytical solution of the quench dynamics for $\mu_1 = 0$, where the
spectrum of the Liouvillian is flat, we show that the entanglement spectrum
dynamical critical exponent takes the exact value $\epsilon = 1/2$ when the
phase boundary of $\Rtop$ is approached along the direction that is indicated
with a black arrow labeled by $\delta \gamma_{\mathrm{x}}$. Numerical evidence
indicates that this value is universal along the phase boundary: In the inset of
Fig.~\ref{fig:phase-diagram-LDM}, we show data which corresponds to approaching
the phase boundaries along the directions $\delta \mu$ and $\delta \gamma$, and
is compatible with $\epsilon = 1/2$. This value is also compatible with
numerical results of Ref.~\cite{Torlai2014}, which studied entanglement dynamics
in the transverse-field Ising model, to which the isolated Hermitian Kitaev chain maps under a
Jordan-Wigner transformation. The universality of the value $\epsilon = 1/2$ is
corroborated further by results for a generalized driven-dissipative Kitaev
chain with quasilocal jump operators, which we present in
Appendix~\ref{sec:quasiloc-jump-operator}. These findings highlight the phase
boundary of $\Rtop$ as a sharp dynamical topological transition at finite
dissipation. The system evolves towards the same trivial infinite-temperature
state for all values of $\mu_1$ and $\gamma$, in contrast to the clear dynamical
distinction between the different phases that we find and show in
Fig.~\ref{fig:phase-diagram-LDM}.

\textit{Entanglement spectrum crossings in the driven-dissipative Kitaev chain
  can be interpreted as a fermion parity pump.} Our exact solution of the quench
dynamics for $\mu_1 = 0$ shows that with each crossing of entanglement
eigenvalues, the fermion parity of many-body entanglement eigenstates is
reversed and, therefore, the topological quench dynamics can be interpreted as a
fermion parity pump. Thus, the exact solution generalizes earlier
results~\cite{Lu2019} for the isolated Kitaev chain to nonzero Markovian
dissipation and non-Hermitian topology. The reversal of the fermion parity is a
robust signature which might be easier to detect than entanglement spectrum
crossings, in particular, in interacting systems.

\textit{The non-Hermitian topological phase transition is associated with a
  distinct type of dynamical criticality: many-body critical damping.} The
dynamical criticality of the entanglement spectrum at the phase boundary between
$\Rtop$ and $\Gedge$ raises the question: Do ``simple'' two-time correlation
functions also become critical at this dynamical topological transition?

To address this question, we focus on the retarded response function in the
steady state. In the non-Hermitian models we consider here, the retarded
response function exhibits two distinct types of dynamical criticality. At the
phase boundary between $\Rtop$ and $\Gedge$, the response function shows what we
dub \textit{many-body critical damping}. We choose this term in analogy to the
paradigmatic classical damped harmonic oscillator, in which \textit{critical
  damping} marks the onset of the overdamped regime. Similarly, as we increase
the rate of dissipation $\gamma$ to approach the phase boundary, the period of
oscillations of the response function diverges. The divergence is governed by
the same value of the exponent of $1/2$ as for the time scale $t_1$ of the first
crossing in the entanglement spectrum. However, the oscillatory behavior and
associated zero crossings of the response function occur in both the topological
and the trivial phases with real line gaps, $\Rtop$ and $\Rtr$,
respectively. Therefore, these zero crossings are not related to topology.

The second and more conventional type of dynamical criticality we find is
\textit{critical relaxation,} which occurs at the boundaries between the gapless
phases $\Gedge$ and $\Gtr$, and the imaginary-line gapped phases $\Iedge$ and
$\Itr$. Upon approaching these phase boundaries, the relaxation time scales of
both the response function and the entanglement spectrum diverge, leading to
power-law decay exactly on the critical line. In stark contrast, the response
function and the entanglement spectrum decay exponentially on the line of
many-body critical damping, i.e., the phase boundary between $\Rtop$ and
$\Gedge$. As for a damped harmonic oscillator, the decay rate is largest at the
onset of the overdamped regime.

The phase boundaries between $\Rtop$ and $\Gedge$ as well as $\Gedge$ and $\Gtr$
coincide for $\gamma \to 0$, when the model becomes Hermitian. Therefore, the
occurrence of two distinct types of dynamical critical behavior---many-body
critical damping and critical relaxation---is unique to non-Hermitian dynamics.

The above results and, in particular, the direct connection between
entanglement spectrum crossings and nontrivial values of the non-Hermitian
winding number $W$, pertain to systems with Hermitian jump operators. Our key
finding for systems with non-Hermitian jump operators reads as follows:

\textit{To reveal the non-Hermitian topology of general quadratic Liouvillians
  through entanglement spectrum crossings, the jump operators have to be
  Hermitianized.} While systems with Hermitian jump operators evolve towards a
trivial infinite-temperature steady state, for non-Hermitian jump operators, the
steady state is in general nontrivial. In particular, a well-defined value of
the fermion parity of the entanglement ground state of the steady state can lead
to the occurrence of entanglement spectrum crossings which are not related to
the topology of the Liouvillian. However, the direct connection between
non-Hermitian topology of $\hat{\mathcal{L}}$ and entanglement spectrum
crossings can be restored by means of Hermitianization of the jump
operators. Hermitianization is a continuous deformation of the jump operators to
make them Hermitian, while the gap and the symmetries of the Liouvillian are
preserved~\cite{Lieu2019}.

\section{Model}
\label{sec:models}

The Kitaev chain of length $N$ is described by the Hamiltonian~\cite{Kitaev2001}
\begin{equation}
  \label{eq:H-Kitaev}
  H = \sum_{i = 1}^N \left[ -J c^{\dagger}_i c_{i+1} + \Delta c_i c_{i+1}
    + \hc - \mu \left( c^{\dagger}_i c_i -\frac{1}{2}
    \right) \right],
\end{equation}
where $c_i^{\dagger}$ and $c_i$ are, respectively, creation and annihilation
operators for spinless Dirac fermions on lattice site $i$. These operators obey
the cannonical anticommutation relations
$\{ c_i, c_j \} =\{ c_i^{\dagger}, c_j^{\dagger} \} = 0$ and
$\{ c_i, c_j^{\dagger} \} = \delta_{i,j}$. We set $c_{N + 1} = 0$ and
$c_{N + 1} = c_1$ for a system with open and periodic boundary conditions,
respectively. The parameters in the Hamiltonian~\eqref{eq:H-Kitaev} are the
hopping matrix element $J$, the $p$-wave pairing amplitude $\Delta$, and the
chemical potential $\mu$. We measure energy and time in units of $J$ and
$\hbar/J$, respectively, and we set $\hbar = 1$.

In what follows, we choose $\Delta = J \in \R_{> 0}$. For this choice, the
Kitaev chain belongs to the Altland-Zirnbauer class BDI~\cite{Altland1997},
which, in one spatial dimension, is characterized by a winding number
$W\in \Z$~\cite{Chui2016}. The ground-state phase diagram of the Kitaev chain
comprises two topologically distinct phases. In the topologically nontrivial
phase, which is realized for $\abs{\mu} < 2 J$, an infinite bulk system is
characterized by a nonvanishing winding number $W = 1$. By virtue of the
bulk-edge correspondence, a finite system with open boundary conditions features
two Majorana zero-energy modes that are localized at the ends of the
chain. These edge modes are absent in the trivial phase at $\abs{\mu} > 2 J$ for
which $W = 0$.

\subsection{Driven-dissipative Kitaev chain}
\label{sec:DDKC}

We are interested in the topological properties of a
driven-dissipative Kitaev chain which is realized by connecting the isolated
Kitaev chain, Eq.~\eqref{eq:H-Kitaev}, to
Markovian baths. For driven-dissipative systems, the notion of a ground state is
ill-defined, and, therefore, we focus on dynamical signatures of topology. Thus,
we proceed to specify the dynamics of the driven-dissipative Kitaev chain, and
the types of Markovian baths we consider in this paper.

The time evolution of the density matrix $\rho$ of the driven-dissipative Kitaev
chain is described by a master equation in Lindblad form,
\begin{equation}
  \label{eq:master-equation}
  \frac{\mathrm{d} \rho}{\mathrm{d} t} = \hat{\mathcal{L}} \rho = - \imag [H, \rho]
  + \hat{\mathcal{D}} \rho.
\end{equation}
In this equation, the Liouvillian superoperator $\hat{\mathcal{L}}$ comprises
both unitary dynamics generated by the Hamiltonian~\eqref{eq:H-Kitaev}, and
Markovian drive and dissipation, which are incorporated in the dissipator
$\hat{\mathcal{D}}$. The latter is given by
\begin{equation}
  \label{eq:dissipator}
  \hat{\mathcal{D}} \rho = \sum_{i=1 }^N \left( 2 L_i \rho L_i^{\dagger} - \{
    L_i^{\dagger} L_i, \rho \} \right),
\end{equation}
where the operators $L_i$, termed Lindblad or quantum jump operators, describe
the coupling between the system and its environment.

In this paper, we study the dynamics of noninteracting open quantum many-body
systems that are described by quadratic Liouvillians. Such Liouvillians are
defined in terms of a quadratic Hamiltonian, together with jump operators $L_i$
that are linear combinations of the fermionic field operators $c_i$ and
$c_i^{\dagger}$. For simplicity, we focus in most of our work on the following
purely local jump operators~\cite{VanCaspel2019}:
\begin{equation}
  \label{eq:jump-operator-LDM}
  L_i^{\mathrm{loc}} = \sqrt{\gamma_{\mathrm{l}}} c_i +
  \sqrt{\gamma_{\mathrm{g}}} c_i^{\dagger}.
\end{equation}
For $\gamma_{\mathrm{g}} = 0$, these jump operators describe loss of particles
at a rate $\gamma_{\mathrm{l}} > 0$, and for $\gamma_{\mathrm{l}} = 0$, particle
gain with rate $\gamma_{\mathrm{g}} > 0$. The jump operators are Hermitian for
balanced gain and loss rates,
$\gamma_{\mathrm{l}} = \gamma_{\mathrm{g}} = \gamma$. As anticipated in
Sec.~\ref{sec:summary-main-results}, nontrivial topology of the Liouvillian
$\hat{\mathcal{L}}$ is reflected in quench dynamics of the driven-dissipative
Kitaev chain only if the jump operators $L_i$ are Hermitian.

The fact that the jump operators~\eqref{eq:jump-operator-LDM} act locally on a
single lattice site simplifies the problem and enables, as described in
Sec.~\ref{sec:sys-hermi-jump-operators}, an exact analytical solution of the
master equation~\eqref{eq:master-equation} in a certain limit. To demonstrate
the validity of our findings beyond the case of purely local jump operators, we
also consider quasilocal and Hermitian jump operators which are given by
\begin{equation}
  \label{eq:jump-operator-QLDM}
  L_i^{\mathrm{qloc}} = \sqrt{\frac{\kappa}{2}} \left( c_i + c_i^{\dagger} - c_{i + 1} - c_{i +
      1}^{\dagger} \right),
\end{equation}
where $\kappa$ is the rate of Markovian dissipation. In the main text, we focus
on the driven-dissipative Kitaev chain with the purely local jump operators
defined in Eq.~\eqref{eq:jump-operator-LDM}. Our results for the quasilocal jump
operators~\eqref{eq:jump-operator-QLDM} are summarized in
Appendix~\ref{sec:quasiloc-jump-operator}.

\subsection{Third quantization}
\label{sec:third-quantization}

Quadratic Liouvillians are equivalent to a class of non-Hermitian Hamiltonians,
and, therefore, the topology of quadratic Liouvillians can be studied by using
tools and concepts of non-Hermitian topological band theory~\cite{Lieu2019}. The
equivalence between quadratic Liouvillians and non-Hermitian Hamiltonians is
established within the formalism of third quantization~\cite{Prosen2008,
  Prosen2010}, which we summarize in the following.

The formalism of third quantization builds on the representation of a quadratic
Hamiltonian $H$ and linear jump operators $L_i$ in terms of $2 N$ Majorana
fermions, which are defined as
\begin{equation}\label{eq:w_from_c}
  w_{2 i - 1} = c_i + c_i^{\dagger}, \qquad
  w_{2 i} = \imag \left( c_i - c_i^{\dagger} \right).
\end{equation}
These definitions lead to the following general expressions for the Hamiltonian
and the jump operators:
\begin{equation}
  \label{eq:H-L-Majorana}
  H = \frac{\imag}{4} \sum_{i, j = 1}^{2 N} w_i A_{i, j} w_j, \qquad L_i =
  \sum_{j = 1}^{2 N} l_{i, j} w_j,
\end{equation}
where $A$ is an antisymmetric real $2 N \times 2 N$ matrix, and $l$ is a complex
matrix with dimensions $N \times 2 N$. For Hermitian jump operators, the matrix
$l$ is real.

More generally, any operator can be expanded in a basis of products of Majorana
operators. Each element $w_1^{s_1} \dotsb w_{2 N} ^{s_{2 N}}$ of this basis is
fully determined by the $2 N$ occupation numbers $s_i = 0, 1$, which appear as
exponents. Therefore, we use the shorthand notation
$\kket{w_{\mathbf{s}}} = \kket{w_1^{s_1} \dotsb w_{2 N} ^{s_{2 N}}}$, where
$\mathbf{s} = \left( s_1, \dotsc, s_{2 N} \right)$, and the double angular
brackets emphasize the interpretation of $\kket{w_{\mathbf{n}}}$ as a superket,
i.e., a vector in the Fock space of operators on which superoperators such as
$\hat{\mathcal{L}}$ act.

We denote fermionic annihilation and creation superoperators over this Fock
space by $\hat{c}_i$ and $\hat{c}_i^{\dagger}$, respectively, where the hat
symbol distinguishes superoperators from ordinary operators. The fermionic
superoperators are defined by their action on basis vectors,
$\hat{c}_i \kket{w_{\mathbf{n}}} = \delta_{n_i, 1} \kket{w_i w_{\mathbf{n}}}$
and
$\hat{c}_i^{\dagger} \kket{w_{\mathbf{n}}} = \delta_{n_i, 0} \kket{w_i
  w_{\mathbf{n}}}$.
These definitions make use of the following property of Majorana operators: The
product $w_i w_i^{s_i}$, which is contained in $w_i w_{\mathbf{s}}$, is given by
$w_i w_i^{s_i} = w_i$ for $s_i = 0$, and by $w_i w_i^{s_i} = w_i^2 = w_i^0 = 1$ for
$s_i = 1$. Further, these definitions imply the canonical anticommutation
relations
$\{ \hat{c}_i, \hat{c}_j \} = \{ \hat{c}^{\dagger}_i, \hat{c}^{\dagger}_j \} =
0$ and $\{ \hat{c}_i, \hat{c}_j^{\dagger} \} = \delta_{i, j}.$

In terms of the fermionic superoperators, the Liouvillian in
Eq.~\eqref{eq:master-equation} can be written as
\begin{equation}
  \label{eq:Liouvillian-fermionic-superops} 
  \hat{\mathcal{L}} = - \frac{1}{2} \sum_{i, j = 1}^{2 N}
  \left( \hat{c}_i^{\dagger}, \hat{c}_i \right)
  \begin{pmatrix}
    X_{i, j} & \imag Y_{i, j} \\
    0 & - X_{i, j}^{\transpose}
  \end{pmatrix}
  \begin{pmatrix}
    \hat{c}_j \\ \hat{c}_j^{\dagger}
  \end{pmatrix}
  - \frac{1}{2} \tr(X) \hat{1},
\end{equation}
where $\transpose$ denotes the transpose of a matrix, and $\hat{1}$ is the
identity superoperator. The real matrices $X$ and $Y$ are defined in terms of
the Hamiltonian matrix $A$ in Eq.~\eqref{eq:H-L-Majorana} and the bath matrix
$M = l^{\transpose} l^{*}$ as
\begin{equation}
  \label{eq:X-Y}
  X = - A + 4 M_R, \qquad Y = - 8 M_I,
\end{equation}
where $M_R = \Re(M)$ and $M_I = \Im(M)$. By definition, the bath matrix $M$ is
Hermitian. Consequently, its real and imaginary parts, $M_R$ and $M_I$, are
symmetric and antisymmetric, respectively.

In view of the representation~\eqref{eq:Liouvillian-fermionic-superops} of the
Liouvillion as a quadratic form in the fermionic superoperators $\hat{c}_i$ and
$\hat{c}_i^{\dagger}$, the master equation~\eqref{eq:master-equation}, rewritten
as
\begin{equation}
  \label{eq:effective-Schrodinger-equation}
  \imag \frac{\mathrm{d}}{\mathrm{d} t} \kket{\rho} = \imag \hat{\mathcal{L}} \kket{\rho},
\end{equation}
can be interpreted as a Schr\"odinger equation for superkets $\kket{\rho}$, with
an effective quadratic Hamiltonian $\imag \hat{\mathcal{L}}$. For an isolated
system with antisymmetric $X = -A$ and $Y = 0$, this effective Hamiltonian is
Hermitian. In contrast, the dynamics of an open system is described by a
non-Hermitian effective Hamiltonian $\imag \hat{\mathcal{L}}$.

Topological properties of the non-Hermitian superoperator
$\imag \hat{\mathcal{L}}$ are encoded in its spectrum and eigenmodes. The
Liouvillian can be diagonalized as
\begin{align}
  \label{eq:Liouvillian-diagonal}
  \imag \hat{\mathcal{L}} = \sum_{i=1}^{2 N} \lambda_i \hat{b}_i' \hat{b}_i,
\end{align} 
where $\hat{b}_{i}$ and $\hat{b}'_{i}$ are fermionic superoperators which obey
the anticommutation relations $\{ \hat{b}_i, \hat{b}_j' \} = \delta_{i, j}$ and
$\{ \hat{b}_i, \hat{b}_j \} = \{ \hat{b}_i', \hat{b}_j' \} = 0$. Further,
$\lambda_i$ are eigenvalues of the matrix
$Z = - \imag X^{\transpose}$~\cite{Lieu2019}. The fact that the spectrum of
$\hat{\mathcal{L}}$ is determined by the matrix $X$ alone follows from the
block-triangular form of Eq.~\eqref{eq:Liouvillian-fermionic-superops}. In
contrast, the matrix $Y$ affects the shape of the eigenmodes of
$\hat{\mathcal{L}}$ as well as the steady state of the master
equation~\eqref{eq:master-equation}. The steady state is the eigenstate of the
Liouvillian with eigenvalue zero,
$\hat{\mathcal{L}} \kket{\rho_{\mathrm{ss}}} = 0$. According to
Eq.~\eqref{eq:Liouvillian-diagonal}, it is the right vaccum state of the modes
$\hat{b}_i$.

Now that we have set out the formal framework in which the equivalence between
quadratic Liouvillians and non-Hermitian Hamiltonians is established, we proceed
to analyze the non-Hermitian topological band structure of the matrix $Z$, which
determines the complex spectrum of the Liouvillian.

\section{Non-Hermitian band theory for the driven-dissipative Kitaev chain}
\label{sec:non-hermitian-band-theory}

Since the matrix $Z$ is derived from a Liouvillian that generates the dynamics
of the density matrix $\rho$ according to Eq.~\eqref{eq:master-equation}, the
spectrum of $Z$ obeys certain conditions. First, to ensure Hermiticity of the
time-evolved density matrix $\rho(t) = \e^{\hat{\mathcal{L}} t} \rho_0$, for
each eigenvalue $\lambda_i$, also its anti-complex-conjugate $- \lambda_i^{*}$
must be an eigenvalue of $Z$~\cite{Lieu2019}. Second, the fact that the
evolution of the density matrix approaches a steady state implies that
$\Im(\lambda_i) \leq 0$, where the negative imaginary parts $- \Im(\lambda_i)$
determine the rate of relaxation to the steady state.

For a translationally invariant system, the spectrum of the Liouvillian can be
determined analytically. In particular, for the driven-dissipative Kitaev chain,
which is described by the Hamiltonian~\eqref{eq:H-Kitaev} and the local jump
operators~\eqref{eq:jump-operator-LDM}, the matrix $Z$ can be decomposed into
translationally invariant $2 \times 2$ blocks,
\begin{equation}
  z_{i - j} =
  \begin{pmatrix}
    Z_{2 i - 1, 2 j - 1} & Z_{2 i - 1, 2 j} \\
    Z_{2 i, 2 j - 1} & Z_{2 i, 2 j}
  \end{pmatrix}.
\end{equation}
Therefore, the matrix $Z$ is block-diagonal in momentum space. Its spectrum
consists of two bands, which are given by the eigenvalues of the Fourier
transform~\cite{Rivas2013}
\begin{equation}
  \label{eq:z-k}
  z_k = \sum_{i = 1}^N \e^{-\imag k i} z_i = - \imag \left( \gamma_{\mathrm{l}}
    + \gamma_{\mathrm{g}} \right) \id + \mathbf{z}_k \cdot \boldsymbol{\sigma},
\end{equation}
where $k \in (-\pi, \pi]$ is the quasimomentum. Further, $\boldsymbol{\sigma} = \left( \sigma_x, \sigma_y, \sigma_z \right)$ is a
vector of Pauli matrices, and
\begin{equation}
  \mathbf{z}_k = \left( 2 \Delta \sin(k), 2 J \cos(k) + \mu, -\imag 2
    \sqrt{\gamma_{\mathrm{l}} \gamma_{\mathrm{g}}} \right).
\end{equation}
The two bands of complex eigenvalues of $\mathbf{z}_k$ are given by
\begin{align}
  \label{eq:lambda-k-LDM}
  \lambda_{\pm, k} =
  & - \imag \left( \gamma_{\mathrm{l}} + \gamma_{\mathrm{g}} \right)
  \nonumber \\
  & \pm \sqrt{4 \Delta^2 \sin(k)^2 + \left( 2 J \cos(k) + \mu \right)^2
    - 4 \gamma_{\mathrm{l}} \gamma_{\mathrm{g}}}.
\end{align}
Before we proceed to analyze the band structure, we note that for $J = \Delta$,
the shifted matrix
\begin{equation}
  \label{eq:z-k-tilde}
  \begin{split}
    \tilde{z}_k & = z_k + \imag \left( \gamma_{\mathrm{l}} + \gamma_{\mathrm{g}}
    \right) \id \\ & =
  \begin{pmatrix}
    - \imag 2 \sqrt{\gamma_{\mathrm{l}} \gamma_{\mathrm{g}}} & 
    - \imag \left( 2 J \e^{\imag k} + \mu \right) \\ \imag \left(
      2 J \e^{-\imag k} + \mu \right) & \imag 2 \sqrt{\gamma_{\mathrm{l}}
      \gamma_{\mathrm{g}}}
  \end{pmatrix},
  \end{split}
\end{equation}
is unitarily equivalent to a non-Hermitian Bloch Hamiltonian studied in
Refs.~\cite{Esaki2011, Liang2013, Lieu2018}. The addition of the constant matrix
$\imag \left( \gamma_{\mathrm{l}} + \gamma_{\mathrm{g}} \right) \id$ corresponds
to a shift of the origin in the complex plane of the eigenvalues
$\lambda_{\pm, k}$. While such a shift does not affect the topology of $z_k$, it
has severe consequences for the dynamics. In particular, as we discuss in
Sec.~\ref{sec:sys-hermi-jump-operators}, dynamical critical behavior is induced
when the spectrum of $z_k$ includes the eigenvalue $\lambda = 0$ in the complex
$\lambda$-plane.

\begin{figure}
  \centering
  \includegraphics[width=\linewidth]{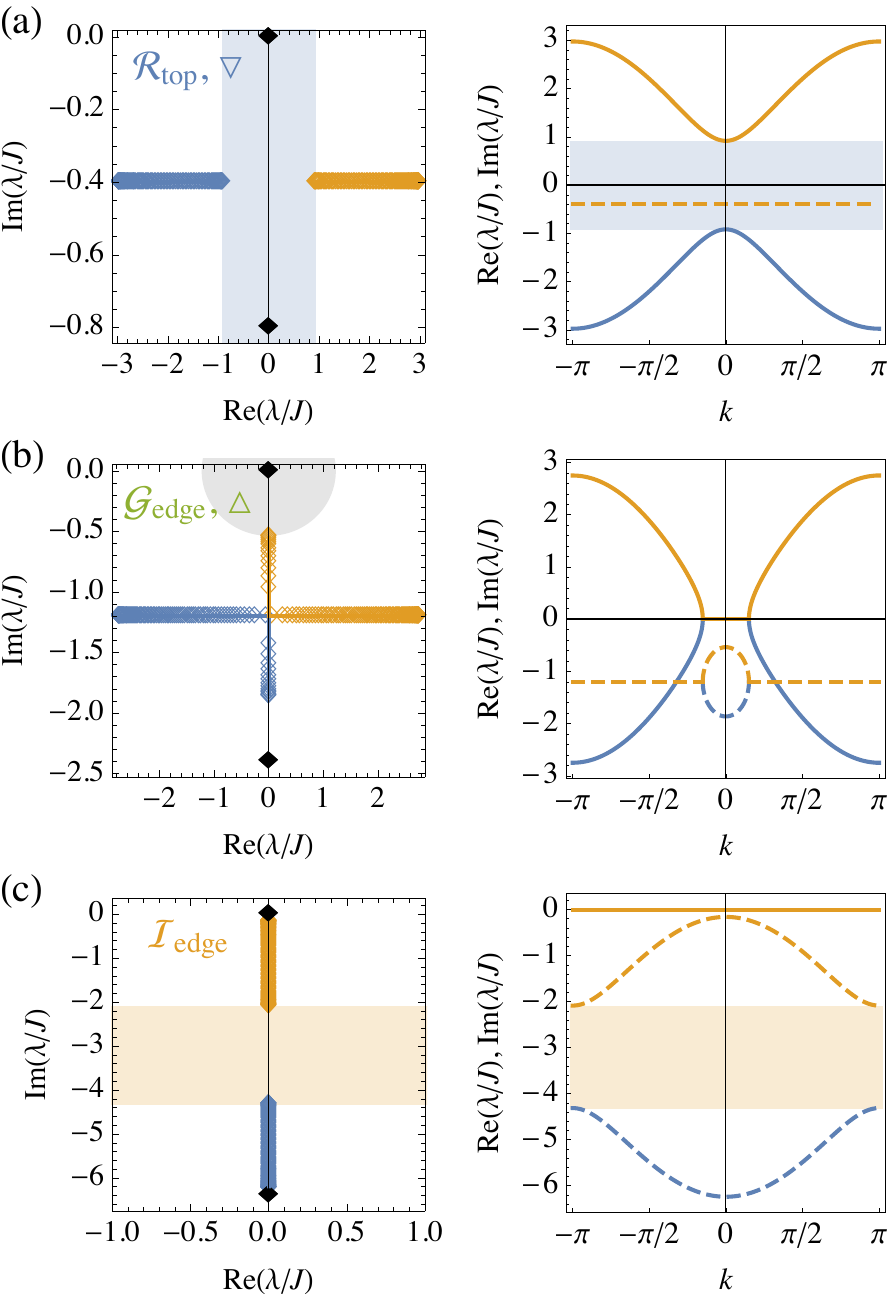}
  \caption{Liouvillian spectra for $J = \Delta = - \mu$ and
    $\gamma_{\mathrm{l}} = \gamma_{\mathrm{g}} = \gamma$. (a) Real line gap in
    the phase $\Rtop$ at weak dissipation with $\gamma = 0.2 J$. Left panel:
    Spectra for an infinite bulk system and a system with $N = 100$ with open
    boundary conditions. The blue and orange solid lines that correspond to the
    bulk system are hidden behind the diamond symbols that indicate the
    eigenvalues of the finite system. Black diamonds at approximately
    $\lambda_L = - \imag 4 \gamma$ and $\lambda_R = 0$ are edge modes that are
    exponentially localized at the left and right edges,
    respectively~\cite{VanCaspel2019}. All other modes are delocalized. Right
    panel: Real and imaginary parts of $\lambda_{\pm, k}$ in
    Eq.~\eqref{eq:lambda-k-LDM} are shown as solid and dashed lines,
    respectively. (b) Gapless spectrum in the phase $\Gedge$ at
    $\gamma = 0.6 J$. The gray disk indicates the point gap at $\lambda = 0$. (c) Imaginary line gap in
    $\Iedge$ at strong dissipation with $\gamma = 1.6 J$.}
  \label{fig:Liouvillian-spectrum-LDM}
\end{figure}

Since the eigenvalues of the Liouvillian are complex, gapped and gapless phases
can be defined by addressing two distinct questions. First, we can ask whether
the two bands $\lambda_{\pm, k}$ are separated in the complex plane. This
question leads to the definition of real and imaginary line
gaps~\cite{Kawabata2019} as specified below. Further, for a complex spectrum,
the point at which two bands touch in a gapless phase does not have to be
$\lambda = 0$.  Therefore, we can introduce another notion of a spectral gap by
asking: Is $\lambda = 0$ contained in the spectrum of $z_k$? If this is not the
case, the system has a point gap at $\lambda = 0$~\cite{Kawabata2019}. As we
detail below and in Sec.~\ref{sec:sys-hermi-jump-operators}, closings of the two
types of gaps, i.e., real or imaginary line gaps that separate the bands in the
complex plane, or the point gap that separates both bands from $\lambda = 0$,
are associated with different types of topological transitions and leave
distinct signatures, in particular, in the dynamics of the system.

We first discuss real and imaginary line gaps. If the expression under the
square root in Eq.~\eqref{eq:lambda-k-LDM} is positive for all values of $k$,
the bands are separated along the real axis by a real line gap, which is
indicated by the blue shading in Fig.~\ref{fig:Liouvillian-spectrum-LDM}(a). In the
phase diagram in Fig.~\ref{fig:phase-diagram-LDM}, a real line gap occurs in
the phases $\Rtop$ and $\Rtr$, in which the rates of dissipation are bounded by
\begin{equation}  
  \label{eq:phase-boundary-R}
  2 \sqrt{\gamma_{\mathrm{l}} \gamma_{\mathrm{g}}} < \min \{ \abs{2 J + \mu},
  \abs{2 J - \mu} \}.
\end{equation}
If the rates $\gamma_{\mathrm{l}}$ and $\gamma_{\mathrm{g}}$ are increased
beyond this bound, the system enters one of the two gapless phases, which are
denoted by $\Gedge$ and $\Gtr$. In these phases, the spectrum features
exceptional points at momenta $\pm k_{*}$, at which the expression under the
square root in Eq.~\eqref{eq:lambda-k-LDM} vanishes and thus the eigenvalues and
the corresponding eigenvectors of $z_k$ coincide. This is exemplified in
Fig.~\ref{fig:Liouvillian-spectrum-LDM}(b). Finally, at strong dissipation, when
the expression under the square root in Eq.~\eqref{eq:lambda-k-LDM} is negative
for all values of $k$, the bands are separated by an imaginary line gap as
indicated by the orange shading in
Fig.~\ref{fig:Liouvillian-spectrum-LDM}(c). The corresponding phases $\Iedge$ and $\Itr$ are
delineated by
\begin{equation}  
  2 \sqrt{\gamma_{\mathrm{l}} \gamma_{\mathrm{g}}} > \max \{ \abs{2 J + \mu},
  \abs{2 J - \mu} \}.
\end{equation}

A non-Hermitian winding number $W$ for a single band can be defined only in the
gapped phases, where the bands are separated, and the definition depends on the
symmetries of the model. The driven-dissipative Kitaev chain belongs to class
$\mathrm{BDI}^{\dagger}$ in the nomenclature of Ref.~\cite{Kawabata2019}, and
obeys the following forms of time-reversal, particle-hole, and chiral symmetries:
\begin{equation}
  \label{eq:symmetries}
  \begin{aligned}
    & \text{TRS$^{\dagger}$:} & \sigma_z z_k^{\transpose}
                                \sigma_z & = z_{-k}, \\    
    & \text{PHS$^{\dagger}$:} & z_k^{*} & = -z_{-k}, \\    
    & \text{CS:} & \sigma_z z_k^{\dagger} \sigma_z & = - z_k.
  \end{aligned}
\end{equation}
For $\gamma_{\mathrm{l}} = \gamma_{\mathrm{g}} = 0$, when $z_k = z_k^{\dagger}$,
the class $\mathrm{BDI}^{\dagger}$ reduces to the class BDI of the isolated Hermitian Kitaev chain~\eqref{eq:H-Kitaev} with $\Delta \in \R$. The definition of the winding
number $W$ for a single band of the isolated system can be generalized to the
entire phases $\Rtop$ and $\Rtr$ with a real line gap~\cite{Kawabata2019}. In
Ref.~\cite{Esaki2011}, the calculation of $W$ for the shifted matrix
$\tilde{z}_k$ in Eq.~\eqref{eq:z-k-tilde} is carried out, and leads to the
result $W = 1$ in the topological phase $\Rtop$ with $\abs{\mu} < 2 J$, and
$W = 0$ in the trivial phase $\Rtr$ with $\abs{\mu} > 2 J$. As we show in
Sec.~\ref{sec:sys-hermi-jump-operators}, non-Hermitian topology as characterized
by the winding number $W = 1$ for an individual band is reflected in the
presence of entanglement spectrum crossings in quench dynamics. In contrast to
the phases $\Rtop$ and $\Rtr$ with a real line gap, the topology of individual
bands of the Liouvillian is always trivial in the phases $\Iedge$ and $\Itr$ with an
imaginary line gap~\cite{Kawabata2019}. Consistently, we observe no entanglement spectrum crossings for quenches into $\Iedge$ and $\Itr$.

A topological classification of the gapless phases can be given in terms of a
global invariant, which is not a property of individual bands, but rather of the
entire Liouvillian. In particular, the global Berry phase $Q$ defined and
calculated for the shifted matrix $\tilde{z}_k$ in Eq.~\eqref{eq:z-k-tilde} in
Ref.~\cite{Liang2013}, takes the values $Q = 1$ for $\abs{\mu} < 2 J$ and
$Q = 0$ for $\abs{\mu} > 2 J$. We stress that the global Berry phase $Q$ is
defined in the entire phase diagram in Fig.~\ref{fig:phase-diagram-LDM},
including the gapless phases $\Gedge$ and $\Gtr$. In contrast, the
classification in terms of the non-Hermitian winding number $W$ applies only to
the phases with real line gaps.

In the driven-dissipative Kitaev chain, a nonzero value of $Q$ implies the
existence of edge modes~\cite{Lieu2018} even within the gapless phase $\Gedge$
and the phase $\Iedge$ with an imaginary line gap~\cite{VanCaspel2019}, and not
only in $\Rtop$. Edge modes are shown in the spectra in
Fig.~\ref{fig:Liouvillian-spectrum-LDM} as black diamonds. A topological
transition occurs at the boundaries between $\Gedge$ and $\Gtr$ as well as
$\Iedge$ and $\Itr$, in which edge modes disappear because the value of $Q$
changes from $Q = 1$ to $Q = 0$. In crossing this transition, and when
$\gamma_{\mathrm{l}} = \gamma_{\mathrm{g}}$, the second type of band gap defined
above, i.e., the point gap at $\lambda = 0$, closes. In
Fig.~\ref{fig:Liouvillian-spectrum-LDM}(b), the point gap is indicated with a
gray disk. Closings of this gap induce critical power-law relaxation in the
dynamics of both the entanglement spectrum and response functions, as we discuss
in Sec.~\ref{sec:sys-hermi-jump-operators} below.

\section{Systems with Hermitian jump operators}
\label{sec:sys-hermi-jump-operators}

The definition of a non-Hermitian winding number $W$ for complex bands
$\lambda_{\pm, k}$, which we discussed in
Sec.~\ref{sec:non-hermitian-band-theory}, is purely formal, and raises the
question about its physical implications. In the following, we show that the
topological transition in the spectrum of the Liouvillian, in which $W$ changes
from $W = 1$ in $\Rtop$ to $W = 0$ or undefined in $\Rtr$ and $\Gedge$,
respectively, is associated with a dynamical phase transition in the time
evolution of the entanglement spectrum.

We begin by introducing the tools which are required to track the time evolution
of the entanglement spectrum, and which have been used to establish the
entanglement spectrum bulk-edge correspondence in quench dynamics of isolated
Hermitian systems~\cite{Gong2017a, Chang2018, Lu2019}. Then, as one of the key
results of our work, we generalize the entanglement spectrum bulk-edge
correspondence to the driven-dissipative Kitaev chain.

In this section, we consider the effective Schr\"odinger
equation~\eqref{eq:effective-Schrodinger-equation} with a non-Hermitian matrix
$Z = - \imag X^{\transpose}$ and $Y = 0$, which corresponds to Hermitian jump
operators, i.e., $\gamma_{\mathrm{l}} = \gamma_{\mathrm{g}} = \gamma$ in
Eq.~\eqref{eq:jump-operator-LDM}. We discuss the general case with
$\gamma_{\mathrm{l}} \neq \gamma_{\mathrm{g}}$ and thus $Y \neq 0$ in
Sec.~\ref{sec:sys-non-hermitian-jump-ops}.

A distinguishing feature of systems with Hermitian jump operators is that they
generically heat up to a fully mixed infinite-temperature state,
$\rho_{\mathrm{ss}} = \rho_{\infty} = \id/D$, where $D$ is the Hilbert-space
dimension. This can be seen by noting that for $L_i = L_i^{\dagger}$, the
dissipator in Eq.~\eqref{eq:dissipator} takes the form of a double commutator,
\begin{equation}  
  \hat{\mathcal{D}} \rho = - \sum_{i=1 }^N [L_i, [L_i, \rho]].
\end{equation}
Therefore, the trivial state $\rho_{\infty} = \id/D$, which satisfies
$[H, \rho_{\infty}] = [L_i, \rho_{\infty}] = 0$, is indeed a steady
state. Typically, this steady state is also unique. Degeneracy of the steady state requires the existence of at least two zero modes of the matrix $Z$~\cite{VanCaspel2019, Lieu2019a}. For the driven-dissipative Kitaev chain with Hermitian jump operators, the spectra in Fig.~\ref{fig:Liouvillian-spectrum-LDM} display only a single zero mode which is localized at the right edge of the chain. Thus, $\rho_{\mathrm{ss}} = \rho_{\infty}$ is the unique steady state.

\subsection{Topology from quench dynamics}
\label{sec:quench-dynamics}

To explore the dynamical topology of the driven-dissipative Kitaev chain, we
study the time evolution of a pure state $\ket{\psi_0}$, the ground state of the
isolated Kitaev chain, after sudden changes of parameters. The quench protocol is
illustrated in Fig.~\ref{fig:schematic}: At time $t=0^+$, the chemical potential
is varied abruptly from its prequench value $\mu_0$ to a postquench value
$\mu_1$, and at the same time the system is connected to Markovian baths.

Due to the coupling to Markovian baths, described by the term
$\hat{\mathcal{D}} \rho$ in Eq.~\eqref{eq:master-equation}, the pure initial
state $\rho_0 = \ket{\psi_0} \bra{\psi_0}$ evolves into a mixed state
$\rho(t) = \e^{\hat{\mathcal{L}} t} \rho_0$. Since $H$ and $\hat{\mathcal{L}}$ are
quadratic, both the initial ground state and the time-evolved density matrix
$\rho(t)$ are Gaussian states, i.e., they can be written as the exponential of a quadratic form in the fermionic operators $c_i$ and $c_i^{\dagger}$. Gaussian states are fully determined by the real and antisymmetric covariance matrix~\cite{Campbell2015, Sergey2012},
\begin{equation}
  \label{eq:cov-mat}
  \Gamma_{i, j}(t) = \frac{\imag}{2} \tr([w_i, w_j] \rho(t)).
\end{equation}
The covariance matrix obeys the equation of motion
$\mathrm{d} \Gamma/\mathrm{d} t = - X \Gamma - \Gamma X^{\transpose} - Y$, which
is solved by~\cite{Prosen2011}
\begin{equation}
  \label{eq:cov-mat-explicit-solution}
  \Gamma(t) = \e^{- X t} \left( \Gamma(0) - \int_0^t
    \mathrm{d} t' \, \e^{X t'} Y \e^{X^{\transpose} t'} \right)  \e^{-
    X^{\transpose} t}.
\end{equation}
Here, $\Gamma(0)$ is the covariance matrix of the initial state. As stated
above, we assume that the system is prepared in the ground state of the isolated
Kitaev chain, and the corresponding covariance matrix is elaborated in
Appendix~\ref{sec:cov-mat-gs}. For $Y = 0$, the second term in
Eq.~\eqref{eq:cov-mat-explicit-solution} vanishes, and $\Gamma(t)$ decays to
zero in the steady state $\rho_{\mathrm{ss}} = \rho_{\infty} = \id/D$.

In the following, we assume that the number of lattice sites $N$ is even, and we
consider an equal bipartition of the system into its left and right halves. There are in total $2 N$ sites of Majorana fermions. We
denote the set of site indices of Majorana fermions which belong to the left
half of the system by $A = \{ 1, \dotsc, N  \}$. Then, the reduced covariance
matrix of subsystem $A$ is defined as
$\Gamma_{i, j}^A = \left( \Gamma_{i, j} \right)_{i,j \in A}$. Its eigenvalues,
which come in pairs $\pm \xi_i$ of bounded values $0 \leq \xi_i \leq 1$ with
$i = 1, \dotsc, N/2$, form the single-particle entanglement
spectrum~\cite{Cheong2004, Peschel2009, Kraus2009}.

For Gaussian states, the equality $\xi_i = \tanh(\varepsilon_i/2)$ relates the
entanglement eigenvalues $\xi_i$ to the single-particle eigenvalues
$\varepsilon_i$ of the entanglement Hamiltonian $H_A$, which is determined by
the reduced density matrix as
$\rho_A = {\tr}_{A^{\mathrm{c}}}(\rho) = \e^{- H_A}/Z_A$ where
$Z_A = {\tr}_A(\e^{-H_A})$~\cite{Cheong2004, Peschel2009, Kraus2009}. The
many-body entanglement spectrum, which is defined as the spectrum of $\rho_A$,
is thus given by~\cite{Fidkowski2010}
\begin{equation}
  \label{eq:mb-ent-spec}
  \Xi_{\mathbf{e}} = \prod_{i = 1}^{N/2} \frac{1}{2} \left[ 1 + \left( -1
    \right)^{e_i} \xi_i \right],
\end{equation}
where $\mathbf{e} = \left( e_1, \dotsc, e_{N/2} \right)$, and where the numbers
$e_i$ are to be understood as occupations of single-particle entanglement levels
that take the values $e_i = 0,1$. Therefore, for Gaussian states, the
single-particle and many-body entanglement spectra contain the same
information. In particular, as detailed below, Eq.~\eqref{eq:mb-ent-spec}
implies that a zero crossing in the single-particle entanglement spectrum leads
to the simultaneous crossing of pairs of many-body entanglement eigenvalues.

\subsubsection{time evolution of entanglement spectra}
\label{sec:phase-diagram}

\begin{figure}
  \centering
  \includegraphics[width=\linewidth]{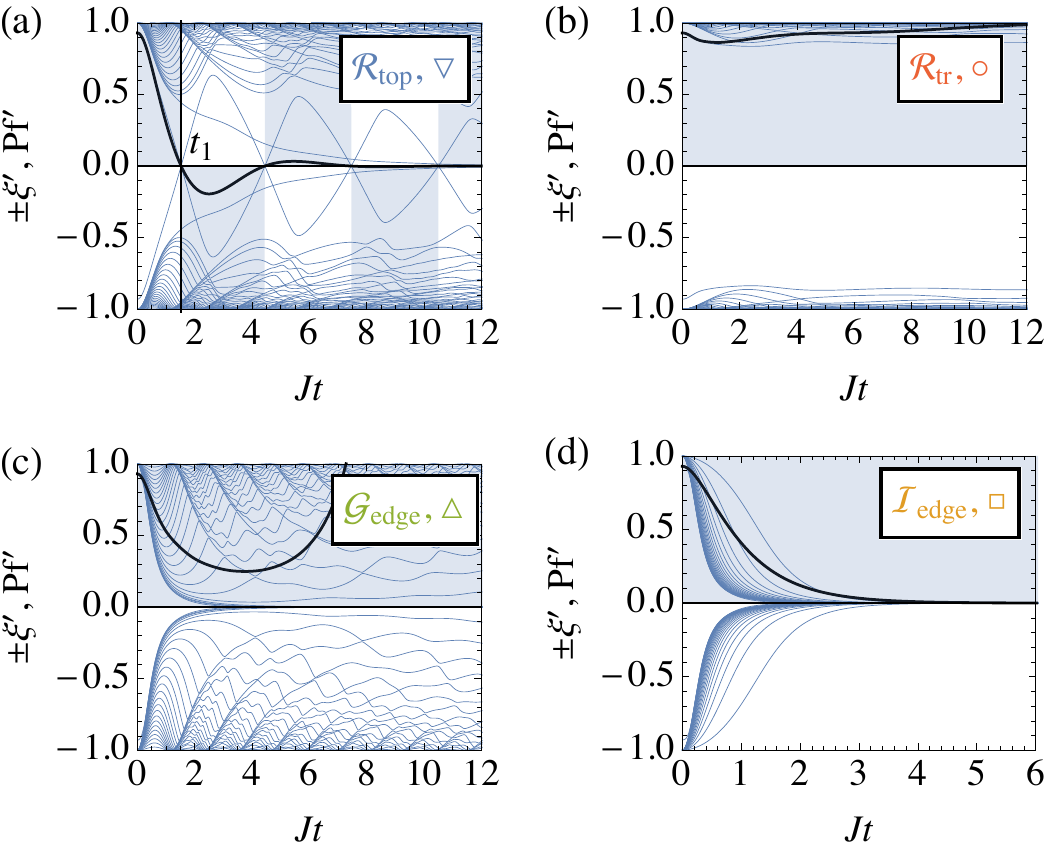}
  \caption{Time evolution of entanglement spectra. Rescaled single-particle
    entanglement spectrum, $\xi' = \xi \e^{4 \gamma t}$, for a quench from the
    trivial phase at $\mu_0 = - 3 J$ to (a) the topological phase
    $\mathcal{R}_{\rm top}$ with $W = 1$ at $\mu_1 = -J$ and $\gamma = 0.2 J$,
    (b) the trivial phase $\Rtr$ with $W = 0$ at $\mu_1 = - 2.5 J$ and
    $\gamma = 0.1 J$, (c) the gapless phase $\Gedge$ at $\mu_1 = - J$ and
    $\gamma = 0.6 J$, and (d) the trivial phase $\Iedge$ at $\mu_1 = 0$ and
    $\gamma = 1.1 J$. Zero crossings in the entanglement spectrum, which occur
    in (a) but not in (b--d), are tracked by the rescaled Pfaffian,
    $\mathrm{Pf}' = \mathop{\mathrm{Pf}} \e^{4 \gamma N t}$, which is shown as a
    black solid line. The sign of the Pfaffian is indicated by the blue
    shading. The parameters in (a), (b), (c), and (d) correspond, respectively,
    to the blue inverted triangle, red circle, green triangle, and orange
    square in Fig.~\ref{fig:phase-diagram-LDM}.}
  \label{fig:quench-dynamics-LDM}
\end{figure}

Before we continue, it is instructive to recall previous results for isolated
systems. Notably, Gong and Ueda~\cite{Gong2017a} connected the occurrence of entanglement
spectrum crossings in the quench dynamics and the topology of the prequench and
postquench Hamiltonians, $H_0$ and $H$, respectively, of a one-dimensional
isolated system. Their key observation is that the time-evolved state
$\ket{\psi(t)} = \e^{-\imag H t} \ket{\psi_0}$, where $\ket{\psi_0}$ is the
ground state of $H_0$, is the ground state of a time-dependent parent
Hamiltonian defined as $H_{\mathrm{p}}(t) = \e^{-\imag H t} H_0 \e^{\imag H t}$.
If $H$ is assumed to be band-flattened, the time dependence of
$H_{\mathrm{p}}(t)$ is periodic, and, therefore, in the first-quantized
Hamiltonian $H_{\mathrm{p}}(k, t)$, the time $t$ can be interpreted as a second
quasimomentum variable. By virtue of the entanglement spectrum bulk-edge
correspondence, the evolution of the entanglement spectrum over one period gives
the exact edge spectrum of the $(1 + 1)$-dimensional band-flattened Bloch
Hamiltonian $H_{\mathrm{p}}(k, t)$~\cite{Fidkowski2010}. In particular,
nontrivial topology of $H_{\mathrm{p}}(k, t)$ is reflected in the presence of
entanglement edge modes which cross the band gap as a function of $t$. For the
Altland-Zirnbauer classes in which nontrivial topology of $H_{\mathrm{p}}(k, t)$
is realized, the two-dimensional topological index of the parent Hamiltonian
$H_{\mathrm{p}}(k, t)$ is the difference between the one-dimensional topological
indices of $H_0$ and $H$~\cite{Gong2017a}. Therefore, entanglement spectrum
crossings occur, in particular, for quenches from a trivial phase of $H_0$ to a
topological phase of $H$, and serve as a proxy for nontrivial topology of
$H$. As we discuss in the following, the connection between entanglement
spectrum crossings and nontrivial topology of the generator of dynamics remains
intact if, instead of unitary time evolution generated by a Hamiltonian $H$, we
consider the driven-dissipative dynamics of an open quantum many-body system as
described by the master equation~\eqref{eq:master-equation} with a non-Hermitian
superoperator $\hat{\mathcal{L}}$.

Throughout this section, we take the system to be initialized in the topologically trivial ground
state of the isolated Kitaev chain with Hamiltonian~\eqref{eq:H-Kitaev}, where
$J = \Delta$ sets the energy scale. Our results do not depend qualitatively on
the precise prequench value of the chemical potential. For concreteness, we
choose $\mu_0 = -3 J$.

Figure~\ref{fig:quench-dynamics-LDM}(a) shows a quench to the non-Hermitian
topological phase $\Rtop$ with $W = 1$ at $\mu_1 = -J$ and $\gamma = 0.2 J$. The
parameters of the postquench Liouvillian are indicated with a blue inverted
triangle in Fig.~\ref{fig:phase-diagram-LDM}, and the spectrum of the
Liouvillian for these parameters is shown in
Fig.~\ref{fig:Liouvillian-spectrum-LDM}(a). As can be seen in the latter figure,
the bulk modes, which are shown as blue and orange diamond symbols, decay with
the same rate $- \Im(\lambda_{\mathrm{bulk}}) = 2 \gamma = 0.4 J$. Then, for the
covariance matrix, and, consequently, its eigenvalues which form the
single-particle entanglement spectrum,
Eq.~\eqref{eq:cov-mat-explicit-solution} implies a decay rate of
$4 \gamma = 0.8 J$. To best illustrate the evolution of the entanglement
spectrum in Fig.~\ref{fig:quench-dynamics-LDM}, we account for this overall
decay by rescaling the entanglement eigenvalues as $\xi' = \xi \e^{4 \gamma t}$.
With this rescalling, in Fig.~\ref{fig:quench-dynamics-LDM}(a), most of the
single-particle entanglement eigenvalues, which are shown as blue lines, stay
close to $\pm 1$. However, a pair of eigenvalues $\pm \xi_i'$ undergoes repeated
zero crossings. To trace these crossings, we employ the Pfaffian of the reduced
covariance matrix,
\begin{equation}
  \label{eq:pfaffian}
  \mathrm{Pf}(t) = \pf(\Gamma_A(t)).
\end{equation}
The Pfaffian, which can be defined for any antisymmetric matrix, has the
following properties: Its absolute value is given by the square root of the
determinant, and, most importantly, it changes sign whenever a pair of
entanglement eigenvalues $\pm \xi_i$ crosses zero. In all panels of
Fig.~\ref{fig:quench-dynamics-LDM}, the rescaled Pfaffian
$\mathrm{Pf}' = \mathop{\mathrm{Pf}} \e^{4 \gamma N t}$ is shown as a black
line, and its sign is indicated with blue shaded areas. At long times, the
covariance matrix~\eqref{eq:cov-mat-explicit-solution}, and, consequently, also
the Pfaffian~\eqref{eq:pfaffian} decay to zero.

The entanglement spectrum crossings disappear when the postquench parameters are
within a trivial phase in which $W = 0$ or $W$ is
undefined. Figure~\ref{fig:quench-dynamics-LDM}(b) shows the entanglement
spectrum dynamics following a quench to the real-line gapped trivial phase
$\Rtr$ with $W = 0$ at $\mu_1 = -2.5J$ and $\gamma = 0.1 J$, as indicated by the
red circle in Fig.~\ref{fig:phase-diagram-LDM}. As above, the decay rates of all
bulk modes are equal to $2 \gamma$. Therefore, rescaling the entanglement
eigenvalues with $\e^{4 \gamma t}$ removes the overall decay of the entanglement
spectrum and leaves all entanglement eigenvalues close to $\pm 1$.

The evolution of the entanglement spectrum shows richer phenomenology for
quenches to the gapless phases $\Gedge$ and $\Gtr$, in which the winding number
$W$ is undefined. Figure~\ref{fig:quench-dynamics-LDM}(c) shows a quench to the
phase $\Gedge$ at $\mu_1 = -J$ and $\gamma = 0.6 J$, as indicated by the green
triangle in Fig.~\ref{fig:phase-diagram-LDM}. The corresponding spectrum of the
postquench Liouvillian is shown in
Fig.~\ref{fig:Liouvillian-spectrum-LDM}(b). Decay rates of bulk modes are within
a finite range of values, and, therefore, the rescaling of the entanglement
spectrum with $\e^{4 \gamma t}$ leads to the occurrence of exponentially growing
values. Thus, the rescaled Pfaffian in Fig.~\ref{fig:quench-dynamics-LDM}(c)
soon exceeds one. Crucially, and in agreement with the fact that in a gapless
phase the postquench Liouvillian is not characterized by a nontrivial
non-Hermitian winding number, the Pfaffian does not change sign.

Figure~\ref{fig:quench-dynamics-LDM}(d) presents the time evolution of the
entanglement spectrum after a quench to the imaginary-line gapped phase $\Iedge$
at $\mu_1 = 0$ and $\gamma = 1.1 J$, as indicated by the orange square in
Fig.~\ref{fig:phase-diagram-LDM}. For these postquench parameters, the two
bands~\eqref{eq:lambda-k-LDM} are purely imaginary as shown in
Fig.~\ref{fig:Liouvillian-spectrum-LDM}(c). This is reflected vividly in the
purely decaying behaviour of the entenglement spectrum.

We proceed with a quantitative analysis of the time evolution of the
entanglement spectrum for quenches to the topological phase $\Rtop$. The color
scale within the phase $\Rtop$ in Fig.~\ref{fig:phase-diagram-LDM} encodes the
time $t_1$ at which the first zero crossing in the single-particle entanglement
spectrum occurs. In Fig.~\ref{fig:quench-dynamics-LDM}(a), the time $t_1$ is
indicated with a vertical black line. The values of $t_1$ diverge at the
boundaries of the phase $\Rtop$. This is illustrated in the inset of
Fig.~\ref{fig:phase-diagram-LDM}, where $t_1$ is shown for variations of the
parameters of the postquench Liouvillian along the black arrows in the main
panel of the figure, i.e., for $\delta \mu = \mu_1 + 2 J \to 0$ and
$\gamma = 0$, and for $\mu_1 = - J$ and $\delta \gamma = 0.5 J - \gamma \to 0$,
respectively. The numerical data is consistent with a square-root singularity,
which is shown as a gray line in the inset of Fig.~\ref{fig:phase-diagram-LDM},
and which corresponds to the behavior $t_1 \sim \delta \mu^{- \epsilon}$ and
$t_1 \sim \delta \gamma^{-\epsilon}$, where $\epsilon = 1/2$ can be interpreted
as a dynamical critical exponent for entanglement spectrum crossings. In
Sec.~\ref{sec:driv-diss-parity}, we show that the value $\epsilon = 1/2$ is
exact for the specific case of a postquench Liouvillian with flat spectrum, in
which an exact solution of the quench dynamics is possible. In the numerical
data shown in the inset of Fig.~\ref{fig:phase-diagram-LDM}, the singularity of
$t_1$ is cut due to the finite system size. The values of $t_1$ saturate earlier
if the phase boundary is approached by varying the strength of dissipation.

\begin{figure}
  \centering
  \includegraphics[width=\linewidth]{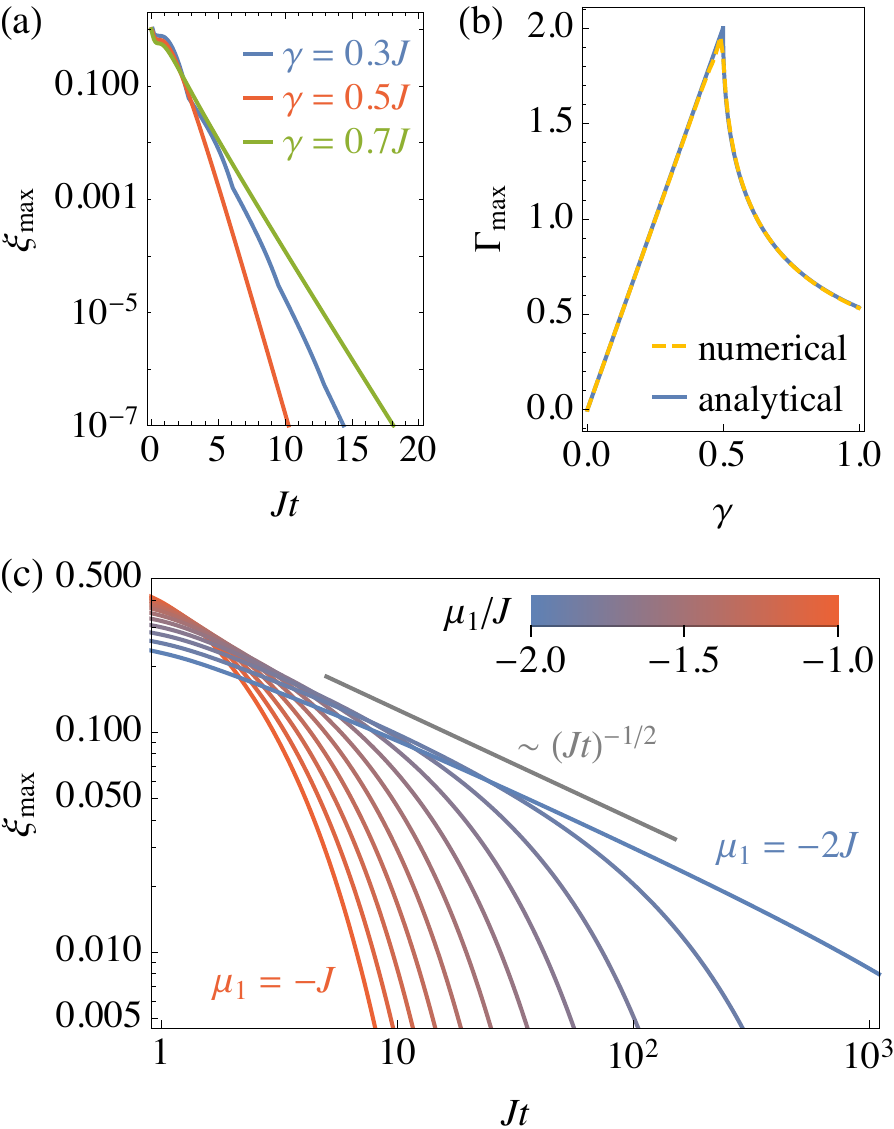}
  \caption{Relaxation dynamics of the largest entanglement eigenvalue
    $\xi_{\mathrm{max}}$. (a) Exponential decay across the phase boundary
    between $\Rtop$ and $\Gedge$ at $\mu_1 = - J$ and for three values of
    $\gamma$ below, at, and above the critical value
    $\gamma_{\mathrm{c}} = 0.5 J$. (b) The decay $\Gamma_{\mathrm{max}}$ of
    $\xi_{\mathrm{max}}$ exhibits a maximum at the phase boundary. The
    numerically determined value of $\Gamma_{\mathrm{max}}$ is shown as an
    orange dashed line, and the blue solid line corresponds to the analytical
    expression~\eqref{eq:gamma-max}. (c) Critical relaxation of
    $\xi_{\mathrm{max}}$ at the phase boundary between $\Gedge$ and $\Gtr$ at
    $\gamma = J$ and for values of $\mu_1$ ranging from $\mu_1 = -J$ to the
    critical value $\mu_1 = -2 J$, shown as red to blue lines. For comparison,
    power-law behavior $\xi_{\mathrm{max}} \sim \left( J t \right)^{-1/2}$ is
    shown as a gray line.}
  \label{fig:critical-entanglement-spectrum-dynamics}
\end{figure}

The singularity of $t_1$ at the boundaries of the topological phase $\Rtop$
raises the question, whether the time evolution of the entanglement spectrum
exhibits also other signatures of dynamical criticality.  To complement our
analysis of oscillatory behavior of entanglement eigenvalues, which is
characterized by the time scale $t_1$, we illustrate in
Fig.~\ref{fig:critical-entanglement-spectrum-dynamics} the decay of the largest
eigenvalue $\xi_{\mathrm{max}}$. It exhibits critical power-law relaxation at
the phase boundary between the gapless phases $\Gedge$ and $\Gtr$, where the
global Berry phase jumps from $Q = 1$ to $Q = 0$ and, concomitantly, the edge
modes of the Liouvillian disappear. This form of decay is also realized
generically for quenches in isolated Hermitian systems~\cite{Jhu2017}. However,
the decay of $\xi_{\mathrm{max}}$ follows a simple exponential form across the
boundary between $\Rtop$ and $\Gedge$.

In particular, in Fig.~\ref{fig:critical-entanglement-spectrum-dynamics}(a) we
show the decay of $\xi_{\mathrm{max}}$ for three different values of the rate of
dissipation $\gamma$ below, at, and above the critical value
$\gamma_{\mathrm{c}} = 0.5 J$ for $\mu_1 = - J$. For each value of $\gamma$, the
approximately linear behavior of $\xi_{\mathrm{max}}$ on the semilogarithmic
scale indicates dominantly exponential relaxation. Indeed, we find good
agreement of the numerical data with
$\xi_{\mathrm{max}} \sim \e^{-\Gamma_{\mathrm{max}} t}$ and
$\xi_{\mathrm{max}} \sim \e^{-\Gamma_{\mathrm{max}} t}/\sqrt{t}$ for
$\gamma < \gamma_{\mathrm{c}}$ and $\gamma > \gamma_{\mathrm{c}}$,
respectively. The numerically determined decay rate $\Gamma_{\mathrm{max}}$ is
shown as a function of $\gamma$ in
Fig.~\ref{fig:critical-entanglement-spectrum-dynamics}(b). It agrees well with
the analytical expression
\begin{equation}
  \label{eq:gamma-max}
  \Gamma_{\mathrm{max}} =
  \begin{cases}
    4 \gamma & \text{for } \gamma < \gamma_{\mathrm{c}}, \\
    2 \left( 2 \gamma - \sqrt{4 \gamma^2 - \left( 2 J + \mu_1 \right)^2} \right)
    & \text{for } \gamma > \gamma_{\mathrm{c}},
  \end{cases}
\end{equation}
which is inspired by the exact result for the decay rate of the retarded
response function as elaborated in Sec.~\ref{sec:dynam-crit-behav} below.
Interestingly, $\Gamma_{\mathrm{max}}$ takes a maximum value of
$\Gamma_{\mathrm{max}} = 2 J$ exactly at the critical point
$\gamma = \gamma_{\mathrm{c}} = 0.5 J$. This observation runs counter to the
common expectation that relaxation to the steady state becomes exceedingly slow
in the vicinity of a phase transition. In Sec.~\ref{sec:dynam-crit-behav}, we
explain this unusual behavior through a mechanism of many-body critical damping.

The dynamics of the entanglement spectrum exhibits rather different behavior
across the boundary between $\Gedge$ and $\Gtr$. As discussed above, there is no
qualitative change in the oscillatory dynamics: For quenches to both $\Gedge$
and $\Gtr$, there are no zero crossings of entanglement eigenvalues. However, as
illustrated in Fig.~\ref{fig:critical-entanglement-spectrum-dynamics}(c), the
decay of the largest entanglement eigenvalue $\xi_{\mathrm{max}}$ shows critical
slowing down as the phase boundary is approached, and power-law relaxation
$\xi_{\mathrm{max}} \sim \left( J t \right)^{-1/2}$ exactly on the phase
boundary. The origin of this behavior is the closing of the point gap at
$\lambda = 0$. We provide a quantitative analysis of dynamical criticality in
terms of two-time correlation functions in Sec.~\ref{sec:dynam-crit-behav}.

\subsubsection{Driven-dissipative fermion parity pump}
\label{sec:driv-diss-parity}

We proceed to complement our numerical results for the time evolution of
entanglement spectra with an exact analytical solution, which is enabled by a
judicious choice of the initial state and the parameters of the postquench
Liouvillian. First, this provides us with an interpretation of topological
quench dynamics as a fermion parity pump; and second, for specific parameter
values, we are able to show explicitly that the entanglement spectrum dynamical
critical exponent takes the exact value $\epsilon = 1/2$.

To facilitate analytical progress, we assume that the system is initialized in
the vacuum of Dirac fermions $c_i$, which is the ground state of the Hamiltonian
of the Kitaev chain for $J=\Delta>0$ and $\mu_0 \to -\infty$. Further, we choose the parameters of the postquench Liouvillian as
$J = \Delta > 0$, $\mu_1 = 0$, and
$\gamma_{\mathrm{l}} = \gamma_{\mathrm{g}} = \gamma$. For this choice of
postquench parameters, the bulk spectrum~\eqref{eq:lambda-k-LDM} for an infinite
chain becomes flat with
\begin{equation}
  \label{eq:lambda-flat}
  \lambda_{\pm} = 2 \left( - \imag \gamma \pm \sqrt{J^2 - \gamma^2} \right).
\end{equation}
The spectrum exhibits a real line gap when $\gamma < \gamma_{\mathrm{c}} = J$,
and the expression under the square root is positive. In contrast, for
$\gamma > J$, the two flat bands are separated by an imaginary line gap.

\begin{figure}
  \centering
  \includegraphics[width=.94\linewidth]{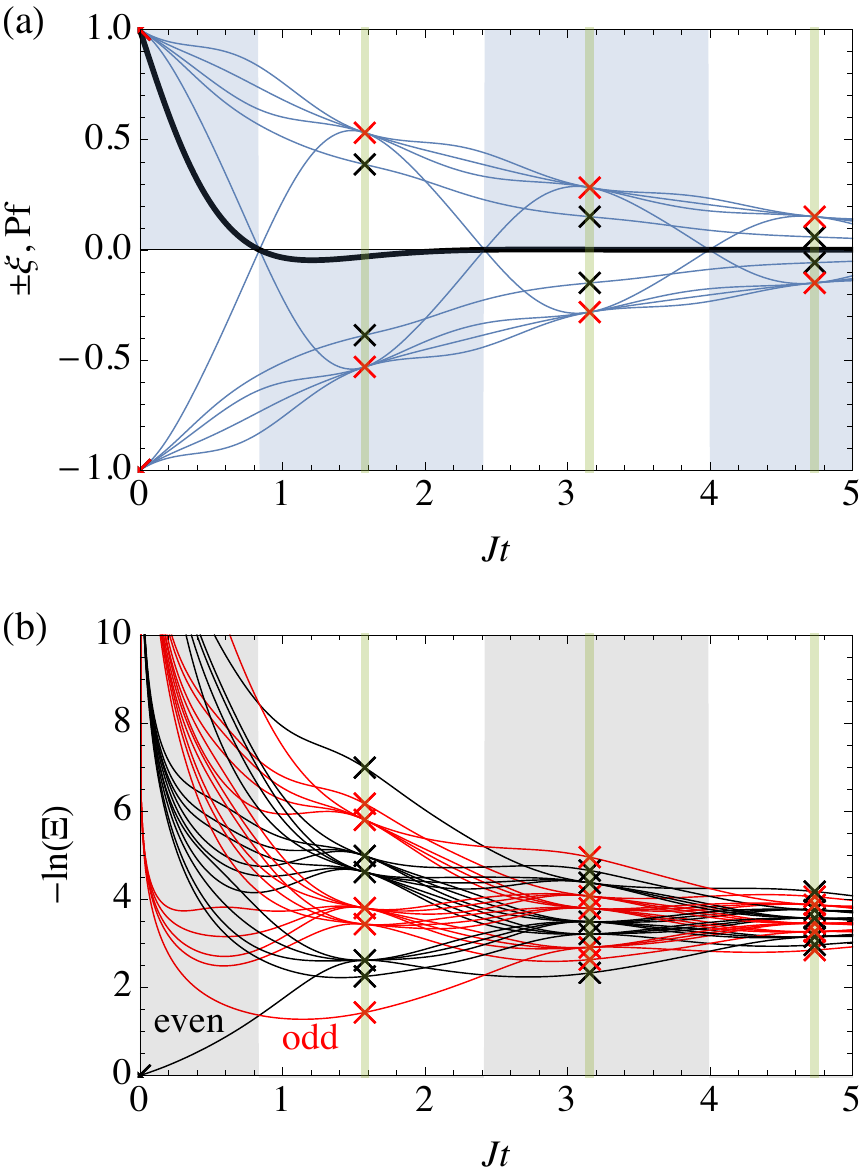}
  \caption{(a) Time evolution of the single-particle entanglement spectrum for a
    quench from $\mu_0 \to - \infty$, to $\mu_1 = 0$, and
    $\gamma_{\mathrm{l}} = \gamma_{\mathrm{g}} = \gamma = 0.1 J$ for $N = 10.$
    The solid black line is the Pfaffian, and the blue shading indicates the
    sign of the Pfaffian. Green vertical lines mark integer multiples of the
    time scale $T$ in Eq.~\eqref{eq:T}. Black and red crosses indicate,
    respectively, $\pm \xi_{1, m}$ and $\pm \xi_{2, m}$, for the analytical
    results given in Eq.~\eqref{eq:exact-single-particle-ES}. (b) Fermion parity
    pump illustrated with the evolution of the many-body entanglement spectrum
    for a quench with the same parameters as in (a). Black and red lines
    correspond to many-body entanglement eigenstates with even and odd parity,
    respectively. The parity of the entanglement ground state is even (odd) in
    the regions with (without) gray shading. Crosses indicate analytical results
    at multiples of $T$ marked by green vertical lines.}
  \label{fig:exact-quench}
\end{figure}

For a Liouvillian with flat bulk bands, also the spectrum for an open finite chain of
length $N$ can be obtained analytically, and is given by
$\{ \lambda_{\pm}, \lambda_L, \lambda_R \}$. In a finite system, the values
$\lambda_{\pm}$ given in Eq.~\eqref{eq:lambda-flat} have degeneracy $N - 1$
each, and $\lambda_L = - \imag 4 \gamma$ and $\lambda_R = 0$ are eigenvalues
corresponding to edge modes which are localized on the left and the right end of
the chain, respectively. For $\gamma < J$, the real part of the eigenvalues
$\lambda_{\pm}$ sets a time scale,
\begin{equation}
  \label{eq:T}
  T = \pm \frac{\pi}{\Re(\lambda_{\pm})} = \frac{\pi}{2 \sqrt{J^2 - \gamma^2}},
\end{equation}
for oscillations of the bulk modes. As we show in the following, this is the
time scale on which the fermion parity of many-body entanglement eigenstates is
reversed and, concomitantly, both single-particle and many-body entanglement
eigenvalues cross. More precisely, as illustrated in Fig.~\ref{fig:exact-quench}
where the times $m T$, with $m \in \N_0$ a nonnegative integer, are indicated
with green vertical lines, entanglement spectrum crossings occur at times $t_m$
which obey $m T < t_m < \left( m + 1 \right) T$.

The key result, based on which we establish the connection between
entanglement spectrum crossings and the reversal of the fermion parity at the
edges of the Kitaev chain, is an exact expression for the reduced density matrix
$\rho_A(m T)$ at times $t = m T$ with $m \in \N_0$, which we derive in
Appendix~\ref{sec:fermion-parity-pump}. As we show there, for the choice of
initial state and postquench Liouvillian specified above, the reduced density
matrix $\rho_A(m T)$ is diagonal in a basis of Fock states with fixed number of
particles on each lattice site, which we denote by $\ket{\mathbf{n}}_A$ where
$\mathbf{n} = \left( n_1, \dotsc, n_{N/2} \right)$ is a vector of lattice-site
occupation numbers with $n_i = 0,1$. In terms of projectors
$P^{\mathbf{n}} = \ket{\mathbf{n}}_A \bra{\mathbf{n}}$ on these basis states,
the reduced density matrix can be written as
\begin{equation}
  \label{eq:reduced-density-matrix-flat-band}
  \rho_A(m T) = \sum_{\mathbf{e}} \Xi_{\mathbf{e}, m} P^{\mathbf{n}_{\mathbf{e}, m}},
\end{equation}
where the sum is over all combinations
$\mathbf{e} = \left( e_1, \dotsc, e_{N/2} \right)$ of single-particle
entanglement occupation numbers $e_i = 0,1$, and the coefficients
$\Xi_{\mathbf{e}, m}$ are the eigenvalues of $\rho_A(m T)$ and determine the
many-body entanglement spectrum. We specify the relation between the
lattice-site occupation numbers $\mathbf{n}_{\mathbf{e}, m}$ and the
entanglement occupation numbers $\mathbf{e}$ below.

For the exact solution~\eqref{eq:reduced-density-matrix-flat-band} derived in
Appendix~\ref{sec:fermion-parity-pump}, the many-body entanglement eigenvalues
$\Xi_{\mathbf{e}, m}$ take the general form given in Eq.~\eqref{eq:mb-ent-spec},
with only two distinct single-particle entanglement eigenvalues,
\begin{equation}
  \label{eq:exact-single-particle-ES}
  \xi_{1, m} = \e^{-6 \gamma m T}, \qquad \xi_{2, m} = \e^{-4 \gamma m T}.
\end{equation}
The first eigenvalue $\xi_{1, m}$ is nondegenerate, whereas $\xi_{2, m}$ has
degeneracy $N/2 - 1$. Both single-particle entanglement eigenvalues decay
exponentially with rates $6 \gamma$ and $4 \gamma$, respectively. The exact
analytical results for $\pm \xi_{1, m}$ and $\pm \xi_{2, m}$ are shown, respectively, as black
and red crosses in Fig.~\ref{fig:exact-quench}(a). We find perfect
agreement with the numerically calculated entanglement spectrum
$\{ \pm \xi_i(t) \}$, which is indicated with blue lines, at times $t = m T$.

The reversal of the fermion parity at the ends of the chain is encoded in the
relation between entanglement occupation numbers $\mathbf{e}$ and the site
occupation numbers $\mathbf{n}_{\mathbf{e}, m}$ in
Eq.~\eqref{eq:reduced-density-matrix-flat-band}. As we show in
Appendix~\ref{sec:fermion-parity-pump},
\begin{equation}
  \label{eq:lattice-site-occupation-numbers}
  \mathbf{n}_{\mathbf{e}, m} = 
  \begin{cases}
    \mathbf{e} & \text{for $m$ even,} \\
    \bar{\mathbf{e}} & \text{for $m$ odd,}
  \end{cases}
\end{equation}
where vectors with an overbar are defined by $\bar{e}_1 = (e_1 + 1) \bmod 2$,
whereas $\bar{e}_i = e_i$ for $i = 2, \dotsc, N/2$. That is, $\mathbf{e}$ and
$\bar{\mathbf{e}}$ differ only in the first component.

To demonstrate the reversal of the fermion parity after each time step of
duration $T$, we follow the evolution of the entanglement ground state, i.e.,
the state which corresponds to the smallest eigenvalue of the entanglement
Hamiltonian $H_A$. According to the relation between the reduced density matrix
$\rho_A$ and the entanglement Hamiltonian, $\rho_A = \e^{- H_A}/Z_A$ where
$Z_A = {\tr}_A(\e^{-H_A})$, the entanglement ground state is the eigenstate of
$\rho_A$ with the largest many-body entanglement eigenvalue $\Xi_{\mathbf{e}}$
in Eq.~\eqref{eq:mb-ent-spec}. The latter equation implies that the largest
many-body entanglement eigenvalue is obtained if all entanglement occupation
numbers are equal to zero, $\mathbf{e} = \mathbf{0}$. The corresponding
lattice-site occupation numbers~\eqref{eq:lattice-site-occupation-numbers} are
$\mathbf{n}_{\mathbf{0}, m} = \mathbf{0}$ for even values of $m$, and
$\mathbf{n}_{\mathbf{0}, m} = \bar{\mathbf{0}} = \left( 1, 0, \dotsc, 0 \right)$
for odd values of $m$. The entanglement ground state is thus given by
$\ket{\mathbf{0}}_A$ when $m$ is even, and by $\ket{\bar{\mathbf{0}}}_A$ when
$m$ is odd. These states have opposite fermion parity:
\begin{equation}  
  P_A \ket{\mathbf{0}}_A = \ket{\mathbf{0}}_A, \qquad
  P_A \ket{\bar{\mathbf{0}}}_A = - \ket{\bar{\mathbf{0}}}_A,
\end{equation}
as follows directly from the definition of the fermion parity operator for
subsystem $A$, i.e., the left half of the chain,
\begin{equation}
  \label{eq:fermion-parity}
  P_A = \e^{\imag \pi \sum_{i = 1}^{N/2} c_i^{\dagger} c_i}.
\end{equation}

We proceed to connect the reversal of the fermion parity of the entanglement
ground state to crossings in the many-body entanglement spectrum. Indeed, as we
show in Appendix~\ref{sec:cons-parity-entangl}, the entanglement eigenstates
have definite parity at all times $t$. Therefore, a change of the fermion parity
of the entanglement ground state between $t = m T$ and
$t = \left( m + 1 \right) T$ implies that there is a crossing in the
entanglement spectrum in which the entanglement ground state and the first
excited state, which have opposite fermion parity, switch places. Numerically,
as shown in Fig.~\ref{fig:exact-quench}(b), we find that there is exactly one
crossing between the entanglement ground state and the first excited state,
which occurs at a time $t_m$ in the interval
$m T < t_m < \left( m + 1 \right) T$.

A crossing between the entanglement ground state and the first excited state in
turn implies a zero crossing in the single-particle entanglement spectrum. To
see this, we consider the difference between the entanglement eigenvalues of the
ground state and first excited state. As we have already noted, the ground state
corresponds to $\mathbf{e} = \mathbf{0}$. The first excited state is obtained by
occupying the single-particle entanglement level with the smallest eigenvalues
$\xi_1(t)$, where we order the single-particle entanglement eigenvalues
$\xi_i(t)$ such that $0 \leq \xi_1(t) \leq \dotsb \leq \xi_{N/2}(t)$. Thus, for
the first excited state, the entanglement occupation numbers are
$\mathbf{e} = \bar{\mathbf{0}}$.  According to Eq.~\eqref{eq:mb-ent-spec}, which
expresses the many-body entanglement spectrum in terms of the single-particle
entanglement spectrum, the difference between the entanglement eigenvalues
$\Xi_{\mathbf{0}}(t)$ and $\Xi_{\bar{\mathbf{0}}}(t)$ is given by
\begin{equation}
  \Xi_{\mathbf{0}}(t) - \Xi_{\bar{\mathbf{0}}}(t) = \xi_1(t) \prod_{i = 2}^{N/2}
  \frac{1}{2} \left( 1 + \xi_i(t) \right).
\end{equation}
This difference is equal to zero when $\xi_1(t) = 0$. That is, when the fermion
parity for the many-body entanglement ground state is reversed, there is a zero
crossing in the single-particle entanglement spectrum.

Interestingly, a zero crossing of $\xi_1(t)$ leads to simultaneous crossings of
all pairs of many-body entanglement eigenstates with entanglement occupation
numbers $\mathbf{e}$ and $\bar{\mathbf{e}}$. The difference between the
corresponding entanglement eigenvalues is given by
\begin{equation}
  \Xi_{\mathbf{e}}(t) - \Xi_{\bar{\mathbf{e}}}(t) = \left( -1 \right)^{e_1}
  \xi_1(t) \prod_{i = 2}^{N/2} \frac{1}{2} \left[ 1 + \left( -1 \right)^{e_i}
    \xi_i(t) \right],
\end{equation}
and vanishes again when $\xi_1(t) = 0$. Figures~\ref{fig:exact-quench}(a)
and~(b) illustrate this connection between zero crossings in the single-particle
entanglement spectrum and simultaneous crossings of pairs of many-body
entanglement eigenvalues. As can be seen in Fig.~\ref{fig:exact-quench}(b),
there are further occasional crossings in the many-body entanglement spectrum
that are not related to zero crossings in the single-particle entanglement
spectrum, and which are not of topological origin.

Finally, we demonstrate numerically that the simultaneous crossings in the
many-body entanglement spectrum occur between pairs of states which have
opposite parity, and that, therefore, a zero crossing in the single-particle
entanglement spectrum leads to the simultaneous reversal of the fermion parity
in all entanglement eigenstates. To this end, we first note that according to
Eq.~\eqref{eq:lattice-site-occupation-numbers}, pairs of entanglement
eigenstates with entanglement spectrum occupation numbers $\mathbf{e}$ and
$\bar{\mathbf{e}}$ have opposite parity at times $t = m T$. As discussed in
Appendix~\ref{sec:cons-parity-entangl}, parity is conserved during the
dynamics. We can thus track the parity of all entanglement eigenstates at all
times, even without calculating the full many-body density matrix $\rho_A$. In
Fig.~\ref{fig:exact-quench}(b), black and red lines correspond to even and odd
parity, respectively. The parity of the entanglement ground state is even in regions with gray shading. At the boundaries of these regions, which agree with the boundaries of the blue-shaded regions in Fig.~\ref{fig:exact-quench}(a) that mark changes of the sign of the Pfaffian due to zero crossings in
the single-particle entanglement spectrum, simultaneous crossings in the many-body entanglement spectrum occur between pairs of states with opposite parity.

The above considerations establish the interpretation of entanglement spectrum
crossings as a fermion parity pump for a particular choice of parameters. In
general, this interpretation remains valid for continuous variations of pre- and
postquench parameters within the trivial phase of the isolated Kitaev chain and
the non-Hermitian topological phase of the driven and open Kitaev chain,
respectively. To convince oneself that this is the case, it is sufficient to note that, as
detailed in Appendix~\ref{sec:cons-parity-entangl}, fermion parity is always a good
quantum number of entanglement eigenstates for the models and dynamics we
consider here, i.e., quench dynamics starting from a state with definite parity, and
where time evolution is generated by a quadratic Liouvillian.

The period $T$ of the fermion parity pump diverges when the rate of dissipation
approaches the critical value of $\gamma_{\mathrm{c}} = J$.  Specifically, by
inserting $\delta \gamma = \gamma_{\mathrm{c}} - \gamma$ in Eq.~\eqref{eq:T}, we
obtain
\begin{equation}
  \label{eq:T-dynamical-critical-behavior}
  T \sim \frac{\pi}{2 J \sqrt{2 \delta \gamma/J}}, \quad \delta \gamma \searrow 0.
\end{equation}
Consequently, also the times $t_m$ at which entanglement spectrum crossings
occur exhibit a square-root singularity, $t_m \sim \delta \gamma^{-\epsilon}$
with $\epsilon = 1/2$. In the inset of Fig.~\ref{fig:phase-diagram-LDM}, we
present numerical evidence for the universality of the value of the entanglement
spectrum critical exponent $\epsilon$. In particular, our numerical results
indicate that $\epsilon = 1/2$ when we (i) consider dispersive bands of the
postquench Liouvillian, (ii) approach the phase boundary of $\Rtop$ from different
directions, by tuning either $\gamma$ or $\mu_1$, (iii) consider a more general
topologically trivial initial state with $J,\Delta\neq0$ and
$2 J< \abs{\mu_0} <\infty$, i.e., a trivial state that is different from the
vacuum of Dirac fermions, and (iv) study different models such as the one with
quasilocal jump operators defined in Eq.~\eqref{eq:jump-operator-QLDM} and
discussed in detail in Appendix~\ref{sec:quasiloc-jump-operator}.

\subsection{Dynamical critical behavior}
\label{sec:dynam-crit-behav}

In Sec.~\ref{sec:phase-diagram}, we have encountered two distinct types of
dynamical critical behavior of the entanglement spectrum: the time $t_1$ of the
first entanglement spectrum crossing diverges with an exponent $\epsilon = 1/2$
at the boundary of $\Rtop$, while the relaxation of the largest entanglement
eigenvalue $\xi_{\mathrm{max}}$ follows a simple exponential form across the
phase boundary; In contrast, there are no entanglement spectrum crossings on
both sides of the phase boundary between $\Gedge$ and $\Gtr$. However,
$\xi_{\mathrm{max}}$ exhibits critical power-law relaxation exactly at this
phase boundary. In the following, we show that a unified perspective on these
different forms of dynamical criticality can be obtained by studying two-time
correlation functions.

We consider the Keldysh Green's function, together with the retarded and advanced response
functions~\cite{Sieberer2016a}, in terms of which all two-time correlation
functions can be specified:
\begin{align}
  \label{eq:Keldysh-GF} \chi^K_{i, j}(t) & = - \frac{\imag}{2} \langle [w_i(t),
w_j] \rangle, \\
  \label{eq:retarded-response} \chi^R_{i, j}(t) & = \frac{1}{2} \theta(t)
\langle \{ w_i(t), w_j \} \rangle, \\
  \label{eq:advanced-response} \chi^A_{i, j}(t) & = \frac{1}{2} \theta(- t)
\langle \{ w_i(t), w_j \} \rangle,
\end{align}
where $\theta(t)$ is the Heaviside step function. We omit the usual factors
$\mp \imag$ in the definitions of the retarded and advanced response functions
to obtain real-valued quantities. The expectation values are taken in the steady
state $\rho_{\mathrm{ss}}$ and, therefore, it is sufficient to include a single
time argument since two-time averages depend only on the time difference. For
dynamics described by a master equation in Lindblad form, two-time averages can
be calculated with the aid of the quantum regression theorem~\cite{Gardiner2014}
as detailed in Appendix~\ref{sec:ret-resp-fct}. Further, we assume here an
infinite bulk system such that the correlation functions are translationally
invariant.

The two-time correlation functions defined above are distinguished from the
covariance matrix Eq.~\eqref{eq:cov-mat} in two aspects: first, the expectation value is taken in the steady
state $\rho_{\mathrm{ss}}$ in the two-time correlation functions, and not in the time-dependent state $\rho(t) = \e^{\mathcal{L} t} \rho_0$ as in the case of the covariance matrix; second, here the Majorana operators $w_i$ and $w_j$ are evaluated at different
times. In the limit $t \to \infty$, the covariance matrix approaches the
equal-time Keldysh Green's function, $\Gamma_{i, j}(t) \to -\chi^K_{i, j}(0)$.

We consider here a system with a trivial steady state
$\rho_{\mathrm{ss}} = \id/D$, as is typically the case for Hermitian jump
operators. Then, the trace of the commutator in Eq.~\eqref{eq:Keldysh-GF}
evaluates to zero, and the Keldysh Green's function vanishes identically. The
retarded and the advanced response functions carry the same qualitative features
in terms of divergent timescales and scaling behavior. For concreteness, in the
following, we focus on the equal-position retarded response function
$\chi^R_{i, i}(t)$. We present results for odd values of $i$. Even values of $i$
lead to the same qualitative behavior.

As in our discussion of the interpretation of topological quench dynamics in
terms of a fermion parity pump in Sec.~\ref{sec:driv-diss-parity}, it is worthwhile to
consider first the case of a flat-band Liouvillian, which enables exact
analytics, and brings out the relation between the dynamics of the retarded
response function and of a classical damped harmonic oscillator most clearly.

The bands $\lambda_{\pm, k}$ in Eq.~\eqref{eq:lambda-k-LDM} are flat when the
chemical potential is equal to zero. Then, the evolution equation for the
retarded response function, as detailed in Appendix~\ref{sec:ret-resp-fct}, can
be written in the form of the equation of motion of a damped harmonic oscillator
which is kicked at $t = 0$:
\begin{equation}
  \frac{\mathrm{d}^2 \chi^R_{i, i}}{\mathrm{d} t^2} + 4 \gamma
  \frac{\mathrm{d} \chi^R_{i, i}}{\mathrm{d} t} + 4 J^2 \chi^R_{i, i} = 4 \gamma \delta(t)
+ \frac{\mathrm{d} \delta(t)}{\mathrm{d} t},
\end{equation}
where $\delta(t)$ is the Dirac delta function. The solution which obeys the
boundary condition $\chi^R_{i, i}(t) = 0$ for $t < 0$ reads
\begin{equation}
  \label{eq:ret-resp-flat-band}
  \chi^R_{i, i}(t) = \theta(t) \e^{-2 \gamma t}
  \left( \cos(\omega t) + \frac{2 \gamma}{\omega} \sin(\omega t) \right),
\end{equation}
where $\omega = 2 \sqrt{J^2 - \gamma^2}$. In the phase $\Rtop$ where
$\gamma < J$, the response function exhibits underdamped oscillations. The
boundary between the phases $\Rtop$ and $\Iedge$ is at the point of critical
damping where $\gamma = J$. As $\gamma$ is increased beyond this value, $\omega$
becomes imaginary, and the response function ceases to oscillate.

\begin{figure}
  \centering
  \includegraphics[width=.98\linewidth]{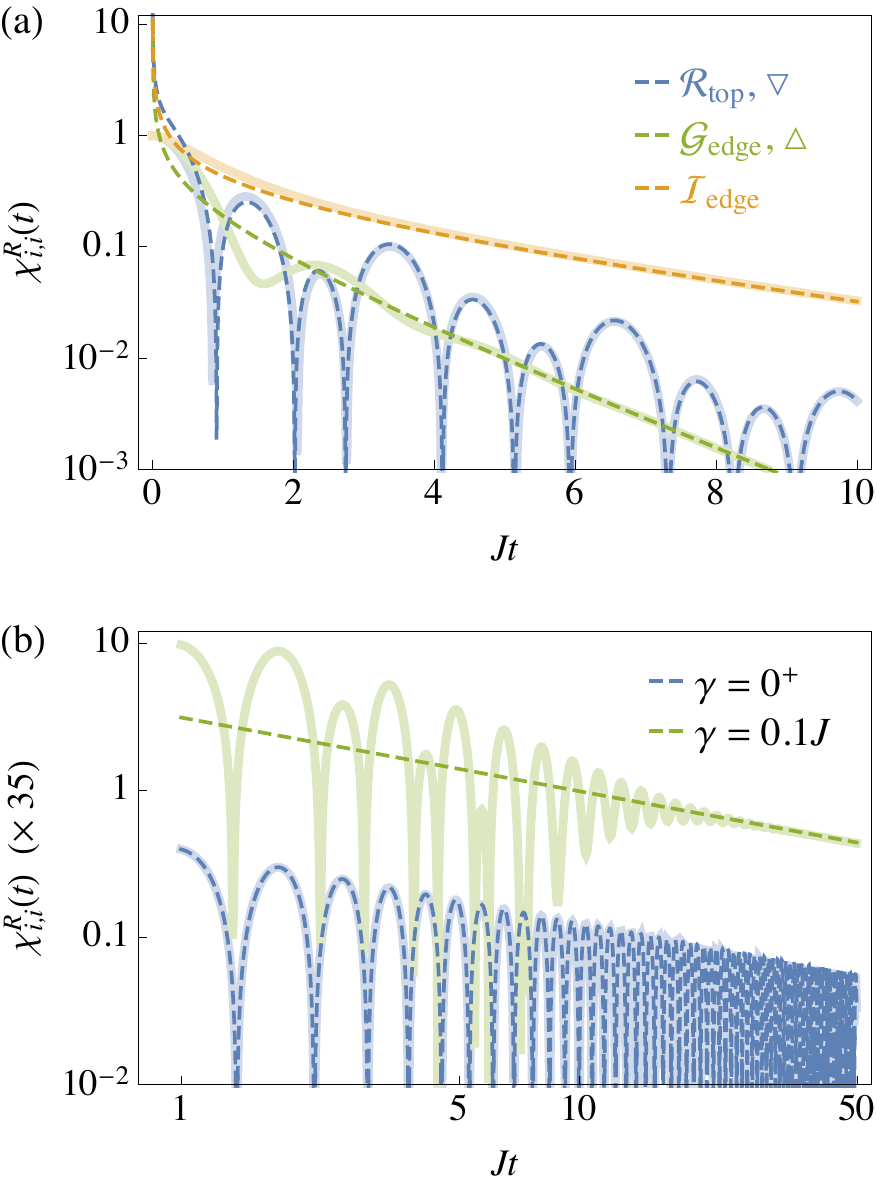}
  \caption{Retarded response function. (a) Generic behavior within various
    phases, specifically in $\Rtop$ at $\mu_1 = - J$ and
    $\gamma = 0.2 J$, in $\Gedge$ at $\mu_1 = - 1.5 J$ and
    $\gamma = 0.4 J$, and in $\Iedge$ at $\mu_1 = -0.5 J$
    and $\gamma = 1.5 J$. The parameters for
    $\Rtop$ and $\Gedge$ are marked
    with, respectively, an inverted blue triangle and a green triangle in
    Fig.~\ref{fig:phase-diagram-LDM}. Thin dashed lines show the asymptotic
    late-time behavior given in Eqs.~\eqref{eq:ret-resp-real-line-gap}
    and~\eqref{eq:ret-resp-gapless-imag-line-gap}, and thick solid lines are exact
    numerical results. (b)~Critical relaxation at $\mu = -2 J$. For better
    visibility, the response function for $\gamma = 0.1 J$ is multiplied by $35$.}
  \label{fig:ret_resp}
\end{figure}

While the dynamics of the response function for $\mu = 0$ is completely
analogous to that of a single damped harmonic oscillator, the many-body
character of the response function is restored when the chemical potential takes
nonzero values and, consequently, the bands $\lambda_{\pm, k}$ in
Eq.~\eqref{eq:lambda-k-LDM} are dispersive. For this case, the response function
cannot be calculated exactly. However, its asymptotic late-time behavior, which
is of main interest with regard to dynamical criticality, can still be obtained
analytically as described in Appendix~\ref{sec:ret-resp-fct}.

The behavior of the retarded response function in the phase $\Rtop$ is illustrated in Fig.~\ref{fig:ret_resp}(a). In both phases with a real line gap, $\Rtop$ and $\Rtr$, the late-time asymptotic form of the response function reads
\begin{multline}
  \label{eq:ret-resp-real-line-gap}
  \chi^R_{i, i}(t) \sim \frac{\e^{-\Gamma_R t}}{\sqrt{2 \pi J t}} \left( A_+
    \cos(\omega_+ t + \phi_+ + \pi/4) \right. \\ \left. + A_- \sin(\omega_- t +
    \phi_- + \pi/4) \right).
\end{multline}
That is, the response function is a superposition of two oscillatory
contributions with frequencies $\omega_{\pm}$, and damps out exponentially with a decay rate $\Gamma_R = 2\gamma$. It is worth stressing that the response
function exhibits oscillations and zero crossings in both the topological and
the trivial phase. The amplitudes, phases, and frequencies of the oscillatory
components with $\sigma = \pm$ are given by
\begin{equation}
  \begin{gathered}  
    A_{\sigma} = \frac{\abs{2 J + \sigma \mu}}{\sqrt{2 \omega_{\sigma}
        \abs{\mu}}}, \quad \phi_{\sigma} = \arg \! \left( \frac{\omega_{\sigma}
        - \imag 2
        \gamma}{\abs{2 J + \sigma \mu}} \right), \\
    \omega_{\sigma} = \sqrt{\left( 2 J + \sigma \mu \right)^2 - 4 \gamma^2}.
  \end{gathered}  
\end{equation}
When $\gamma$ is increased towards the critical value
$\gamma_{\mathrm{c}} = J + \mu/2$ which marks the boundary between the phases
$\Rtop$ and $\Gedge$ for $\mu < 0$, the amplitude $A_+$ diverges and, therefore,
the asymptotic behavior of the response function is dominated by the
contribution which oscillates at the frequency $\omega_+$. This frequency
vanishes at the phase boundary. In particular, for
$\gamma = \gamma_{\mathrm{c}} - \delta \gamma$, we obtain
$\omega_+ \sim 2 \sqrt{\left( 2 J + \mu\right) \delta \gamma}$. Thus, upon
approaching $\delta \gamma = 0$, the oscillation period
$2 \pi/\omega_+ \sim \delta \gamma^{-\epsilon}$ diverges with the same exponent
$\epsilon = 1/2$ as $t_1$. For $\mu > 0$, the roles of $\sigma = +$ and
$\sigma = -$ are interchanged: $A_-$ diverges and $\omega_-$ vanishes at the
phase boundary, while $A_+$ and $\omega_+$ remain finite.

For both a flat and a dispersive spectrum of the Liouvillian, which lead
to Eqs.~\eqref{eq:ret-resp-flat-band}
and~\eqref{eq:ret-resp-real-line-gap} for the response function, respectively, the period of
oscillations of the response function diverges when the real line gap closes. However, there are notable differences between Eqs.~\eqref{eq:ret-resp-flat-band}
and~\eqref{eq:ret-resp-real-line-gap}. First, the latter equation
does not reduce to Eq.~\eqref{eq:ret-resp-flat-band} in the limit $\mu \to 0$,
which shows that this limit does not commute with the late-time limit
$t \to \infty$ in which Eq.~\eqref{eq:ret-resp-real-line-gap} is valid. The
continuum of modes, which contribute to the response function in a many-body
system, results in additional structure of Eq.~\eqref{eq:ret-resp-real-line-gap}
in comparison to Eq.~\eqref{eq:ret-resp-flat-band}. In particular, the
exponential decay is augmented by an algebraic factor $1/\sqrt{t}$ in
Eq.~\eqref{eq:ret-resp-real-line-gap}. Further, while the entire spectrum
becomes purely damped with $\Re(\lambda) = 0$ for critical damping with
$\gamma = J$ at $\mu = 0$, for a generic point on the boundary between a real-line gapped phase and a gapless phase, the behavior of the many-body response
function is determined by a continuum of modes $\lambda$. Even within the
gapless phases there are modes for which $\Re(\lambda) \neq 0$, as shown in
Fig.~\ref{fig:Liouvillian-spectrum-LDM}(b). These modes give subleading
oscillatory contributions with frequency $\Re(\lambda)$ to the response
function, while the leading asymptotic late-time behavior is determined by modes
with the smallest values of the damping rate $- \Im(\lambda)$. For these modes,
the oscillation frequency vanishes, $\Re(\lambda) = 0$. The damped oscillatory
contribution is clearly visible for the line in Fig.~\ref{fig:ret_resp}(a) which
corresponds to $\Gedge$. Due to these differences between
Eqs.~\eqref{eq:ret-resp-flat-band} and~\eqref{eq:ret-resp-real-line-gap}, we
refer to the divergence of the period of oscillations of the response function
at finite values of the chemical potential as many-body critical
damping. Exactly on the boundary between $\Rtop$ and $\Gedge$, the retarded
response function behaves as
\begin{equation}
  \label{eq:ret-resp-critical-mb-damping}
  \chi^R_{i, i}(t) \sim \frac{\gamma_{\mathrm{c}}}{\sqrt{2 J \abs{\mu}}} \e^{- 2
    \Gamma_R t},  
\end{equation}
where $\gamma_{\mathrm{c}} = J - \abs{\mu}/2$ and $\Gamma_R = 2\gamma_c$.

The late-time form of the response function in the gapless phase $\Gedge$ is shown in Fig.~\ref{fig:ret_resp}(a). 
In fact, the leading asymptotic behavior of the response
function is the same in two gapless phases
$\Gedge$ and $\Gtr$ as well as the imaginary-line gapped phases
$\Iedge$ and $\Itr$. This is because the modes with the
smallest decay rates $- \Im(\lambda)$, which determine the late-time
asymptotics, have the same behavior $\Re(\lambda) = 0$ and
$-\Im(\lambda) \sim a + b k^2$ for $k \to 0$ with positive real coefficients
$a$ and $b$, as illustrated in Figs.~\ref{fig:Liouvillian-spectrum-LDM}(b)
and~\ref{fig:Liouvillian-spectrum-LDM}(c) for $\Gedge$ and $\Iedge$,
respectively. As a result, we find
\begin{equation}
  \label{eq:ret-resp-gapless-imag-line-gap}
  \chi^R_{i, i}(t) \sim \frac{\e^{- \Gamma_R t}}{4 \sqrt{\pi J t \abs{\mu}}}
  \left( \frac{2 \gamma}{\sqrt{2 \gamma - \Gamma_R}} + \sqrt{2 \gamma -
      \Gamma_R} \right).
\end{equation}
Here, the decay rate is given by
\begin{equation}
  \label{eq:gamma-sigma}
  \Gamma_R = 2 \gamma - \sqrt{4 \gamma^2 - \left( 2 J - \abs{\mu} \right)^2}.
\end{equation}
As can be seen in Fig.~\ref{fig:ret_resp}(a), the behavior of the response function in the
gapless phase $\Gedge$ and the imaginary-line gapped phase $\Iedge$ differs by the presence of subleading oscillatory contributions in the
former case.

The above results for the late-time asymptotics of the retarded response
function allow us to justify our finding in Sec.~\ref{sec:phase-diagram}, that
the numerically determined decay rate $\Gamma_{\mathrm{max}}$ of the largest
entanglement eigenvalue $\xi_{\mathrm{max}}$ agrees with the analytical
expression given in Eq.~\eqref{eq:gamma-max}. Indeed, by combining
Eqs.~\eqref{eq:ret-resp-real-line-gap}, \eqref{eq:ret-resp-critical-mb-damping},
and~\eqref{eq:ret-resp-gapless-imag-line-gap}, we see that decay rate $\Gamma_R$ 
of the retarded response function is related to $\Gamma_{\mathrm{max}}$ by
\begin{equation} \label{eq:Gamma_max-Gamma_R}
    \Gamma_{\mathrm{max}} = 2 \Gamma_R.
\end{equation}
The factor of two can be explained by comparing
Eqs.~\eqref{eq:cov-mat-explicit-solution} and~\eqref{eq:ret-resp-k} for the
covariance matrix and the retarded response function, respectively. In the
former equation, the matrix $X$ appears twice in the exponent, and, therefore,
the covariance matrix decays twice as fast. The maximum of $\Gamma_R$ and
$\Gamma_{\mathrm{max}}$ at $\gamma = \gamma_{\mathrm{c}}$ is again in agreement
with the physics of a damped harmonic oscillator, which decays fastest for
critical damping.

Upon approaching the critical lines at $\mu = \pm 2 J$ which separate the
gapless phases $\Gedge$ and $\Gtr$, the decay rate~\eqref{eq:gamma-sigma}
vanishes as $\Gamma_R \sim \delta \mu^2/(4 \gamma)$, where we set
$\mu = \mu_{\mathrm{c}} - \delta \mu$ for $\mu_{\mathrm{c}} = 2 J$ and
$\delta \mu > 0$. That is, the response function exhibits critical slowing down
with a divergent relaxation time scale. Exactly on the critical lines at
$\mu = \pm 2 J$, the response function decays as a power law,
\begin{equation}
  \chi^R_{i, i}(t) \sim \frac{1}{2 J} \sqrt{\frac{\gamma}{\pi t}}.
\end{equation}
As is illustrated in Fig.~\ref{fig:ret_resp}(b), subleading oscillations of the
response function damp out for $t \to \infty$.

In stark contrast to the critical relaxation of the response function at the
boundary between the gapless phases $\Gedge$ and $\Gtr$, according to
Eq.~\eqref{eq:ret-resp-critical-mb-damping}, the response function exhibits
exponential decay on the line of many-body critical damping, i.e., the phase
boundary between $\Rtop$ and $\Gedge$. This difference of exponential
\textit{vs.} power-law decay can be traced backed to the different types of gap
closings at the boundaries between $\Rtop$ and $\Gedge$, and $\Gedge$ and
$\Gtr$, respectively. In the latter case, the point gap at $\lambda = 0$ closes,
and critical relaxation results from the existence of a continuum of modes which
includes $\lambda = 0$: Technically, scaling behavior is due to the integral in Eq.~\eqref{eq:ret-resp-Fourier-representation} over a
continuum of modes, which develops a singularity when the continuum includes
zero. In contrast, although the real line gap closes at the phase boundary
between $\Rtop$ and $\Gedge$, the point gap at $\lambda = 0$ remains intact.

Finally, we note that the retarded response function can be calculated exactly
at $\mu = \pm 2 J$ and in the limit $\gamma = 0^+$ when we assume that the
system reaches the trivial infinite-temperature steady state even for
vanishingly small dissipation. We find
\begin{equation}
  \chi^R_{i, i}(t) = \frac{1}{2} \theta(t) J_0(4 J t),    
\end{equation}
where $J_n(z)$ is a Bessel function of the first kind with late-time asymptotic
behavior
\begin{equation}
  \chi^R_{i, i}(t) \sim \sin(4 J t + \pi/4)/\sqrt{2 \pi J t}.
\end{equation}
Thus, for $\gamma = 0^+$, oscillations of the response function
persist indefinitely, as can be seen in Fig.~\ref{fig:ret_resp}(b).

\section{Systems with non-Hermitian jump operators}
\label{sec:sys-non-hermitian-jump-ops}

So far, we have mainly focused on a special case of the
driven-dissipative Kitaev chain: For
$\gamma_{\mathrm{l}} = \gamma_{\mathrm{g}}$, the jump
operators~\eqref{eq:jump-operator-LDM} are Hermitian, and the steady state of
the time evolution described by the master equation~\eqref{eq:master-equation}
is a trivial infinite-temperature state. We showed that for Hermitian jump
operators, non-Hermitian topology of the Liouvillian $\hat{\mathcal{L}}$ is
revealed through entanglement spectrum crossings in quench dynamics. Now we
proceed to discuss the general case of non-Hermitian jump operators with
$\gamma_{\mathrm{l}} \neq \gamma_{\mathrm{g}}$, where the interplay of unitary
and dissipative dynamics results in a nontrivial steady state. As we explain in
the following, in this general case, the presence or absence of entanglement
spectrum crossings is determined not only by the topology of
$\hat{\mathcal{L}}$, but also by properties of the initial and steady states.

In particular, as discussed in Sec.~\ref{sec:driv-diss-parity}, each
zero crossing in the single-particle entanglement spectrum is accompanied by a
reversal of the fermion parity of the entanglement ground state. Therefore, if
the initial and steady states have opposite parities, an entanglement spectrum
crossing must occur even if the topology of $\hat{\mathcal{L}}$ is
trivial. However, as we show below, the connection between non-Hermitian
topology of $\hat{\mathcal{L}}$ and entanglement spectrum crossings can be
restored by means of Hermitianization of the jump operators~\cite{Lieu2019}.

\subsection{Fermion parity and Pfaffian in the initial and steady states}
\label{sec:ferm-parity-pfaff-0-ss}

For Hermitian jump operators, which lead to a trivial steady state
$\rho_{\mathrm{ss}} = \rho_{\infty} = \id/D$, also the reduced density matrix is
trivial,
$\rho_{A, \mathrm{ss}} = {\tr}_{A^{\mathrm{c}}}(\rho_{\mathrm{ss}}) = \id_A/D_A$
with $D_A = 2^{N/2}$. The corresponding many-body entanglement spectrum is flat,
and states with even and odd fermion parity are degenerate. Consequently, the
parity of the entanglement ground state is undetermined. Further, the covariance
matrix of the steady state vanishes identically. Thus, also the sign of the
steady-state Pfaffian $\mathrm{Pf}_{\mathrm{ss}}$ is undetermined.

In contrast, for non-Hermitian jump operators with
$\gamma_{\mathrm{l}} \neq \gamma_{\mathrm{g}}$, the entanglement spectrum of the
steady state is nondegenerate, and the corresponding entanglement ground state
has definite fermion parity $p_{A, \mathrm{ss}}$. The sign of the steady-state
Pfaffian is given by
\begin{equation}
  \label{eq:sgn-Pf-ss}
  \sgn(\mathrm{Pf}_{\mathrm{ss}}) = p_{A, \mathrm{ss}} = \left( \sgn(\delta)
 \right)^{N/2}
\end{equation}
with $\delta = \gamma_{\mathrm{l}} - \gamma_{\mathrm{g}}$. As we show in
Appendix~\ref{sec:even-odd-half-sys-size-effect}, this relation holds for
$J = \Delta > 0$, and arbitrary postquench values of the chemical potential
$\mu_1$. Notably, if $\delta < 0$, Eq.~\eqref{eq:sgn-Pf-ss} exhibits an even-odd
half-system-size effect for $\sgn(\mathrm{Pf}_{\mathrm{ss}})$. Here, as in
Sec.~\ref{sec:sys-hermi-jump-operators}, we restrict our discussion to systems
with an even number of lattice sites, $N \in 2 \N$, such that $N/2 \in \N$ is an
integer.

Equation~\eqref{eq:sgn-Pf-ss} should be compared with the corresponding relation
for the initial state of the quench protocol, i.e., the ground state of the
Hamiltonian~\eqref{eq:H-Kitaev} of the Kitaev chain. In the trivial phase with
$\abs{\mu_0} > 2 J$, we find
\begin{equation}
  \label{eq:sgn-Pf-0}
  \sgn(\mathrm{Pf}_0) = p_{A, 0} = \left( - \sgn(\mu_0) \right)^{N/2}.
\end{equation}
For values of $\mu_0$, $\delta$, and $N/2$ that lead to disagreements between
the right-hand sides of Eqs.~\eqref{eq:sgn-Pf-ss} and~\eqref{eq:sgn-Pf-0}, there
must be a reversal of the sign of the Pfaffian in the course of the time
evolution, and, consequently, also a crossing in the entanglement spectrum. This
conclusion does not rely on the topology of the Liouvillian, and can even be in
conflict with expectations based on the latter. In particular, for $\delta < 0$
and $\mu_0 < 0$, Eqs.~\eqref{eq:sgn-Pf-ss} and~\eqref{eq:sgn-Pf-0} imply that an
entanglement spectrum crossing must occur if $N/2$ is odd. By contrast, we
should expect no crossing if $N/2$ is even. This even-odd half-system-size
effect cannot be captured by the non-Hermitian winding number $W$, which is
defined for the bands of an infinite bulk system.

We proceed to demonstrate the occurrence of the even-odd effect explicitly in
the time evolution of entanglement spectra, and we then show how the discrepancy
between the entanglement spectrum crossings and the topology of the Liouvillian
can be resolved through Hermitianization of the jump operators.

\subsection{Quench dynamics with non-Hermitian and Hermitianized jump operators}

\begin{figure}
  \centering
  \includegraphics[width=\linewidth]{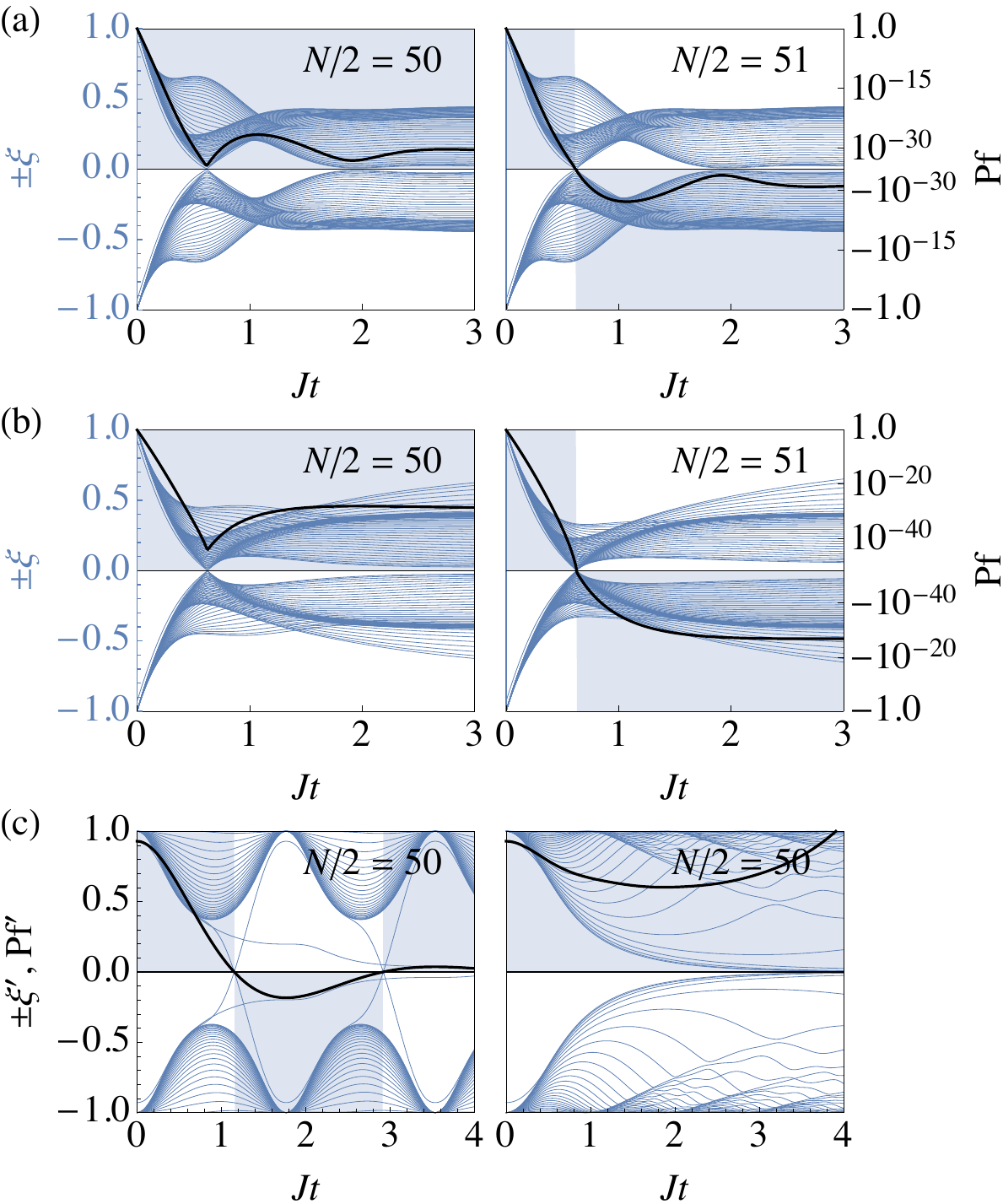}
  \caption{Quench dynamics for non-Hermitian jump operators. (a) Quench to the
    topological phase $\Rtop$ with $\gamma_{\mathrm{l}} = 0.3 J$ and
    $\gamma_{\mathrm{g}} = 0.7 J$ such that
    $\delta = \gamma_{\mathrm{l}} - \gamma_{\mathrm{g}} = - 0.4 J$, and
    $\mu_1 = 0$. Left panel: For $N/2 = 50$,
    $\sgn(\mathrm{Pf}_0) = \sgn(\mathrm{Pf}_{\mathrm{ss}}) = 1$, and the
    Pfaffian stays positive at all times. Right panel: For $N/2 = 51$,
    $\sgn(\mathrm{Pf}_{\mathrm{ss}}) = -1$, and the Pfaffian exhibits a single
    zero crossing. (b) Quench to the trivial phase $\Gedge$ for
    $\mu_1 = - 1.5 J$ and the same values of $\gamma_{\mathrm{l}}$
    and $\gamma_{\mathrm{g}}$ as in (a). Again, a single entanglement spectrum
    crossing occurs when $N/2$ is odd. (c) The connection between entanglement
    spectrum crossings and topology is restored for the deformed Liouvillian
    with Hermitianized jump operators and a trivial steady state. The left and
    right panels correspond to quenches with the parameters of (a) and (b),
    respectively. We show the rescaled entanglement spectrum
    $\xi' = \xi \e^{2 \left( \gamma_{\mathrm{l}} + \gamma_{\mathrm{g}} \right)
      t}$
    and Pfaffian
    $\mathrm{Pf}' = \mathop{\mathrm{Pf}} \e^{2 \left( \gamma_{\mathrm{l}} +
        \gamma_{\mathrm{g}} \right) N t}$
    for $N/2 = 50$. The corresponding plots for $N/2 = 51$ look essentially the
    same.}
  \label{fig:quench-dynamics-LDM-non-Hermitian}
\end{figure}
The impact of the even-odd effect of the Pfaffians of the initial and steady
states on the time evolution of the entanglement spectrum is illustrated in
Fig.~\ref{fig:quench-dynamics-LDM-non-Hermitian}. In particular,
Fig.~\ref{fig:quench-dynamics-LDM-non-Hermitian}(a) shows the evolution of the
single-particle entanglement spectrum for a quench to the topological phase
$\Rtop$ with $\gamma_{\mathrm{l}} = 0.3 J$, $\gamma_{\mathrm{g}} = 0.7 J$, and
$\mu_1 = 0$. For these parameter values, the bulk
spectrum~\eqref{eq:lambda-k-LDM} of the Liouvillian becomes flat, and we might
expect periodically recurring entanglement spectrum crossings. However, as shown
in the figure, there is no entanglement spectrum zero crossing for $N/2 = 50$,
and only a single crossing for $N/2 = 51$. These findings are in agreement with
our expectations based on Eqs.~\eqref{eq:sgn-Pf-ss} and~\eqref{eq:sgn-Pf-0}.

In the course of the time evolution, the Pfaffian drops rapidly to very small
values. Indeed, as discussed in Appendix~\ref{sec:pfaff-init-steady}, the
steady-state Pfaffian $\mathrm{Pf}_{\mathrm{ss}}$ is exponentially small in
$N$. However, there we also show that the parity $p_{A, \mathrm{ss}}$, and,
therefore, the sign of $\mathrm{Pf}_{\mathrm{ss}}$ are well-defined for
arbitrarily large systems.

Quenches to the gapless trivial phase $\Gedge$ with the same values of $\gamma_{\mathrm{l}}$ and
$\gamma_{\mathrm{g}}$ as above and $\mu_1 = -1.5 J$ are shown in
Fig.~\ref{fig:quench-dynamics-LDM-non-Hermitian}(b). Again, there is no
entanglement spectrum zero crossing for $N/2 = 50$, and only a single zero crossing
for $N/2 = 51$.

The presence or absence of entanglement spectrum crossings can be reconciled
with the topological properties of the Liouvillian $\hat{\mathcal{L}}$ through
Hermitianization of the jump operators, i.e., by continuously deforming the jump
operators to make them Hermitian as described in
Ref.~\cite{Lieu2019}. Hermitianization of the jump operators corresponds to a
transformation of the matrices $X$ and $Y$ that determine the Liouvillian
through Eq.~\eqref{eq:Liouvillian-fermionic-superops}, in the course of which
$X = \mathrm{const.}$, while $Y \to 0$. That is to say, the spectral gap and the
symmetries of the Liouvillian are preserved, whereas the steady state becomes
trivial with $\mathrm{Pf}_{\mathrm{ss}} = 0$. Consequently, for a master
equation with Hermitianized jump operators, the even-odd half-system size effect
is resolved.

In particular, for the quenches shown in
Figs.~\ref{fig:quench-dynamics-LDM-non-Hermitian}(a)
and~\ref{fig:quench-dynamics-LDM-non-Hermitian}(b), the time evolution of the
entanglement spectrum with Hermitianized jump operators is shown in,
respectively, the left and right panel of
Fig.~\ref{fig:quench-dynamics-LDM-non-Hermitian}(c). Both for $N/2 = 50$ as
shown in the figure and for $N/2 = 51$, entanglement spectrum crossings occur
only when the non-Hermitian winding number of the postquench Liouvillian takes
the value $W = 1$.

\begin{figure}
  \centering  
  \includegraphics[width=.9\linewidth]{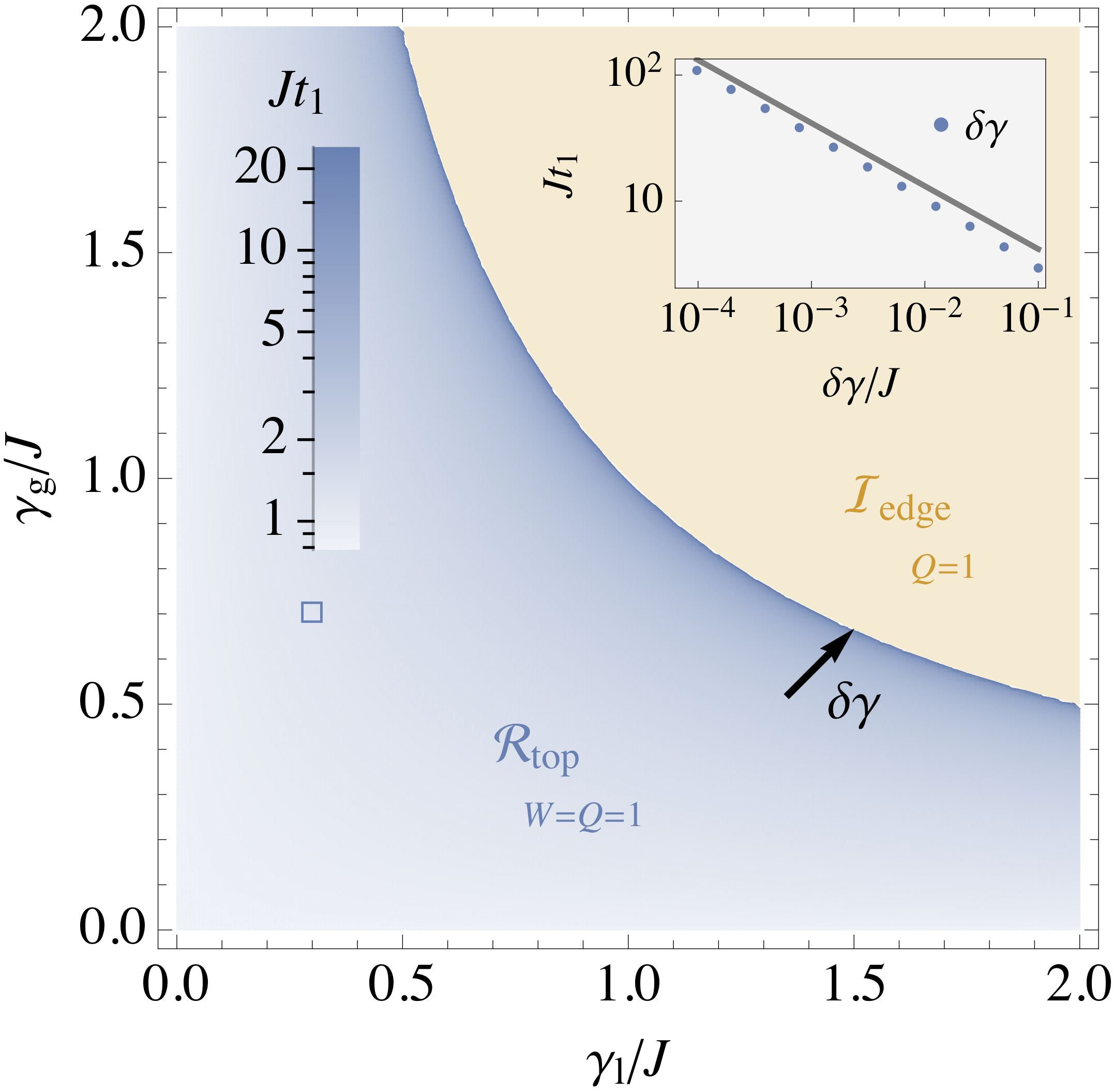}
  \caption{Dynamical topological phase diagram for Hermitianized jump operators
    and with $\mu_1 = 0$. Labels of the phases and the color scale are the same as in
    Fig.~\ref{fig:phase-diagram-LDM}. Inset: As for the case of Hermitian jump
    operators, $t_1$ diverges at the boundary of $\Rtop$ with a critical
    exponent of $\epsilon = 1/2$. The data shown corresponds to approaching the
    phase boundary along the black arrow labeled by $\delta \gamma$ in the main
    panel, i.e.,
    $\gamma_{\mathrm{l}} = \gamma_{\mathrm{l}, \mathrm{c}} - \delta \gamma$ and
    $\gamma_{\mathrm{g}} = \gamma_{\mathrm{g}, \mathrm{c}} - \delta \gamma$,
    where $\gamma_{\mathrm{l}, \mathrm{c}} = 1.5 J$ and, according to
    Eq.~\eqref{eq:phase-boundary-R},
    $\gamma_{\mathrm{g}, \mathrm{c}} = J^2/\gamma_{\mathrm{l}, \mathrm{c}}$. For
    comparison, the gray solid line shows a square-root singularity. The system
    size is $N = 100$ both in the inset and the main panel.}
  \label{fig:phase-diagram-LDM-gammal-gammag}
\end{figure}

Hermitianization of the jump operators allows us to explore the entire
topological phase diagram for $\gamma_{\mathrm{l}} \neq \gamma_{\mathrm{g}}$ in
terms of entanglement spectrum zero crossings. This is illustrated in
Fig.~\ref{fig:phase-diagram-LDM-gammal-gammag}. For the value $\mu_1 = 0$ shown
in the figure, the real-line gapped phase $\Rtop$ borders the phase $\Iedge$
with an imaginary line gap. The time scale $t_1$, at which the first
entanglement spectrum crossing occurs, diverges at the boundary of the
topological phase $\Rtop$. In particular, as shown in the inset of
Fig.~\ref{fig:phase-diagram-LDM-gammal-gammag}, $t_1$ exhibits power-law
behavior, $t_1 \sim \delta \gamma^{- \epsilon}$. For the value of the exponent,
we find $\epsilon = 1/2$ in agreement with our results for Hermitian jump
operators.

\section{Conclusions and Outlook}
\label{sec:outlook}

Our work establishes the entanglement spectrum as a tool to study non-Hermitian
dynamical topology of driven and open quantum many-body systems. We circumvent
the problem that complex spectra do not allow to define ground
states~\cite{Herviou2019} by considering the time evolution of entanglement
spectra after a quench.

We show that for systems with Hermitian or Hermitianized jump operators, the
presence of entanglement spectrum crossings reflects nontrivial topology of
individual complex bands of the Liouvillian. The topology of individual bands is
characterized by the non-Hermitian winding number $W$. In contrast, as discussed
in Ref.~\cite{Lieu2018}, the presence of edge modes is tied to the global Berry
phase $Q$, which pertains to the full spectrum, and is well-defined also in
gapless phases. The existence of two distinct topological transitions, at which,
respectively, the value of the non-Hermitian winding number $W$ and the global
Berry phase $Q$ changes, is unique to non-Hermitian systems. Indeed, as shown in
the dynamical topological phase diagram in Fig.~\ref{fig:phase-diagram-LDM}, the
corresponding phase boundaries between $\Rtop$ and $\Gedge$ as well as $\Gedge$
and $\Gtr$, coincide in the Hermitian limit where the rate of dissipation
$\gamma$ vanishes.

The characterization of non-Hermitian dynamical topology through the
entanglement spectrum opens up interesting directions for future research. While
we focus here on a driven and dissipative version of the Kitaev chain, which
belongs to the non-Hermitian symmetry class
$\mathrm{BDI}^{\dagger}$~\cite{Kawabata2019}, our approach can be applied
directly to Liouvillian generalizations~\cite{Lieu2019} of the Altland-Zirnbauer
symmetry classes that accommodate nontrivial $(1 + 1)$-dimensional dynamical
topology~\cite{Gong2017a}.

Apart from different one-dimensional and noninteracting models, as well as other
symmetry classes, a promising prospect is to generalize the
entanglement spectrum bulk-edge correspondence to quench dynamics of
driven-dissipative systems in higher spatial dimensions, and of interacting open
quantum many-body systems. Analogous generalizations were obtained for isolated
Hermitian systems in Refs.~\cite{Chang2018, Gong2017a}. For noninteracting Hermitian and non-Hermitian systems, the existence of a band structure is sufficient to define topological invariants, even without referring to a particular state. By contrast, for interacting Hermitian systems, a well-defined ground state commonly forms the basis of any discussion of topology~\cite{Chui2016}. This raises the following questions: How can the topology of the complex spectra of {\it quartic} Liouvillians, which describe interacting open quantum many-body systems, be classified? Do sharp dynamical topological transitions exist also for such systems? Our approach based on the entanglement spectrum provides a way to address, in particular, the latter question. For example, we expect that also for a generalization of an interacting Kitaev chain~\cite{Fidkowski2010}, which includes Markovian drive and dissipation, the time evolution of the many-body entanglement spectrum carries signatures of non-Hermitian dynamical topology of the quartic Liouvillian.

A substantial part of our work is concerned with a detailed investigation of dynamical criticality at the two distinct topological phase transitions which are related to the non-Hermitian winding number $W$ and the global Berry phase $Q$. We find strikingly different behavior: While the
transition that is described by $Q$ exhibits conventional critical relaxation,
we observe the phenomenon of many-body critical damping at the
boundaries of the topological phase $\Rtop$ with $W = 1$.

Our analysis of dynamical criticality, and, in particular, of many-body critical
damping, opens up new avenues for future research, which bear relevance beyond
the field of non-Hermitian topology. An obvious question is whether it is
possible to identify different universality classes of many-body critical
damping. To address this question, it will be interesting to study, e.g., a
driven and open Kitaev chain with long-range hopping and
pairing~\cite{Vodola2014, Alecce2017, Maity2020}, or to explore whether the dynamical critical exponent for entanglement spectrum crossings is renormalized due to interactions.

The focus of the present work is to explore theoretically signatures of
non-Hermitian topology in the time evolution of the entanglement spectrum. A
natural next step is to devise experimental implementations of the physics
discussed in our work. The required building blocks are available in synthetic
quantum systems: First, realizations of Kitaev chain have been proposed in
systems of cold atoms~\cite{Jiang2011, Nascimbene2013, Hu2015a,
  Kraus2013}. Alternatively, the transverse field Ising model, to which the
Kitaev chain is mapped through the Jordan-Wigner
transformation~\cite{Fendley2012}, can be implemented in quantum simulators of
many-body spin systems such as trapped atomic ions~\cite{Monroe2019}, Rydberg
atom arrays~\cite{Browaeys2020}, or superconducting qubits~\cite{Viehmann2012,
  Barends2016}. Further, Markovian dissipation, which is described by a quantum
master equation with Hermitian jump operators, can be induced by subjecting the
system to classical noise. This approach has been used, e.g., to implement local
dephasing of individual spins in a recent experiment with trapped
ions~\cite{Maier2019}.  Finally, the measurement of entanglement spectra in
various experimental platforms has been developed in Refs.~\cite{Pichler2016,
  Dalmonte2018, Kokail2020}.

These developments open up the prospect to experimentally explore non-Hermitian
topology of driven and open quantum many-body systems, and thus go fundamentally
beyond the paradigm of classical non-Hermitian wave physics. A comprehensive
understanding of the robust signatures of non-Hermitian topology in quantum
many-body systems is central for this endeavour. Our work, which identifies
entanglement spectrum crossings as such a signature, represents an important
step in this direction.

\section*{Acknowledgements}

We thank Loic Herviou, Max McGinley, and Frank Pollmann for helpful
discussions. AGG\ and SS\ acknowledge funding from the project
ANR-18-CE30-0001-01 (TOPODRIVE). JY acknowledges support from the European
Union’s Horizon 2020 research and innovation programme under Grant Agreement
No. 731473 (FWF QuantERA via QTFLAG I03769). LMS\ acknowledges support from the
Austrian Science Fund (FWF) through the project P 33741-N.

\appendix

\section{Driven-dissipative Kitaev chain with quasilocal Hermitian jump
  operators}
\label{sec:quasiloc-jump-operator}

Our key findings, which are summarized in Sec.~\ref{sec:summary-main-results},
are derived for a driven-dissipative Kitaev chain with strictly local jump
operators. In the following, we provide evidence for the general validity of our
conclusions by presenting results for the quasilocal Hermitian jump operators
given in Eq.~\eqref{eq:jump-operator-QLDM}.

\begin{figure}
  \centering  
  \includegraphics[width=\linewidth]{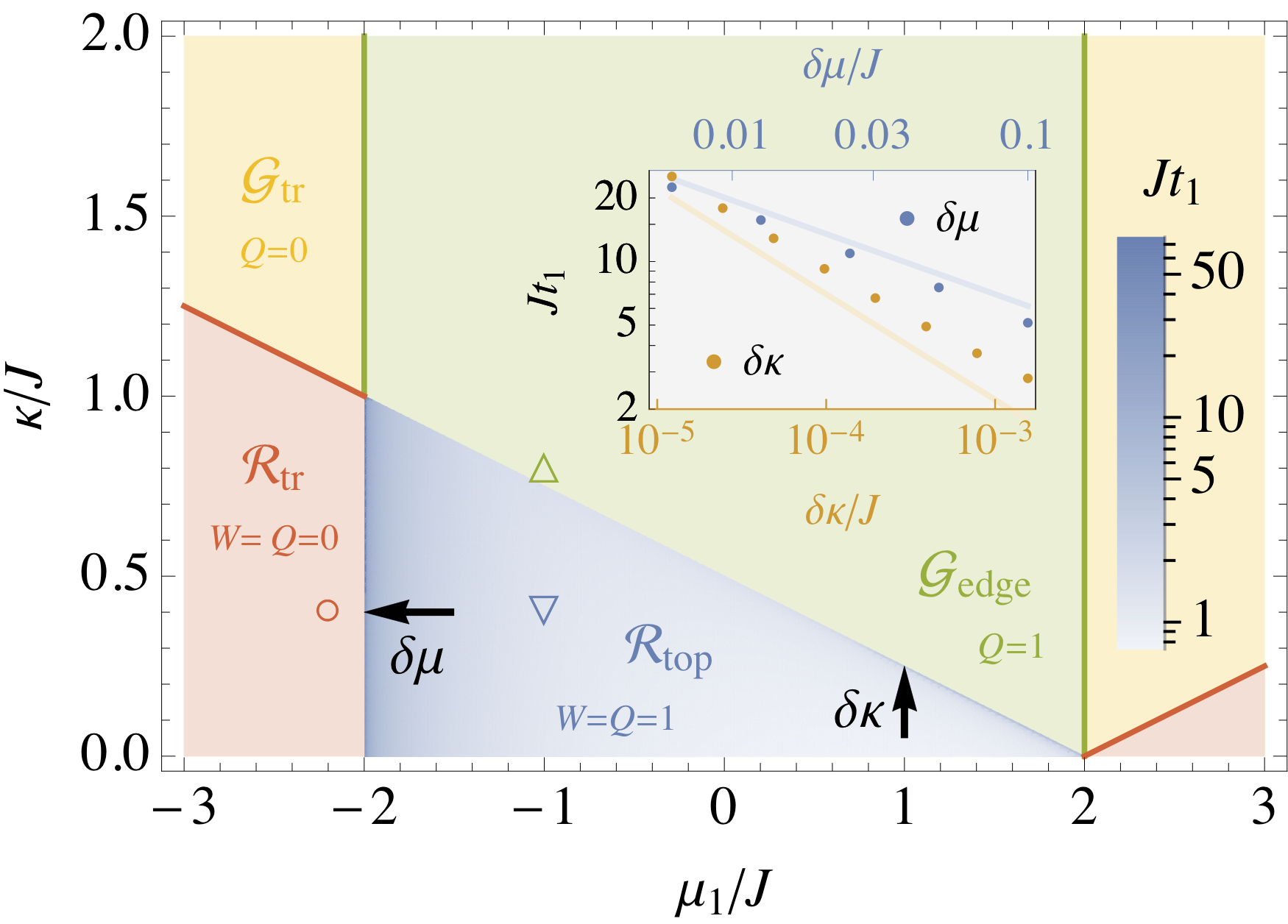}
  \caption{Dynamical topological phase diagram of a driven-dissipative Kitaev
    chain with quasilocal jump operators. The designations of different phases
    and the logarithmic color scale in $\Rtop$ are the same as in
    Fig.~\ref{fig:phase-diagram-LDM}. The values of $t_1$ are obtained for a
    quench with a half-flattened Liouvillian as explained in
    Sec.~\ref{sec:non-hermitian-band-QLDM}. Light blue and orange lines in the
    inset correspond to square-root singularities and have different slopes due
    to different scales for $\delta \mu$ and $\delta \kappa$. The system size is
    $N = 100$ both in the inset and the main panel.}
  \label{fig:phase-diagram-QLDM}
\end{figure}

The dynamical topological phase diagram of the driven-dissipative Kitaev chain
with quasilocal jump operators is shown in Fig.~\ref{fig:phase-diagram-QLDM}. A
comparison with Fig.~\ref{fig:phase-diagram-LDM} reveals two important
qualitative differences: First, for quasilocal jump operators, there is a direct
phase boundary between the real-line gapped topological $\Rtop$ and trivial $\Rtr$ phases, even at finite values of the dissipation rate $\kappa$. Second, the model
with quasilocal jump operators does not exhibit phases with an imaginary line
gap. We proceed with a concise summary of the key properties of the
driven-dissipative Kitaev chain with quasilocal jump operators.

\subsection{Non-Hermitian band theory}
\label{sec:non-hermitian-band-QLDM}

\begin{figure}
  \centering
  \includegraphics[width=\linewidth]{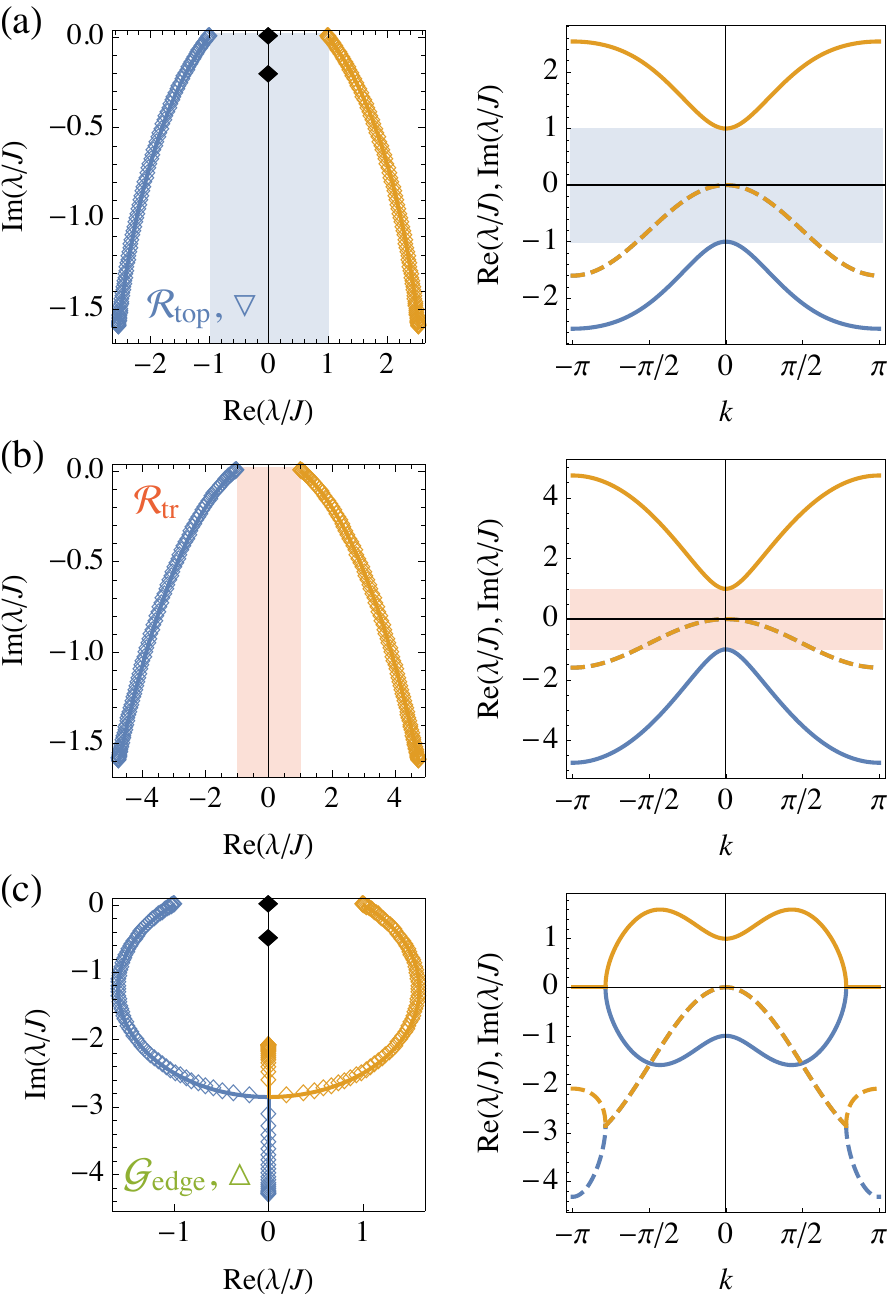}
  \caption{Liouvillian spectra of a driven-dissipative Kitaev chain with
    quasilocal jump operators and for $J = \Delta$. The notation and symbols are
    the same as those in Fig.~\ref{fig:Liouvillian-spectrum-LDM}. The spectrum
    exhibits a real line gap in (a) the topological phase $\Rtop$, shown here
    for $\kappa = 0.4 J$ and $\mu = - J$, and (b) the trivial phase $\Rtr$,
    shown for $\mu = -3 J$ and $\kappa = 0.4 J$. (c) Gapless spectrum in
    $\Gedge$ at $\mu = - J$ and $\kappa = 0.8 J$.}
  \label{fig:Liouvillian-spectrum-QLDM}
\end{figure}

The bulk spectrum of the Liouvillian is determined by the eigenvalues of the
matrix
\begin{equation}
  z_k = - \imag 2 \kappa \left( 1 - \cos(k) \right) \id + \mathbf{z}_k \cdot
  \boldsymbol{\sigma},
\end{equation}
where
\begin{equation}
  \mathbf{z}_k = \left( 2 \Delta \sin(k), 2 J \cos(k) + \mu, - \imag 2 \kappa
    \left( 1 - \cos(k)  \right) \right). 
\end{equation}
We obtain two complex bands which are given by
\begin{multline}
  \lambda_{\pm, k} = -\imag 2 \kappa \left( 1 - \cos(k) \right) \pm \left[ 4 \Delta^2
    \sin(k)^2 \vphantom{\left( 1 - \cos(k) \right)^2} \right. \\ \left. + \left(
      2 J \cos(k) + \mu \right)^2 - 4 \kappa^2 \left( 1 - \cos(k) \right)^2
  \right]^{1/2}.
\end{multline}
For $4 \kappa < \abs{2 J - \mu}$, the expression in brackets is positive for all
values of the quasimomentum $k$, and the spectrum exhibits a real line
gap. Spectra in the two phases $\Rtop$ and $\Rtr$ with a real line gap are shown
in Figs.~\ref{fig:Liouvillian-spectrum-QLDM}(a) and~\ref{fig:Liouvillian-spectrum-QLDM}(b), respectively.

As illustrated in Fig.~\ref{fig:Liouvillian-spectrum-QLDM}(c), the spectrum is
gapless for $4 \kappa > \abs{2 J - \mu}$. In the phase diagram in
Fig.~\ref{fig:phase-diagram-QLDM}, the two gapless phases are designated by
$\Gedge$ and $\Gtr$. Unlike the case of local jump operators, the spectrum
remains gapless even for high values of $\kappa$, and the present model does not
have phases with an imaginary line gap.

The symmetries stated in Eq.~\eqref{eq:symmetries} remain intact for quasilocal
jump operators, and the system still belongs to the class
$\mathrm{BDI}^{\dagger}$. We omit details of the straightforward calculation of
the non-Hermitian winding number $W$ for brevity. This calculation yields
$W = 1$ in $\Rtop$ and $W = 0$ in $\Rtr$. Further, the global Berry phase is
given by $Q = 1$ for $\abs{\mu} < 2 J$ and $Q = 0$ for $\abs{\mu} > 2 J$. Edge
modes, which occur in $\Rtop$ and $\Gedge$ where $Q = 1$, are indicated by black
diamond symbols in Fig.~\ref{fig:Liouvillian-spectrum-QLDM}. At the critical
lines $\mu = \pm 2 J$, where the edge modes disappear, the point gap of the bands
$\lambda_{\pm, k}$ at $\lambda = 0$ closes.

A notable difference between the spectra shown in
Figs.~\ref{fig:Liouvillian-spectrum-LDM} and~\ref{fig:Liouvillian-spectrum-QLDM}
is the wide dispersion of imaginary parts $\Im(\lambda_{\pm, k})$ which occurs
in the case of quasilocal jump operators. The coexistence of modes with widely
disparate decay rates hampers the numerical analysis of the present model. In
particular, the time evolution of entanglement spectra, which we discuss in the
next section, is hard to resolve for values of $\kappa$ close to $\kappa = J$
within the topological phase $\Rtr$ in Fig.~\ref{fig:phase-diagram-QLDM}. To
mitigate this problem, we consider dynamics generated by a band-flattened
Liouvillian. Band-flattening can be achieved through a continuous deformation of
the original Liouvillian, and, therefore, does not affect its
topology~\cite{Kawabata2019}.

Our starting point is the third-quantized form of the Liouvillian
$\hat{\mathcal{L}}$ given in Eq.~\eqref{eq:Liouvillian-fermionic-superops}. As
explained in Sec.~\ref{sec:third-quantization}, the spectrum of
$\imag \hat{\mathcal{L}}$ is determined by the matrix
$Z = - \imag X^{\transpose}$. To resolve the problem of disparate decay rates,
it is sufficient to consider a half-flatttened Liouvillian, which we obtain by
replacing $Z = V \Lambda V^{-1}$, with
$\Lambda = \mathop{\mathrm{diag}}(\lambda_1, \dotsc, \lambda_{2 N})$ a diagonal
matrix, by $Z_{\mathrm{hf}} = V \Re(\Lambda) V^{-1}$, i.e., by removing the
decay rates $- \Im(\Lambda)$ by hand. We note that the spectrum becomes
completely flat if $Z$ is replaced by
$Z_{\mathrm{fl}} = J V \sgn(\Re(\Lambda)) V^{-1}$, where the prefactor $J$ is
introduced as an arbitrary overall energy scale. However, $Z_{\mathrm{fl}}$ is
well-defined only if there are no eigenvalues of $Z$ with vanishing real
part. The time-evolved entanglement spectra which we discuss below, as well as
the data for $t_1$ shown in Fig.~\ref{fig:phase-diagram-QLDM}, are obtained with
a half-flattened Liouvillian.

\subsection{time evolution of entanglement spectra}
\label{sec:time-evol-entangl-QLDM}

\begin{figure}
  \centering
    \includegraphics[width=\linewidth]{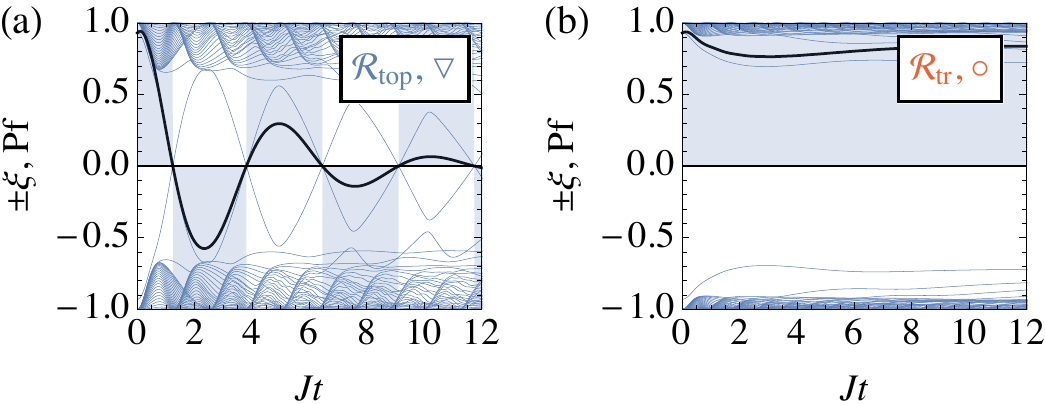}
    \caption{Time evolution of entanglement spectra. The dynamics is generated
      by a half-flattened Liouvillian as explained in the main text. (a) Zero
      crossings in the entangement spectrum occur for a quench to $\Rtop$ with
      $W = 1$ at $\mu_1 = -J$ and $\kappa = 0.4 J$. (b) The entanglement
      spectrum remains gapped for a quench to $\Rtr$ with $W = 0$ at
      $\mu_1 = - 2.2 J$ and $\kappa = 0.4 J$.}
  \label{fig:quench-dynamics-QLDM}
\end{figure}

Entanglement spectrum crossings reveal non-Hermitian topology of the Liouvillian
also for a driven-dissipative Kitaev chain with quasilocal Hermitian jump
operators. This is illustrated in Figs.~\ref{fig:quench-dynamics-QLDM}(a) and~\ref{fig:quench-dynamics-QLDM}(b)
for quenches to $\Rtop$ and $\Rtr$, respectively. For time evolution with a
half-flattened Liouvillian, the entangement spectrum does not show significant
decay, and most entanglement eigenvalues remain close to $\pm 1$.

A quantitative analysis of the time $t_1$ of the first entanglement spectrum
zero crossing within the entire phase $\Rtop$ is shown in
Fig.~\ref{fig:phase-diagram-QLDM}. The logarithmic color scale indicates that
$t_1$ diverges at the boundaries of $\Rtop$. This is illustrated further in the
inset, which shows the increase of $t_1$ upon approaching the phase boundary
along the black arrows in the main panel, i.e., for
$\delta \mu = \mu_1 + 2 J \to 0$ with $\kappa = 0.4 J$, and for $\mu_1 = J$ with
$\delta \kappa = J/4 - \kappa \to 0$. The numerical data is consistent with a
square-root singularity, and, therefore, corroborates the universality of the
value $\epsilon = 1/2$ of the dynamical critical exponent for
entanglement spectrum crossings.

\section{Covariance matrix in the ground state}
\label{sec:cov-mat-gs}

To find the covariance matrix of the ground state of the Kitaev chain, we start
from the Majorana representation of the Hamiltonian given in
Eq.~\eqref{eq:H-L-Majorana}. The matrix $A$ in this equation is real and
antisymmetric. Therefore, there exists a real special orthogonal matrix $O$
which brings $A$ to the following canonical block-diagonal form:
\begin{equation}
  \label{eq:A-canoncial-form}
  O A O^{\transpose} = \bigoplus\limits_{i = 1}^N
  \begin{pmatrix}
    0 & - \epsilon_i \\ \epsilon_i & 0
  \end{pmatrix},
\end{equation}
where $\epsilon_i \geq 0$. The matrix $O$ can be found by employing the
algorithm described in Ref.~\cite{Wimmer2012}. Then, the covariance matrix of
the ground state is given by~\cite{Kraus2009}
\begin{equation}
  \Gamma_{A, 0} = O^{\transpose} \left( \bigoplus\limits_{i = 1}^N \imag
    \sigma_y \right) O.
\end{equation}

\section{Exactly solvable quench dynamics}
\label{sec:fermion-parity-pump}

Here we derive the exact solution
in Eq.~\eqref{eq:reduced-density-matrix-flat-band} for the time evolution of the
reduced density matrix. We obtain this exact solution for a specific choice of
pre- and postequench parameters as detailed in the following.

\subsection{Initial state}
\label{sec:initial-state}

We assume that the system is initialized in the ground state $\ket{\psi_0}$ of
the Hamiltonian of the Kitaev chain for $J = \Delta > 0$ and $\mu_0 \to - \infty$, which is the vacuum of Dirac fermions $c_i$, i.e.,
$\ket{\psi_0} = \ket{\Omega}$ with $c_i \ket{\Omega} = 0$ for all
$i = 1, \dotsc, N$. The corresponding density matrix is
\begin{equation}
  \label{eq:rho-0}
  \rho_0 = \ket{\psi_0} \bra{\psi_0} = \ket{\Omega} \bra{\Omega} = \prod_{i
    = 1}^N \rho_{i, 0}, \quad \rho_{i, 0} = P^0_i.
\end{equation}
Here and below, we use the following notation for projectors on states for which
the lattice site $i$ is empty and occupied, respectively:
\begin{equation}
  \label{eq:P-0-1}
  P^0_i = 1 - n_i = \ket{0}_i \bra{0}, \qquad P^1_i = n_i = \ket{1}_i \bra{1},
\end{equation}
where the site occupation number operator is $n_i = c_i^{\dagger} c_i$.

\subsection{Postquench Liouvillian}
\label{sec:flat-band-Liouvillian}

For the choice of postquench parameters $J = \Delta$, $\mu_1 = 0$, and
$\gamma_{\mathrm{l}} = \gamma_{\mathrm{g}} = \gamma$, the third-quantized
Liouvillian~\eqref{eq:Liouvillian-fermionic-superops} takes the form
\begin{equation}
  \label{eq:flat-band-L}
  \hat{\mathcal{L}} = - 4 \gamma \hat{n}_1 - 2 \sum_{i = 1}^{N - 1} \left(
    \hat{c}_{2 i}^{\dagger}, \hat{c}_{2 i + 1}^{\dagger} \right)
    \begin{pmatrix}
      0 & J \\ - J & 2 \gamma
    \end{pmatrix}
    \begin{pmatrix}
      \hat{c}_{2 i} \\ \hat{c}_{2 i + 1}
    \end{pmatrix},
\end{equation}
where the site occupation number superoperator is defined as
$\hat{n}_i = \hat{c}_i^{\dagger} \hat{c}_i$. Left and right edge modes are
created by the superoperators $\hat{c}_L^{\dagger} = \hat{c}_1^{\dagger}$ and
$\hat{c}_R^{\dagger} = \hat{c}_{2 N}^{\dagger}$, respectively. The corresponding
eigenvalues of $\imag \hat{\mathcal{L}}$ are $\lambda_L = - \imag 4 \gamma$ and
$\lambda_R = 0$. Since $\Re(\lambda_L) = \Re(\lambda_R) = 0$, the dynamics of
these modes is purely dissipative. In contrast, the dynamics of bulk modes has
both oscillatory and dissipative components. The bulk spectrum is flat and
consist of two values $\lambda_{\pm}$, which are determined by the eigenvalues
of the matrix in the second term in Eq.~\eqref{eq:flat-band-L}, and are given in
Eq.~\eqref{eq:lambda-flat}.

To separate the dissipative and oscillatory contributions to the dynamics of the
bulk, we decompose the Liouvillian as
$\hat{\mathcal{L}} = \hat{\mathcal{L}}_0 + \hat{\mathcal{L}}_1$, where
\begin{equation}
  \label{eq:L-0-1}
  \hat{\mathcal{L}}_0 = \sum_{i = 1}^N \hat{\mathcal{L}}_{0, i}, \qquad \hat{\mathcal{L}}_1 =
  \sum_{i = 1}^{N-1} \hat{\mathcal{L}}_{1, i} 
\end{equation}
with
\begin{equation}
  \label{eq:L-0-i}
  \hat{\mathcal{L}}_{0, i} =
  \begin{cases}
    - 2 \gamma \left( 2 \hat{n}_1 + \hat{n}_2 \right) & \text{for } i = 1, \\
    - 2 \gamma \left( \hat{n}_{2 i - 1} + \hat{n}_{2 i} \right) & \text{for } i
    = 2, \dotsc N-1, \\
    - 2 \gamma \hat{n}_{2 N-1} & \text{for } i = N ,
  \end{cases}
\end{equation}
and
\begin{equation}
  \label{eq:L-1-i}  
  \hat{\mathcal{L}}_{1, i} = 2 \left[ \gamma \left( \hat{n}_{2 i} - \hat{n}_{2 i + 1} \right) - J
    \left( \hat{c}_{2 i}^{\dagger} \hat{c}_{2 i + 1} - \hat{c}_{2 i +
        1}^{\dagger} \hat{c}_{2 i}
    \right) \right].
\end{equation}
With the aid of anticommutation relation of superoperators stated in
Sec.~\ref{sec:third-quantization}, it is straightforward to check that
$\hat{\mathcal{L}}_0$ and $\hat{\mathcal{L}}_1$ commute. Therefore, the
time evolution superoperator factorizes as
\begin{equation}
  \e^{\hat{\mathcal{L}} t} = \e^{\hat{\mathcal{L}}_1 t} \e^{\hat{\mathcal{L}}_0 t}.
\end{equation}
We proceed to analyze first the dissipative, and then the oscillatory component
of the dynamics.

\subsection{Dissipative dynamics}
\label{sec:dissipative-dynamics}

The Liouvillian $\hat{\mathcal{L}}_0$ in Eq.~\eqref{eq:L-0-1} is a sum of
commuting terms $\hat{\mathcal{L}}_{0, i}$, which describe the evolution of
fermions on lattice site $i$. Further, since the initial density
matrix~\eqref{eq:rho-0} factorizes in real space, we obtain
\begin{equation}
  \label{eq:exp-L-0-t-rho-0}
  \e^{\hat{\mathcal{L}}_0 t} \rho_0 = \prod_{i = 1}^N \e^{\hat{\mathcal{L}}_{0, i} t}
  \rho_{0, i},
\end{equation}
and, therefore, we can consider each site $i$ independently.

We begin by collecting a few useful relations: First, by expanding the
exponential in a power series and using that $\hat{n}_i^2 = \hat{n}_i$, we find
\begin{equation}
  \e^{-2 \gamma t \hat{n}_i} = \hat{1} - \left( 1 - \e^{- 2 \gamma t} \right)\hat{n}_i.
\end{equation}
Next, from the Majorana representation of the site occupation number operator,
$n_i = \frac{1}{2} \left( 1 - \imag w_{2 i - 1} w_{2 i} \right)$, it follows
that
\begin{equation}
  \label{eq:n-hat-n}
  \hat{n}_{2 i - 1} n_i = \hat{n}_{2 i} n_i = - \frac{\imag}{2} w_{2 i - 1}
  w_{2 i} = n_i - \frac{1}{2}.
\end{equation}
With these relations, it is straightforward to show that
\begin{multline}
  \e^{-2 \gamma t \hat{n}_{2 i - 1}} P^0_i = \e^{-2 \gamma t
    \hat{n}_{2 i}} P^0_i \\ = \frac{1}{2} \left[ \left( 1 +
      \e^{-2 \gamma t} \right) P^0_i + \left( 1 - \e^{-2 \gamma
        t} \right) P^1_i \right],
\end{multline}
and
\begin{multline}  
  \e^{-2 \gamma t \hat{n}_{2 i - 1}} P^1_i = \e^{-2 \gamma t \hat{n}_{2 i}} P^1_i \\
  = \frac{1}{2} \left[ \left( 1 - \e^{-2 \gamma t} \right) P^0_i + \left( 1 +
      \e^{-2 \gamma t} \right) P^1_i \right],
\end{multline}
where the projectors $P^0_i$ and $P^1_i$ are defined in
Eq.~\eqref{eq:P-0-1}. Thus, the evolution~\eqref{eq:exp-L-0-t-rho-0} of the
density matrix under the local Liouvillians $\hat{\mathcal{L}}_{0, i}$ is given by
\begin{equation}
  \label{eq:exp-L-0-i}
  \e^{\hat{\mathcal{L}}_{0, i} t} \rho_{0, i} = \frac{1}{2} \sum_{n = 0}^1 
  \left[ 1 + \left( -1 \right)^n \e^{- 2 \gamma_i t} \right] P^n_i,
\end{equation}
where $\gamma_1 = 3 \gamma$, $\gamma_N = \gamma$, and $\gamma_i = 2 \gamma$ for
$i = 2, \dotsc, N - 1$.

\subsection{Oscillatory dynamics}
\label{sec:oscillatory-dynamics}

We next consider the oscillatory component $\hat{\mathcal{L}}_1$. The local
contributions $\hat{\mathcal{L}}_{1, i}$ in Eq.~\eqref{eq:L-1-i} commute, and we
can again consider the evolution which is generated by each of them
separately. Since $\hat{\mathcal{L}}_{1, i}$ commutes with the sum of number
superoperators $\hat{N}_i = \hat{n}_{2 i} + \hat{n}_{2 i + 1}$, it can be
decomposed into three contributions that correspond to the eigenspaces of
$\hat{N}_i$,
\begin{multline}
  \hat{\mathcal{L}}_{1, i} = \kket{1} \hat{\mathcal{L}}_{1, i}^{(0)} \bbra{1} \\ + \left(
    \kket{w_{2 i}}, \kket{w_{2 i + 1}} \right) \hat{\mathcal{L}}_{1, i}^{(1)}
  \begin{pmatrix}
    \bbra{w_{2 i}} \\ \bbra{w_{2 i + 1}}
  \end{pmatrix}
  \\ + \kket{w_{2 i} w_{2 i + 1}} \hat{\mathcal{L}}_{1, i}^{(2)} \bbra{w_{2 i} w_{2 i + 1}},
\end{multline}
where
\begin{equation}
  \begin{split}
    \hat{\mathcal{L}}_{1, i}^{(0)} & = \bbraket{1 | \hat{\mathcal{L}}_{1, i} | 1} = 0, \\
    \hat{\mathcal{L}}_{1, i}^{(2)} & = \bbraket{w_{2 i} w_{2 i + 1} | \hat{\mathcal{L}}_{1, i} |
      w_{2 i} w_{2 i + 1}} = 0,
  \end{split}
\end{equation}
and
\begin{equation}
  \begin{split}
    \hat{\mathcal{L}}_{1, i}^{(1)} & =
    \begin{pmatrix}
      \bbraket{w_{2 i} | \hat{\mathcal{L}}_{1, i} | w_{2 i}} & \bbraket{w_{2 i} |
        \hat{\mathcal{L}}_{1, i} | w_{2 i + 1}} \\
      \bbraket{w_{2 i + 1} | \hat{\mathcal{L}}_{1, i} | w_{2 i}} & \bbraket{w_{2 i +
          1} | \hat{\mathcal{L}}_{1, i} | w_{2 i + 1}} 
    \end{pmatrix},
    \\ & = 2
    \begin{pmatrix}
      \gamma & - J \\ J & - \gamma
    \end{pmatrix}
    = 2 \left( \gamma \sigma_z - \imag J \sigma_y \right).
  \end{split}
\end{equation}
To calculate $\e^{\hat{\mathcal{L}}_{1, i}^{(1)} t}$, we use the following relation which
holds for $\mathbf{a} \in \C^3$ and where
$\boldsymbol{\sigma} = \left( \sigma_x, \sigma_y, \sigma_z \right)$:
\begin{equation}
  \e^{- \imag \mathbf{a} \cdot \boldsymbol{\sigma} t} = \cos(a t) \id -
  \imag \sin(a t) \frac{\mathbf{a} \cdot
    \boldsymbol{\sigma}}{a}, \quad a^2 = \sum_{i = 1}^3
  a_i^2.
\end{equation}  
For $\mathbf{a} = 2 \left( 0, J, \imag \gamma \right)$, we find $a = \pi/T$
where $T$ is given in Eq.~\eqref{eq:T}, and thus
\begin{equation}
  \e^{\hat{\mathcal{L}}_{1, i}^{(1)} t} = \cos(\pi t/T) \id + \frac{2 T}{\pi}
  \sin(\pi t/T) \left( \gamma \sigma_z - \imag J \sigma_y \right).
\end{equation}
In particular, for $t = m T$, where $m \in \N_0$ is a nonnegative integer,
we obtain
\begin{equation}
  \e^{\hat{\mathcal{L}}_{1, i}^{(1)} m T} = \cos(\pi m) \id = \left( -1 \right)^m \id,
\end{equation}
and, therefore,
\begin{multline}
  \e^{\hat{\mathcal{L}}_{1, i} m T} = \kket{1} \bbra{1} + \kket{w_{2 i} w_{2 i + 1}}
  \bbra{w_{2 i} w_{2 i + 1}} \\ + \left( -1 \right)^m \left( \kket{w_{2 i}}
    \bbra{w_{2 i}} + \kket{w_{2 i + 1}} \bbra{w_{2 i + 1}} \right),
\end{multline}
which can also be written in the compact form
\begin{equation}
  \e^{\hat{\mathcal{L}}_{1, i} m T} = \e^{\imag \pi m \left( \hat{n}_{2 i} +
      \hat{n}_{2 i + 1} \right)}.
\end{equation}
The oscillatory evolution of the entire chain is thus given by the product
\begin{equation}
  \e^{\hat{\mathcal{L}}_1 m T} = \prod_{i = 1}^{N-1} \e^{\hat{\mathcal{L}}_{1, i} m T} = \e^{\imag \pi m \left(
      \hat{n}_1 + \hat{n}_{2 N} \right)} \hat{P}^m,
\end{equation}
where $\hat{P}$ is the parity superoperator which is defined as
\begin{equation}
  \label{eq:parity-superoperator}
  \hat{P} = \e^{\imag \pi \sum_{i = 1}^{2 N} \hat{n}_i} = \prod_{i = 1}^{2 N} \left(
    1 - 2 \hat{n}_i \right).
\end{equation}
Finally, we note that the evolution superoperator can be simplified for even and
odd values of $m$, respectively:
\begin{equation}
  \label{eq:exp-L-1-k-T}
  \e^{\hat{\mathcal{L}}_1 m T} =
  \begin{cases}
    \hat{1} & \text{for $m$ even,} \\ \left( \hat{1} - 2 \hat{n}_1
    \right) \left( \hat{1} - 2 \hat{n}_{2 N} \right) \hat{P} & \text{for $m$
      odd.}
  \end{cases}
\end{equation}

\subsection{Time evolution of the density matrix}
\label{sec:parity-pump}

We now combine Eq.~\eqref{eq:exp-L-0-i} with Eq.~\eqref{eq:exp-L-1-k-T} to find
the time-evolved density matrix $\rho(m T)$. According to
Eq.~\eqref{eq:exp-L-1-k-T}, at times $m T$, not only the dissipative evolution
under $\hat{\mathcal{L}}_0$, but also the oscillatory dynamics generated by
$\hat{\mathcal{L}}_1$ factorizes in real space:
\begin{equation}
  \rho(m T) = \e^{\hat{\mathcal{L}}_1 m T} \e^{\hat{\mathcal{L}}_0 m T} \rho_0 = \prod_{i =
    1}^N \rho_i(m T).
\end{equation}
To determine $\rho_i(m T)$ explicitly, we have to apply superoperators
$\hat{1} - 2 \hat{n}_i$, which occur in Eqs.~\eqref{eq:parity-superoperator}
and~\eqref{eq:exp-L-1-k-T}, to the projectors $P^n_i$ in
Eq.~\eqref{eq:exp-L-0-i}. This can be done with the aid of the relations
\begin{equation}
  \left( \hat{1} - 2 \hat{n}_i \right) P^0_i = P^1_i, \quad \left( \hat{1} - 2
    \hat{n}_i \right) P^1_i = P^0_i,
\end{equation}
which follow from Eqs.~\eqref{eq:P-0-1} and~\eqref{eq:n-hat-n}, and which
immediately lead to
\begin{equation}
  \left( 1 - 2 \hat{n}_{2 i - 1} \right) \left( 1 - 2 \hat{n}_{2 i} \right)
  P^n_i = P^n_i
\end{equation}
for $n = 0, 1$. We thus find
\begin{equation}
  \rho_i(m T) = \frac{1}{2} \sum_{n = 0}^1 \left[ 1 + \left( - 1 \right)^{m_i + n}
    \e^{-2 \gamma_i m T} \right] P^n_i,
\end{equation}
where $m_i = m$ for $i = 1, N$, and $m_i = 0$ for $i = 2, \dotsc, N-1$. Each of
the reduced density matrices for a single lattice site $i$ has trace one,
$\tr_i(\rho_i(m T)) = 1$. Therefore, the reduced density matrix for the left
half of the chain is simply given by the product
\begin{multline}
  \label{eq:almost-reduced-density-matrix-flat-band}
  \rho_A(m T) = \prod_{i = 1}^{N/2} \rho_i(m T) \\ = \sum_{\mathbf{n}}
  \frac{1}{2} \left[ 1 + \left( -1 \right)^{m + n_1} \xi_{1, m} \right]
  P_1^{n_1} \\ \times \prod_{i = 2}^{N/2} \frac{1}{2} \left[ 1 + \left( -1
    \right)^{n_i} \xi_{2, m} \right] P_i^{n_i},
\end{multline}
where the sum in the second equality is over all combinations
$\mathbf{n} = \left( n_1, \dotsc, n_{N/2} \right)$ of site occupation numbers,
and $\xi_{m, 1}$ and $\xi_{m, 2}$ are given in
Eq.~\eqref{eq:exact-single-particle-ES}. Finally, we obtain
Eq.~\eqref{eq:reduced-density-matrix-flat-band} by noting that
$P^{\mathbf{n}} = \ket{\mathbf{n}}_A \bra{\mathbf{n}} = \prod_{i = 1}^{N/2}
P_i^{n_i}$,
and by replacing the sum over site occupation numbers $\mathbf{n}$ in
Eq.~\eqref{eq:almost-reduced-density-matrix-flat-band} by a sum over
entanglement occupation numbers
$\mathbf{e} = \left( e_1, \dotsc, e_{N/2} \right)$ with
$e_1 = (n_1 + m) \bmod 2$ and $e_i = n_i$ for $i = 2, \dotsc, N/2$. Inverting
the relation between $\mathbf{e}$ and $\mathbf{n}$ leads to
Eq.~\eqref{eq:lattice-site-occupation-numbers}.

\section{Conservation of parity of entanglement eigenstates}
\label{sec:cons-parity-entangl}

In the following, we show that the fermion parity of any individual entanglement
eigenstate is conserved. To this end, we note first that the fermion parity
operator $P = \e^{\imag \pi \sum_{i = 1}^{2 N} n_i}$ is related to the parity
superoperator~\eqref{eq:parity-superoperator} via $\hat{P} \rho = P \rho P$.
This can be checked by expanding the density matrix $\rho$ in the basis of
products of Majorana operators specified in
Sec.~\ref{sec:third-quantization}. The relation between $\hat{P}$ and $P$
implies that the density matrix $\rho = \ket{\psi} \bra{\psi}$ for any pure
state $\ket{\psi}$ which has definite positive or negative parity has positive
superparity. In particular, this applies to the ground state of the Hamiltonian
of the Kitaev chain~\eqref{eq:H-Kitaev}, which is the initial state of the
quench protocol we consider in the present work.

The superparity of the initial state is conserved under time evolution with a
quadratic Liouvillian~\cite{Prosen2008}. This follows from the equality
$[\hat{P}, \hat{\mathcal{L}}] = 0$, which further implies that the density
matrix can be diagonalized in a basis of states with definite fermion parity at
all times,
\begin{equation}
  \label{eq:rho-parity-basis}
  \rho = \sum_n c_n \ket{\psi_n} \bra{\psi_n}, \qquad P \ket{\psi_n} = p_n
  \ket{\psi_n},    
\end{equation}
with $p_n = \pm1$. Consequently, also the reduced density matrix
$\rho_A = {\tr}_{A^{\mathrm{c}}}(\rho)$ for a subsystem $A$ can be diagonalized
in a basis of eigenstates of the corresponding fermion parity operator $P_A$,
which is defined in Eq.~\eqref{eq:fermion-parity},
\begin{equation}
  \label{eq:rho-L-parity-basis}
  \rho_A = \sum_l \Xi_l \ket{\phi_l} \bra{\phi_l}, \qquad P_A \ket{\phi_l} =
  p_{A, l} \ket{\phi_l}.
\end{equation}
This can be seen as follows: A single state $\ket{\psi_n}$, which occurs in
Eq.~\eqref{eq:rho-parity-basis}, can be written as
\begin{equation}
  \ket{\psi_n} = \sum_{\alpha, \beta} c_{\alpha \beta} \ket{e^A_{\alpha},
    e^{A^{\mathrm{c}}}_{\beta}} + \sum_{\alpha, \beta} d_{\alpha \beta} \ket{o^A_{\alpha},
    o^{A^{\mathrm{c}}}_{\beta}},
\end{equation}
where the states $\ket{e^{A, A^{\mathrm{c}}}_{\alpha}}$ and
$\ket{o^{A, A^{\mathrm{c}}}_{\alpha}}$ form a basis of states which contain,
respectively, an even and odd number of fermions in $A$ and its complement
$A^{\mathrm{c}}$, and for concreteness we assume that $\ket{\psi_n}$ has even
parity. Upon taking the trace over $A^{\mathrm{c}}$, we find
\begin{multline}
  \rho_A^n = {\tr}_{A^{\mathrm{c}}}(\ket{\psi_n} \bra{\psi_n}) = \sum_{\alpha,
    \alpha', \beta} c_{\alpha \beta} c_{\alpha' \beta}^{*} \ket{e^A_{\alpha}}
  \bra{e^A_{\alpha'}} \\ + \sum_{\alpha, \alpha', \beta} d_{\alpha \beta}
  d_{\alpha' \beta}^{*} \ket{o^A_{\alpha}} \bra{o^A_{\alpha'}}.
\end{multline}
That is, the reduced density matrix $\rho_A^n$ is composed of two disconnected
blocks corresponding to states with even and odd particle number,
respectively. It follows that $P_A \rho_A^n P_A = \rho_A^n$. An analogous
argument applies to states $\ket{\psi_n}$ with odd parity and, therefore, to all
states in the sum in Eq.~\eqref{eq:rho-parity-basis}. This proves
Eq.~\eqref{eq:rho-L-parity-basis}, according to which entanglement eigenstates
have definite parity at all times.

\section{Retarded response function}
\label{sec:ret-resp-fct}

Here we provide details on the calculation of the retarded response function
defined in Eq.~\eqref{eq:retarded-response}. Two-time averages can be obtained
with the aid of the quantum regression theorem~\cite{Gardiner2014}, which yields
for $t > 0$
\begin{equation}
  \label{eq:quantum-regression-theorem}
  \begin{split}
    \langle w_i(t) w_j \rangle & = \tr( w_i \e^{\mathcal{L} t} ( w_j
    \rho_{\mathrm{ss}} )), \\
    \langle w_j w_i(t) \rangle & = \tr( w_i \e^{\mathcal{L} t} (
    \rho_{\mathrm{ss}} w_j )), \\
  \end{split}
\end{equation}
and, therefore,
\begin{equation}
  \label{eq:ret-resp-quantum-regression-theorem}
  \chi^R_{i, j}(t) = \frac{1}{2} \theta(t) \tr( w_i \e^{\mathcal{L} t} \{ w_j,
  \rho_{\mathrm{ss}} \}) = \theta(t) \tr( w_i \e^{\mathcal{L} t} w_j).
\end{equation}
In the second equality, we specified the trivial steady state
$\rho_{\mathrm{ss}} = \id/D$. A closed equation of motion can be derived for the
response matrix, which collects all possible combinations of odd and even
position indices,
\begin{equation}
  \label{eq:R}
  R_{i, j}(t) =
  \begin{pmatrix}
    \chi^R_{2 i - 1, 2 j - 1}(t) & \chi^R_{2 i - 1, 2 j}(t) \\
    \chi^R_{2 i, 2 j - 1}(t) & \chi^R_{2 i, 2 j}(t)
  \end{pmatrix}.
\end{equation}
The equation of motion, which results from taking the derivative of
Eq.~\eqref{eq:ret-resp-quantum-regression-theorem}, takes a particularly simple
form in momentum space:
\begin{equation}
  \label{eq:R-EOM}
  \frac{\mathrm{d} R_k}{\mathrm{d} t} = \delta(t) \id + R_k \widetilde{x}_{-k},
\end{equation}
where $R_k(t) = \sum_{j \in \Z} \e^{-\imag k j} R_{i + j, i}(t)$,
$\widetilde{x}_k = x_k - 4 \tr(M)$, and $x_k = \imag z_k^{\transpose}$. For the
local dissipative model, $z_k$ is given in Eq.~\eqref{eq:z-k}, and
$\tr(M) = \gamma$. The solution to Eq.~\eqref{eq:R-EOM} reads
\begin{equation}
  \label{eq:ret-resp-k}
  R_k(t) = \theta(t) \e^{\widetilde{x}_{-k} t},
\end{equation}
which yields the equal-position retarded response function as
\begin{equation}
  \label{eq:ret-resp-Fourier-representation}
  R_{i, i}(t) = \int_{-\pi}^{\pi} \frac{\mathrm{d} k}{2 \pi} \, R_k(t).
\end{equation}
To obtain the late-time asymptotics of the retarded response function quoted in
Sec.~\ref{sec:dynam-crit-behav} and illustrated in Fig.~\ref{fig:ret_resp}, we
evaluate the integral over quasimomenta using standard techniques of asymptotic
analysis~\cite{Bender1999}. The numerical results shown in Fig.~\ref{fig:ret_resp} result from a
direct numerical integration of Eq.~\eqref{eq:ret-resp-Fourier-representation}.

\section{Even-odd half-system-size effect}
\label{sec:even-odd-half-sys-size-effect}

In this appendix, we derive Eqs.~\eqref{eq:sgn-Pf-ss} and~\eqref{eq:sgn-Pf-0},
which describe the even-odd half-system-size effect of the fermion parity of the
entanglement ground state, and the sign of the Pfaffian.

\subsection{Even-odd effect of the fermion parity}
\label{sec:even-odd-effect-parity}

Before we confirm Eqs.~\eqref{eq:sgn-Pf-ss} and~\eqref{eq:sgn-Pf-0} numerically
for arbitrary values of $\mu_0$ and $\delta$, it is instructive to consider
certain limiting cases, in which the fermion parities $p_{A, 0}$ and
$p_{A, \mathrm{ss}}$ of the entanglement ground states of the initial and steady
states, respectively, can be determined in a straightforward manner.

For $J = \Delta > 0$ and $\mu_0 \to - \infty$, the ground state of the
Hamiltonian~\eqref{eq:H-Kitaev} of the Kitaev chain is the vacuum state
$\ket{\Omega}$, which obeys $c_i \ket{\Omega} = 0$ for all $i = 1, \dotsc, N$;
for $\mu_0 \to \infty$, the ground state is the completely filled state
$\prod_{i = 1}^N c_i^{\dagger} \ket{\Omega}$. In the lattice-site occupation
representation introduced in Sec.~\ref{sec:driv-diss-parity}, the corresponding
entanglement ground states are $\ket{\mathbf{0}}_A$ and $\ket{\mathbf{1}}_A$,
where $\mathbf{1} = \left( 1, \dotsc, 1 \right)$. For the vacuum state, the
fermion parity $p_{A, 0}$ of the entanglement ground state is even,
$p_{A, 0} = 1$.  In contrast, for the completely filled state, the parity
depends on the size of the system as $p_{A, 0} = \left( -1 \right)^{N/2}$, i.e.,
the parity is even or odd if half of the system size is even or odd,
respectively. As in the main text, we consider systems with an even number of
lattice sites, $N \in 2 \N$, such that $N/2 \in \N$ is an integer.

With regard to the steady state, we consider the following limits of the value
of $\delta = \gamma_{\mathrm{l}} - \gamma_{\mathrm{g}}$: For $\delta \to
\infty$, the steady state is the vacuum state with $p_{A, \mathrm{ss}} = 1$; for
$\delta \to - \infty$, the steady state is the completely filled state with
$p_{A, \mathrm{ss}} = \left( -1 \right)^{N/2}$.

If the difference of parities is finite, $p_{A, 0} - p_{A, \mathrm{ss}} \neq 0$,
in the course of the quench dynamics, there has to be at least one reversal of
the parity of the entanglement ground state, and, concomitantly, a zero crossing
in the single-particle entanglement spectrum. Our findings for $p_{A, 0}$ and
$p_{A, \mathrm{ss}}$, and the resulting expectations for the presence or absence
of entanglement spectrum crossings, are summarized in
Table~\ref{tab:parity-gs-ss}.

\renewcommand\arraystretch{1.2}
\begin{table}
  \centering
  \begin{tabular}{|c|c|c|}
    \hline
    \multirow{3}{*}{\diagbox[height=3.6\line]{initial}{steady}}
    & vacuum & filled \\
    & $\delta \to \infty$ & $\delta \to - \infty$ \\
    & $p_{A, \mathrm{ss}} = 1$ & $p_{A, \mathrm{ss}} = \left( -1 \right)^{N/2}$ \\
    \hline    
    vacuum & & crossings are \\
    $\mu_0 \to - \infty$ & no crossing & present/absent if \\
    $p_{A, 0} = 1$ &  & $N/2$ is odd/even \\ \hline
    filled & crossings are & \\
    $\mu_0 \to \infty$ & present/absent if & no crossing \\
    $p_{A, 0} = \left( -1 \right)^{N/2}$ & $N/2$ is odd/even & \\ \hline
  \end{tabular}
  \caption{Even-odd half-system-size effect for the parity in the initial and
    steady state, and, consequently, for the presence or absence of
    entanglement spectrum crossings. The initial and steady states are the
    vacuum state or the completely filled state, $\ket{\Omega}$ or $\prod_{i =
      1}^N c_i^{\dagger} \ket{\Omega}$, respectively, for the given limiting
    values of $\mu_0$ and $\delta$.}
  \label{tab:parity-gs-ss}
\end{table}

\subsection{Pfaffian in the initial and steady state}
\label{sec:pfaff-init-steady}

\begin{figure}
  \centering
  \includegraphics[width=\linewidth]{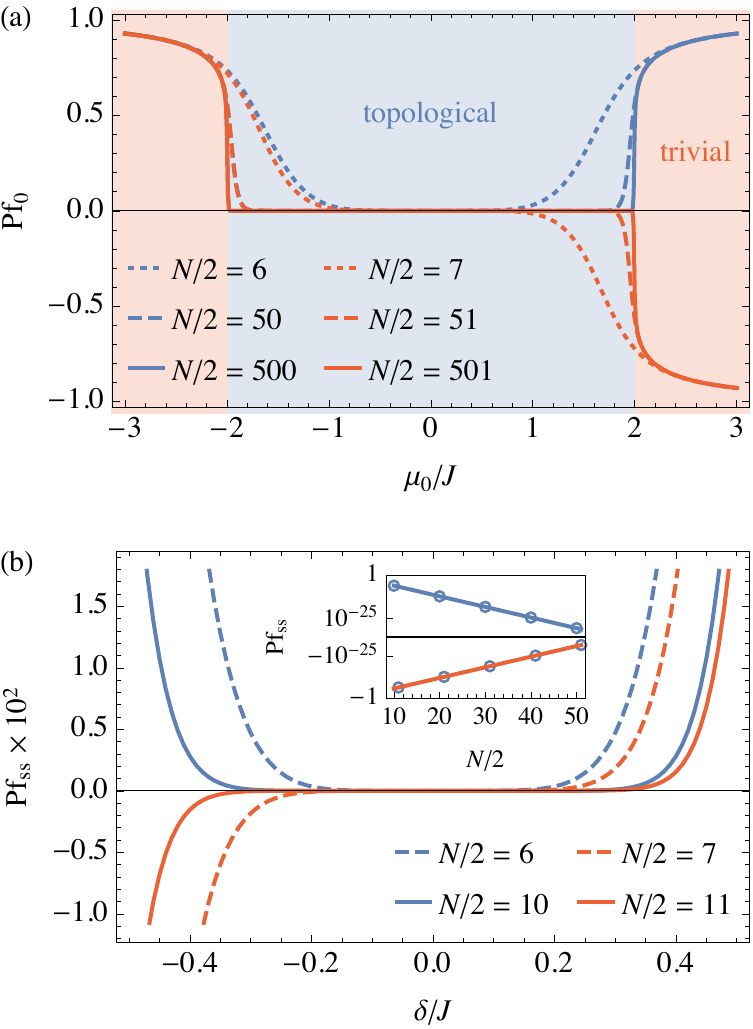}
  \caption{Even-odd effect of the Pfaffian $\mathrm{Pf}$ and the parity $p_A$ of
    the entanglement ground state. (a) The Pfaffian in the ground state
    $\mathrm{Pf}_0$ vanishes in the topological phase for $\abs{\mu_0} < 2 J$ in
    the thermodynamic limit, and the parity is undetermined. In the trivial
    phase, $\sgn(\mathrm{Pf}_0)$ obeys Eq.~\eqref{eq:sgn-Pf-0}. (b) The sign of
    the Pfaffian and the parity in the steady state with
    $\gamma_{\mathrm{l}} + \gamma_{\mathrm{g}} = J$ are given by
    Eq.~\eqref{eq:sgn-Pf-ss}. Inset: $\mathrm{Pf}_{\mathrm{ss}}$ vanishes
    exponentially for $N \to \infty$ as indicated by the linear behavior on the
    logarithmic scale. For the value $\delta = - 0.4 J$ chosen here,
    $\sgn(\mathrm{Pf}_{\mathrm{ss}})$ alternates for even and odd half-system
    sizes $N/2 = 10, 11, 20, 21, \dotsc, 50, 51$ shown in the figure. In (b),
    the chemical potential is $\mu_1 = 0$. In both panels, $J = \Delta > 0$.}
  \label{fig:Pfaffian-0-ss}
\end{figure}

We proceed to confirm numerically and for generic values of pre- and postquench
parameters, that the sign of the Pfaffian $\mathrm{Pf}$~\eqref{eq:pfaffian} of
the reduced covariance matrix of the initial and steady states exhibits the same
half-system-size dependence as the fermion parity $p_A$ of the corresponding
entanglement ground state. Indeed, in all cases we consider, we find
$\sgn(\mathrm{Pf}) = p_A$. The parity can be calculated as described in
Appendix~\ref{sec:ferm-parity-entanglement-gs}.

The even-odd half-system-size effect for the Pfaffian is illustrated in
Fig.~\ref{fig:Pfaffian-0-ss}. In particular, Fig.~\ref{fig:Pfaffian-0-ss}(a)
shows the Pfaffian of the reduced covariance matrix of the ground state of the
Kitaev chain with open boundary conditions as a function of the chemical
potential $\mu_0$. For increasing system size, the Pfaffian $\mathrm{Pf}_0$
develops a sharp drop to zero at the critical values $\mu_0 = \pm 2 J$. In the
topological phase of the Kitaev chain, for $\abs{\mu_0} < 2 J$, the Pfaffian
$\mathrm{Pf}_0$ is equal to zero, which reflects the presence of a zero-energy
edge mode at the entanglement cut. Indeed, since the square of the Pfaffian
equals the determinant of the reduced covariance matrix, $\mathrm{Pf}_0 = 0$ if
one of the single-particle entanglement eigenvalues $\xi_i$ is equal to zero. In
the trivial phase with $\abs{\mu_0} > 2 J$, the sign of $\mathrm{Pf}_0$ is given
by Eq.~\eqref{eq:sgn-Pf-0}.

In Appendix~\ref{sec:even-odd-effect-parity}, we argued that the fermion parity
of the entanglement ground state in the steady state is always positive for
$\delta \to \infty$, whereas it exhibits an even-odd half-system-size effect in
the limit $\delta \to - \infty$. As shown in Fig.~\ref{fig:Pfaffian-0-ss}(b),
the Pfaffian in the steady state exhibits an analogous even-odd half-system-size
effect also for finite values of the imbalance of dissipation rates
$\delta = \gamma_{\mathrm{l}} - \gamma_{\mathrm{g}}$. This confirms
Eq.~\eqref{eq:sgn-Pf-ss} for the values
$\gamma_{\mathrm{l}} + \gamma_{\mathrm{g}} = J$ and $\mu_1 = 0$, chosen in the
figure. We checked that Eq.~\eqref{eq:sgn-Pf-ss} holds also for nonzero values
of $\mu_1$.

\begin{figure}
  \centering  
  \includegraphics[width=\linewidth]{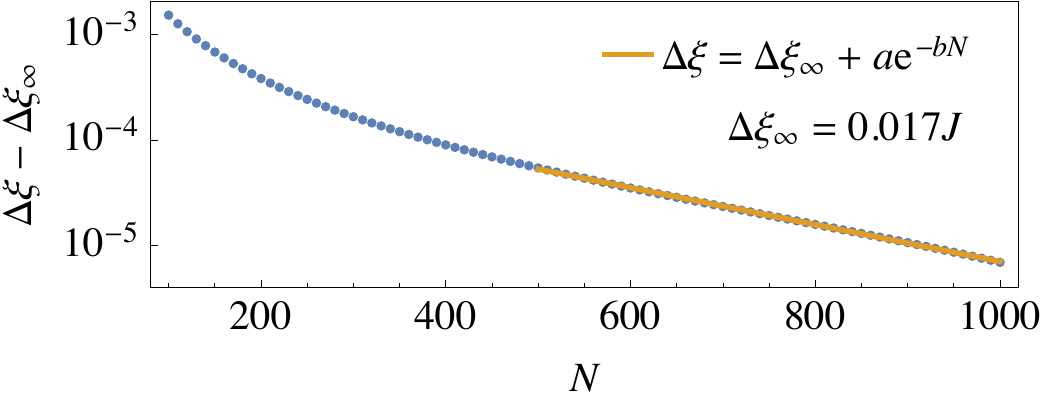}
  \caption{The gap $\Delta \xi$ of the single-particle entanglement spectrum
    approaches its thermodynamic value $\Delta \xi_{\infty}$ exponentially. The
    orange line is a fit to the numerical data shown as blue dots. Parameters
    are $\gamma_{\mathrm{l}} = 0.3 J$, $\gamma_{\mathrm{g}} = 0.7 J$, and
    $\mu_1 = 0$.}
  \label{fig:entanglement-gap}
\end{figure}

The dependence of $\mathrm{Pf}_{\mathrm{ss}}$ on the half-system size $N/2$ is
illustrated further in the inset of Fig.~\ref{fig:Pfaffian-0-ss}(b), which shows
$\mathrm{Pf}_{\mathrm{ss}}$ for $\delta = - 0.4 J$ and a series of half-system
sizes $N/2 = 10, 11, 20, 21, \dotsc, 50, 51$. While the linear behavior of
$\mathrm{Pf}_{\mathrm{ss}}$ on the logarithmic scale clearly indicates
exponential decay of the magnitude of the steady-state Pfaffian, the fermion
parity of the entanglement ground state, and, therefore, the sign of the
Pfaffian, is well-defined for arbitrarily large values of $N$. This follows from
the fact that for $\delta \neq 0$, the single-particle entanglement spectrum
retains a finite gap $\Delta \xi > 0$ for $N \to \infty$, and that a finite gap
implies a unique entanglement ground state with well-defined parity. In the
limits $\delta \to \pm \infty$, the entanglement gap is $\Delta \xi = 1$ for the
vacuum and completely filled state, respectively, as follows from the explicit
expressions
$\Gamma_A^{\mathrm{vac}} = \imag \bigoplus_{i = 1}^{N/2} \sigma_y = -
\Gamma_A^{\mathrm{filled}}$.
For generic parameter values, we present numerical evidence for a finite
entanglement gap in Fig.~\ref{fig:entanglement-gap}.

\section{Fermion parity of the entanglement ground state}
\label{sec:ferm-parity-entanglement-gs}

A numerical evaluation of Eq.~\eqref{eq:cov-mat-explicit-solution} yields the
reduced covariance matrix $\Gamma_A$ at arbitrary times $t$. Here we describe
how the parity of the corresponding entanglement ground state can be
determined. To this end, we use the following representation of the parity
operator~\eqref{eq:fermion-parity}
\begin{equation}
  \label{eq:fermion-parity-Majorana}
  P_A = \e^{\imag \pi \sum_{i = 1}^{N/2} c_i^{\dagger} c_i} = \prod_{i =
    1}^{N/2} \imag w_{2 i - 1} w_{2 i}.
\end{equation}
Our goal is to calculate the expectation value of $P_A$ in the entanglement
ground state. Since the entanglement Hamiltonian is quadratic, all of its
eigenstates are Slater determinants. Therefore, expectations values of products
of fermionic operators as in Eq.~\eqref{eq:fermion-parity-Majorana} can be found
by employing Wick's theorem, which yields~\cite{Kraus2009}
$\left\langle P_A \right\rangle = \pf(\Gamma_A^0)$, where $\Gamma_A^0$ is the
covariance matrix of the entanglement ground state, i.e., the ground state of
the entanglement Hamiltonian. The latter can be written as~\cite{Kraus2009}
\begin{equation}
  \label{eq:H-A}
  H_A = \frac{\imag}{4} \sum_{i, j = 1}^N w_i G_{i, j} w_j,
\end{equation}
where the real and antisymmetric matrix $G$ can be expressed in the form
\begin{equation}
  G = O^{\transpose} \left( \bigoplus\limits_{i = 1}^N
  \begin{pmatrix}
    0 & - \varepsilon_i \\ \varepsilon_i & 0
  \end{pmatrix}
  \right) O.
\end{equation}
Here, $O$ is the real special orthogonal matrix which brings the reduced
covariance matrix $\Gamma_A$ to its canonical block-diagonal form,
\begin{equation}
  O \Gamma_A O^{\transpose} = \bigoplus\limits_{i = 1}^N
  \begin{pmatrix}
    0 & \xi_i \\ - \xi_i & 0
  \end{pmatrix},
\end{equation}
and $\varepsilon_i = 2 \atanh(\xi_i)$. The covariance matrix $\Gamma_A^0$ of the
ground state of the entanglement Hamiltonian~\eqref{eq:H-A} can be found as
described in Appendix~\ref{sec:cov-mat-gs} for the original
Hamiltonian~\eqref{eq:H-Kitaev}, and yields the parity of the entanglement ground
state as described above.

% \bibliography{bibliography}

%Control: production of eprint (-1) disabled
%

\end{document}